\documentclass[useAMS,usenatbib,usegraphicx]{mn2e}
\usepackage[T1]{fontenc}
\usepackage[latin1]{inputenc}
\usepackage{color}
\newcounter{subfigure}

\title[IFU spectroscopy of 10 ETG nuclei: I]
  {Integral field unit spectroscopy of 10 early type galactic nuclei: I - Principal component analysis Tomography and nuclear activity}
\author[Ricci et al.]
  {T.V.~Ricci,$^1$\thanks{tvricci@iag.usp.br}
  J.E.~Steiner,$^1$ R.B.~Menezes$^1$ \\
  $^1$Instituto de Astronomia, Geof\'isica e Ci\^encias Atmosf\'ericas, Universidade de S\~ao Paulo, 05508-900, S\~ao Paulo, Brazil }
\date{Released 2002 Xxxxx XX}

\pagerange{\pageref{firstpage}--\pageref{lastpage}} \pubyear{2002}

\def\LaTeX{L\kern-.36em\raise.3ex\hbox{a}\kern-.15em
    T\kern-.1667em\lower.7ex\hbox{E}\kern-.125emX}

\begin{document}

\label{firstpage}

\maketitle

\begin{abstract}

Most massive galaxies show emission lines that can be characterized as LINERs. To what extent this emission is related to AGNs or to stellar processes is still an open question. In this paper, we analysed a sample of such galaxies to study the central region in terms of nuclear and circumnuclear emission lines, as well as the stellar component properties. For this reason, we selected 10 massive ($\sigma > $200 km $s^{-1}$) nearby (d $<$ 31 Mpc) galaxies and observed them with the IFU/GMOS (integral field unit/Gemini Multi-Object Spectrograph) spectrograph on the Gemini South Telescope. The data were analysed with principal component analysis (PCA) Tomography to assess the main properties of the objects. Two spectral regions were analysed: a yellow region (5100-5800 \AA), adequate to show the properties of the stellar component, and a red region (6250-6800 \AA), adequate to analyse the gaseous component.
We found that all objects previously known to present emission lines have a central AGN-type emitting source. They also show gaseous and stellar kinematics typical of discs. Such discs may be co-aligned (NGC 1380 and ESO 208 G-21), in counter-rotation (IC 1459 and NGC 7097) or misaligned (IC 5181 and NGC 4546). We also found one object with a gaseous disc but no stellar disc (NGC 2663), one with a stellar disc but no gaseous disc (NGC 1404), one with neither stellar nor gaseous disc (NGC 1399) and one with probably ionization cones (NGC 3136). PCA Tomography is an efficient method for detecting both the central AGN and gaseous and stellar discs. In the two cases (NGC 1399 and NGC 1404) in which no lines were previously reported, we found no evidence of either nuclear or circumnuclear emission, using PCA Tomography only.

\end{abstract}


\section{Introduction} \label{sec:intro}

Active galactic nuclei (AGNs) are commonly associated with the capture of matter by a supermassive black hole (SMBH), located at the central region of galaxies \citep{1998AJ....115.2285M}. In the optical, they are characterized by intense and broad emission lines. Quasars are the most luminous type of AGNs, both in optical and in radio. On the other hand, Seyfert galaxies, with prominence of high ionization lines, and low ionization nuclear emission regions (LINERs), whose low ionization emission lines tend to be more intense, belong to the low luminosity regime of AGNs.

Low luminosity AGNs (LLAGNs) are found in $\sim$ 1/3 of the galaxies in the local Universe \citep{2008ARA&A..46..475H}. If one considers only early-type galaxies (ETGs), this ratio raises to 2/3, whereby most AGNs are classified as LINERs \citep{2008ARA&A..46..475H}. Seyferts and transition objects (TOs) are predominantly found in late-type galaxies \citep{2008ARA&A..46..475H}. TOs were proposed by \citet{1993ApJ...417...63H} as AGNs whose light is contaminated by H II regions. \citet{2003ApJ...583..159H} showed that TO hosting galaxies have higher inclination than those containing LINERs or Seyferts, which increases the contamination of H II regions in the nuclear spectrum, apart from having higher FIR luminosities, which is related to a more intense stellar formation. 

LINERs were originally defined by \citet{1980A&A....87..152H} as objects with [O II]$\lambda$3727/[O III]$\lambda$5007 $>$ 1 and [O I]$\lambda$6300/[O III]$\lambda$5007 $>$ 1/3. However, because of the difficulty of observing a large spectral range and, mainly, the high extinction of the [O II]$\lambda$3727 line, \citet{1987ApJS...63..295V} proposed the line ratios [O III]$\lambda$5007/H$\beta$ $<$ 3, [O I]$\lambda$6300/H$\alpha$ $>$ 0.05, ([S II]$\lambda$6716 + [S II]$\lambda$6731)/H$\alpha$ $>$ 0.4 and [N II]$\lambda$6583/H$\alpha$ $>$ 0.5 as a better way to separate LINERs from Seyferts and H II regions. The main reason to use these line ratios is that they are reasonably insensitive to reddening effects. Bi-dimensional plots that compare [O I]$\lambda$6300/H$\alpha$, ([S II]$\lambda$6716 + [S II]$\lambda$6731)/H$\alpha$ or [N II]$\lambda$6583/H$\alpha$ with [O III]$\lambda$5007/H$\beta$ are known as diagnostic diagrams, or BPT diagrams \citep{1981PASP...93....5B}. The [O I]$\lambda$6300 line is emitted in partially ionized zones, which are produced by high-energy photons and are larger in regions photoionized by AGNs compared to those photoionized by starbursts. [S II]$\lambda\lambda$6716, 6731 lines also show considerable emission from these zones. The reasons why [N II]/H$\alpha$ are greater in galaxies with an AGN are more complex, although a fraction of the [N II]$\lambda$6548, 6583 doublet also comes from the partially ionized zone (see \citealt{1987ApJS...63..295V}; \citealt{2006agna.book.....O} for a revision of the origin of each previously discussed line ). In short, BPT diagrams are very useful in distinguishing Seyferts, LINERs and H II regions.  

Several mechanisms have been proposed to explain the LINER emission. \citet{1980A&A....87..152H} associated LINER spectra with shock waves. \citet{1983ApJ...264..105F} and \citet{1983ApJ...269L..37H} showed that typical LINER line ratios may be produced by AGNs with a low-ionization parameter (log U $\sim$ -3.5 - \citealt{1983ApJ...264..105F}). The detection rates of X-rays and radio cores, typical features of AGNs, in LINER and Seyfert nuclei are quite similar \citep{2008ARA&A..46..475H}. In the optical, the detection of a broad line component in LINERs \citep{1997ApJS..112..391H} also supports the existence of an AGN in these objects. Aside from AGNs and shockwaves, instantaneous starburst models, with ages between 3 and 5 Myr and a UV continuum dominated by Wolf Rayet clusters, were proposed by \citet{2000PASP..112..753B} to explain TOs nuclei. Photoionization by old stellar populations, more specifically post asymptotic giant branch (pAGBs) stars, was proposed by \citet{1994A&A...292...13B} to account for ionized gas emission in elliptical galaxies. 

LINERs may also be divided in types 1 and 2, like Seyfert nuclei, within the framework of a unified model for AGN activity \citep{1985ApJ...297..621A,1993ARA&A..31..473A}. An example of a LINER with a broad component in the H$\alpha$ line is NGC 7213 \citep{1979ApJ...227L.121P,1984ApJ...285..458F}. At first sight, there are no reasons to doubt that the unified model could also be applied to LINERs. In fact, \citet{1999ApJ...525..673B} detected a hidden broad line region (BLR) in the LINER of NGC 1052. \citet{1997ApJS..112..391H} have shown that $\sim$ 20\% of the LLAGNs detected in their survey are type 1, with more than half belonging to the LINER category. Sometimes, the broad H$\alpha$ component is not detected in the first attempt, like in M 104. However, \textit{Hubble Space Telescope (HST)} observations \citep{1996ApJ...473L..91K}, as well as careful analysis with high-signal-to-noise spectra \citep{2041-8205-765-2-L40}, do reveal such a feature. However, some type 2 LINERs do not seem to possess BLRs \citep{2008ARA&A..46..475H}. This may be related to the fact that some LINERs are not photoionized by an AGN. Nevertheless, in some cases where an AGN is confirmed by X-ray or radio emission, a BLR is not detected directly or with the use of spectropolarimetry \citep{2008ARA&A..46..475H}. Indeed, BLR formation models proposed in the literature are based on the condensation of clouds caused by winds originated in the accretion discs of the SMBHs \citep{2000ApJ...530L..65N,2006ApJ...648L.101E,2011A&A...525L...8C}. Depending on the model, the formation of those structures may be controlled either by the bolometric luminosity or by the Eddington ratio of the AGNs. Both parameters are lower in LINERs when compared to Seyfert nuclei \citep{2008ARA&A..46..475H}. Hence, some AGN photoionized LINERs just may not be able to produce BLR, i.e. they are genuine type 2 LINERs. 

Nuclear emission lines are often seen in elliptical and lenticular galaxies. \citet{1986AJ.....91.1062P}, using long-slit spectra, detected an extended LINER-type emission in $\sim$ 12\% objects from a sample of 203 ETGs. In all cases, these researchers concluded that the extended region has a disc-like rotation. In a sample of 26 ETGs observed with a long-slit spectrograph, \citet{1989ApJ...346..653K} also suggested that the extended emission of ionized gas has a disc-like geometry, with line ratios typical of LINERs. \citet{2006MNRAS.366.1151S}, in a sample of 48 ETGs observed with the SAURON integral field spectrograph, revealed that a fraction of these objects have a circular disc of gas, and that there is a second group characterized by an integral-sign pattern in the ionized gas distribution, aside from twists in the velocity maps. In the second group, the photometry and kinematics of every object is misaligned, which may indicate that the twist in gas distribution is a consequence of non-axisymmetric potentials \citep{2006MNRAS.366.1151S}. Besides gas kinematics, \citet{1989ApJ...346..653K} also revealed that the H$\alpha$ luminosity did not match X-ray emission in galaxies from which he had both kinds of information. He argued that, aside from the mechanism that produced the X-ray in his sample of ETGs (he associated this emission with cooling flow effects), another photoionization source would be necessary to explain the extended emission in those objects. In other words, if these sources follow a power-law spectra originated at the centre (AGN case), the ionizing photons would have to infiltrate in distances of an order of kpc. Recent works have revealed that photoionization caused only by an AGN is not enough to account for the extended emission of LINERs \citep{2010ApJ...711..796E,2010MNRAS.402.2187S,2010A&A...519A..40A,2012ApJ...747...61Y}. In fact, AGNs' contribution seems to be relevant only on nuclear scales \citep{2010ApJ...711..796E,2010MNRAS.402.2187S}. Several authors (e.g. \citealt{2008ARA&A..46..475H,2008MNRAS.391L..29S,2010ApJ...711..796E,2010MNRAS.402.2187S,2011MNRAS.413.1687C}) recently provided various arguments supporting the initial idea of \citet{1994A&A...292...13B} that this additional source of ionizing photons is pAGBs stars, which are abundant in the passively evolving stellar systems such as ETGs. For instance, \citet{2011MNRAS.413.1687C} showed that the number of ionizing photons is proportional to the stellar continuum in the H$\alpha$ region (constant H$\alpha$ equivalent width) and, with a sample of 700,000 galaxies from the Sloan Digital Sky Survey (SDSS), they proposed a diagram that compares the line ratio [N II]/H$\alpha$ to the equivalent width of H$\alpha$ in order to distinguish regions photoionized by an AGN from those whose emission is accounted for old stellar populations. In this diagram, denominated WHAN by \citet{2011MNRAS.413.1687C}, objects with central gas emission showing $EW(H\alpha)$ $<$ 3\AA\ are mainly photoionized by pAGBs stars and were defined as retired galaxies. Indeed, most retired galaxies from their sample have $EW(H\alpha)$ $\sim$ 1\AA. 

This is the first of a series of papers whose goal is the detection and characterization of the nuclear and circumnuclear (scales of $\sim$ 100 pc) gas components from a sample of 10 ETGs, observed with the integral field spectrograph located at the Gemini South Telescope. With this instrument, one is able to obtain two-dimensional spectra from the central regions of these galaxies within a field of view (FOV) of 3.5 arcsec x 5 arcsec and a spatial resolution $<\sim$ 1 arcsec. Although the SAURON project \citep{2002MNRAS.329..513D} corresponds to a representative sample of ETGs with a larger FOV, which allows information in scales of an order of kpc, their spatial resolution is $\sim$ 3 arcsec, whereas our sample was obtained with a spatial resolution good enough to study nuclear and circumnuclear regions of ETGs in great detail. Our sample, even if not statistically complete, covers a wide range of ETGs, from low ellipticity and low stellar rotation ellipticals to fast rotator lenticular galaxies. 

In this work, we show that nuclear and circumnuclear regions of gas may be detected through the principal component analysis (PCA) Tomography technique (\citealt{1997ApJ...475..173H,2009MNRAS.395...64S}, see a brief review in Section \ref{pca_tomography}). In the following papers, we will validate the results from this work as they relate to the nuclear (Paper II) and circumnuclear (Paper III) regions, apart from additional results associated with these regions. This will be done using gas-only spectra (i.e. with the starlight component properly subtracted from each spectrum of the data cubes using stellar population synthesis techniques) of the central regions of the galaxies.

\section{Galaxy sample}

\begin{figure*}
\begin{center}
\includegraphics[scale=0.40]{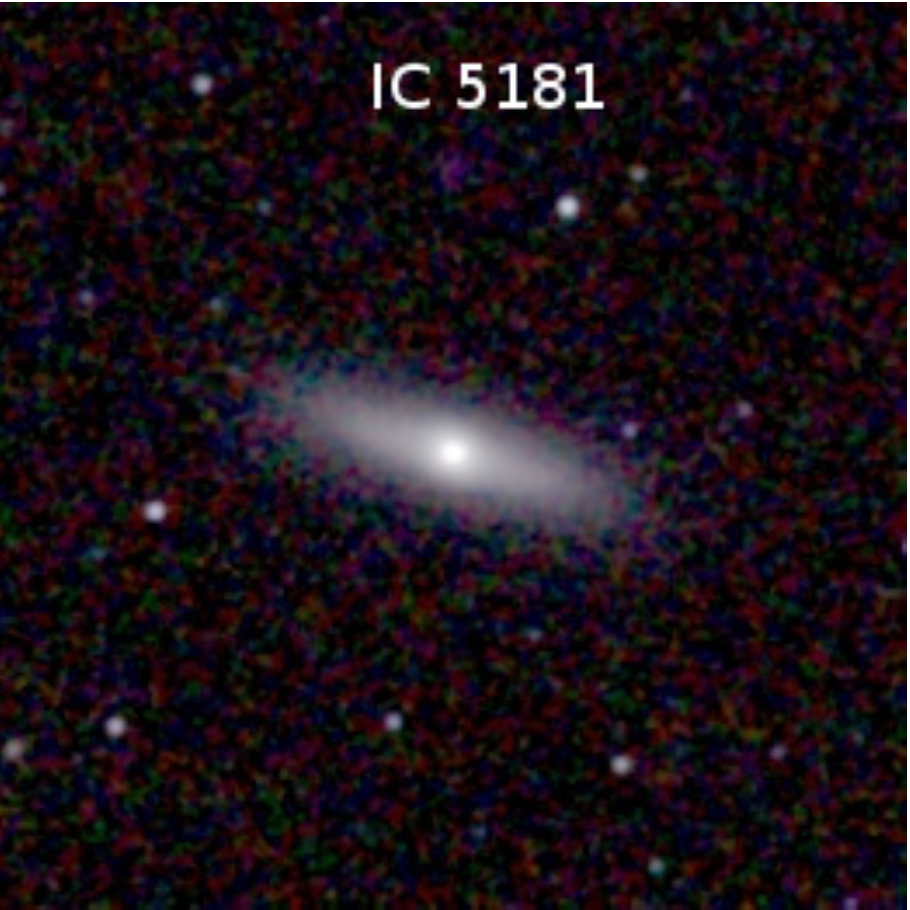}
\hspace{-0.15cm}
\vspace{0.3cm}
\includegraphics[scale=0.345]{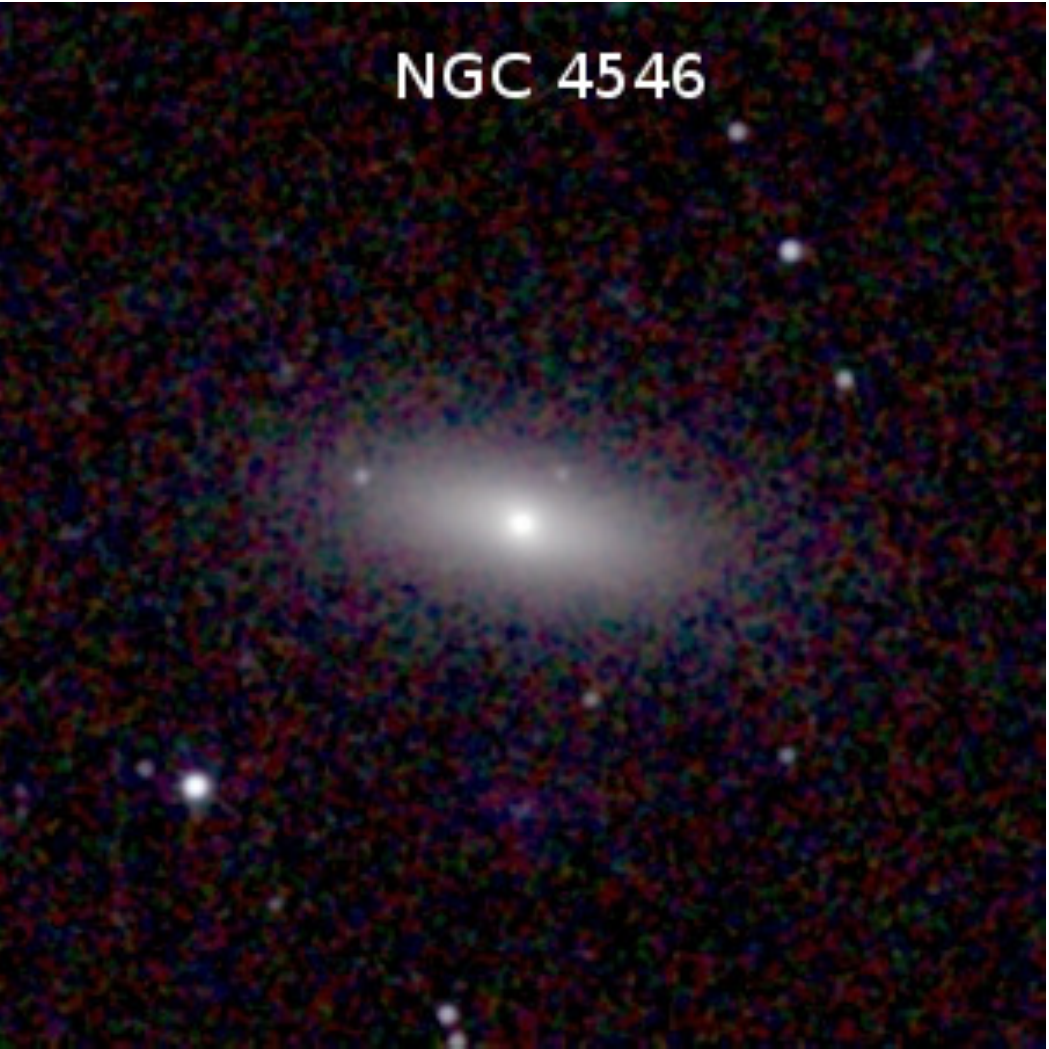}
\includegraphics[scale=0.43]{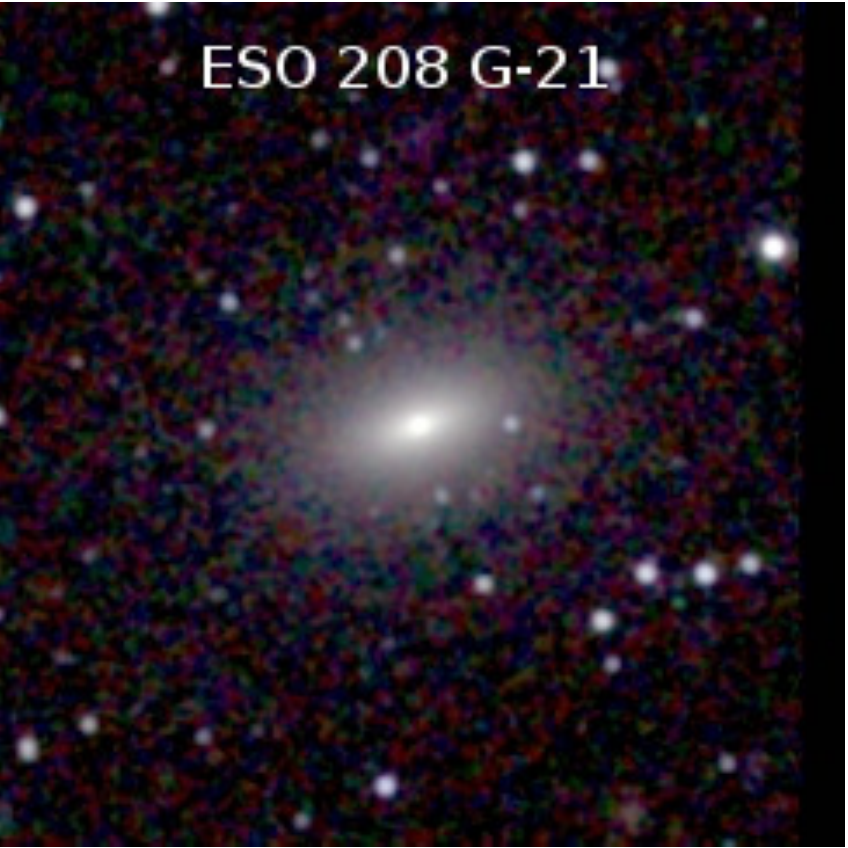}
\hspace{-0.15cm}
\vspace{0.3cm}
\includegraphics[scale=0.28]{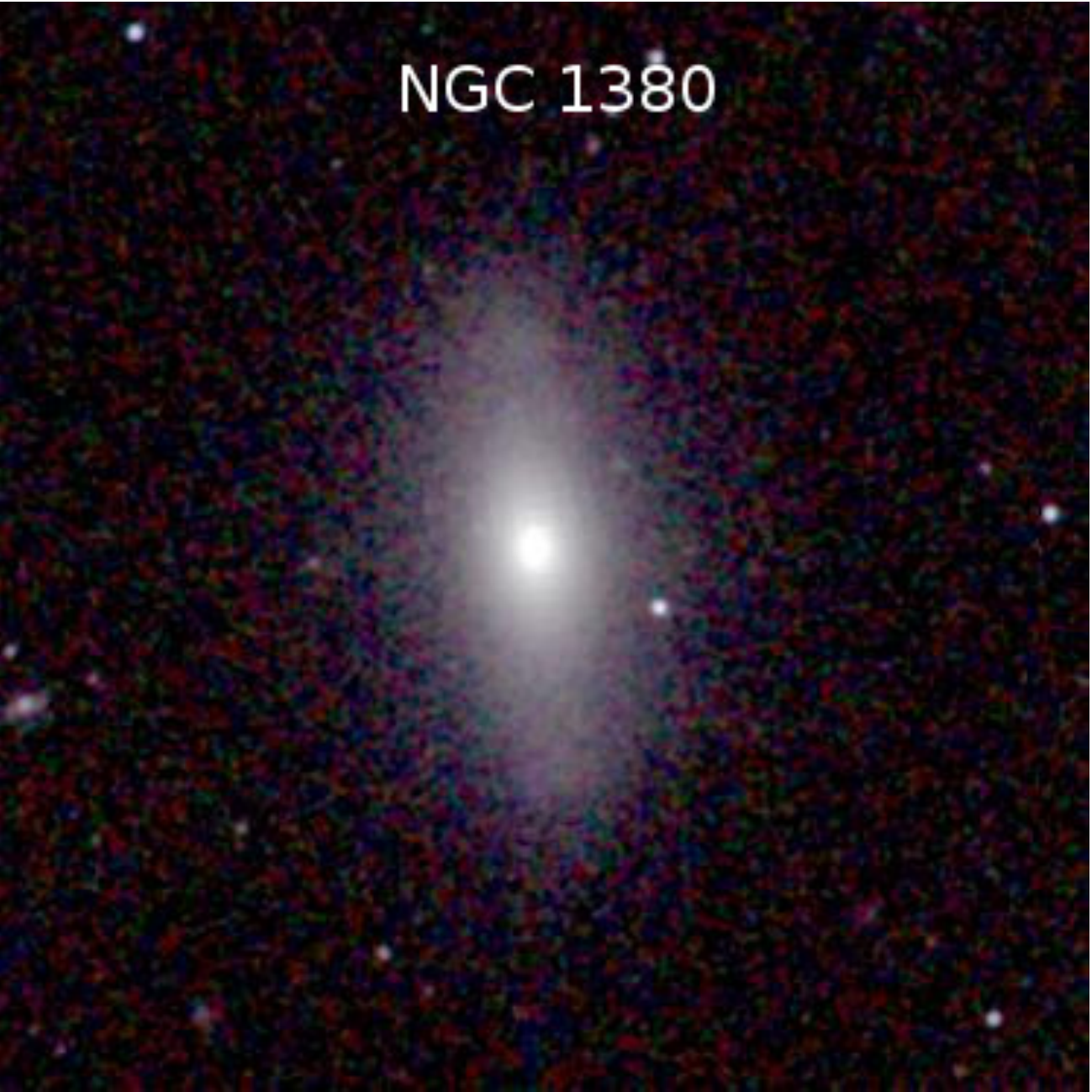}
\includegraphics[scale=0.16]{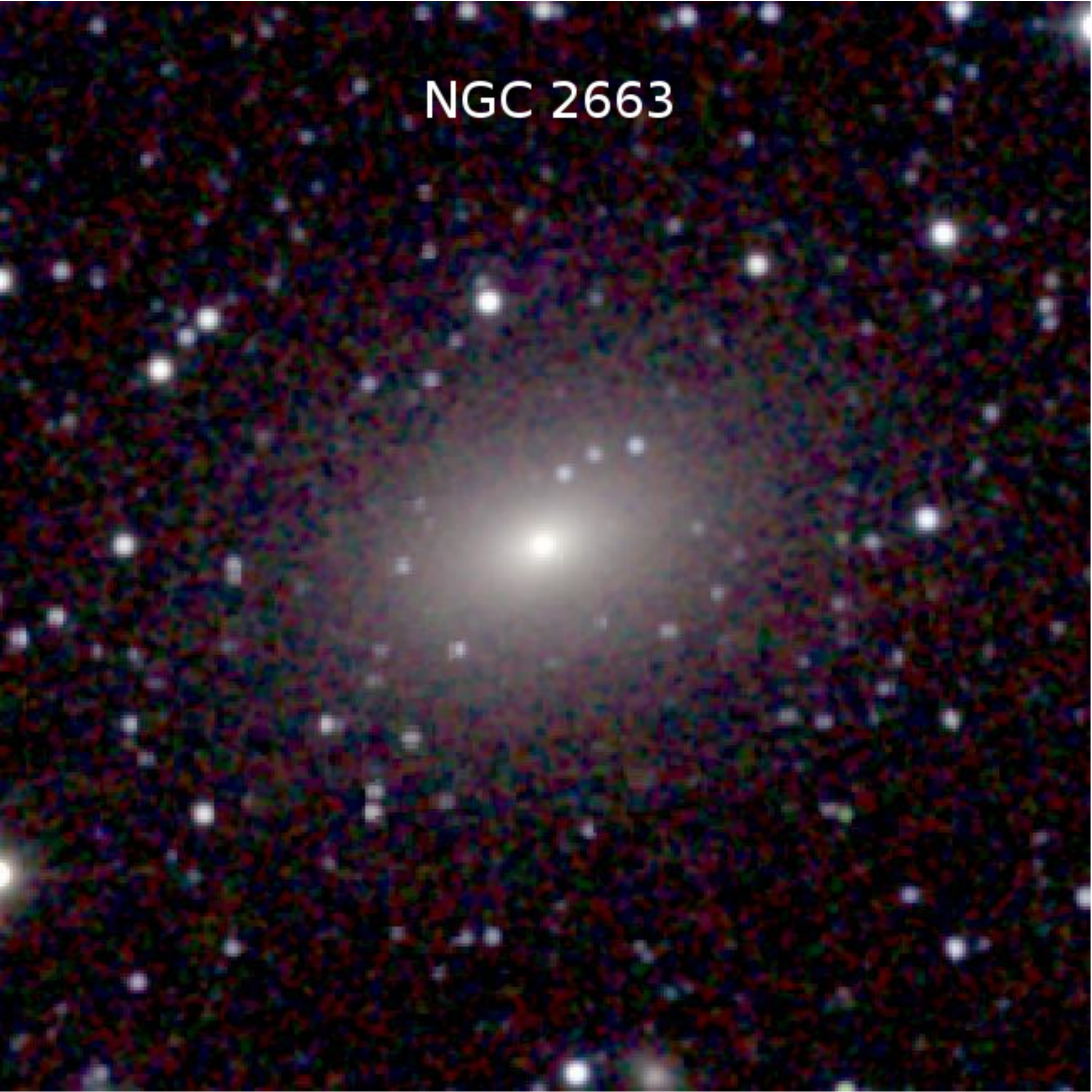}
\hspace{-0.15cm}
\vspace{0.3cm}
\includegraphics[scale=0.31]{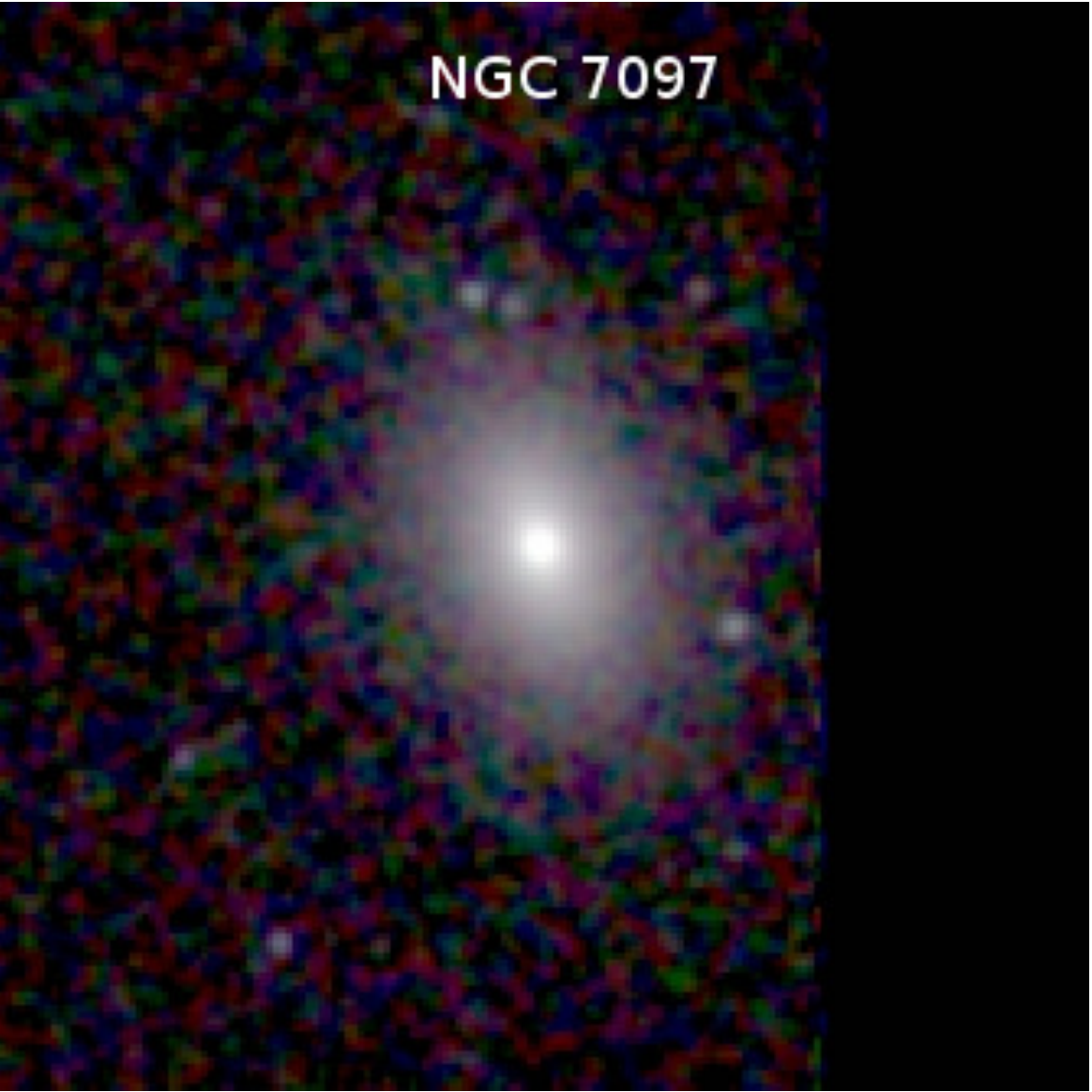}
\includegraphics[scale=0.275]{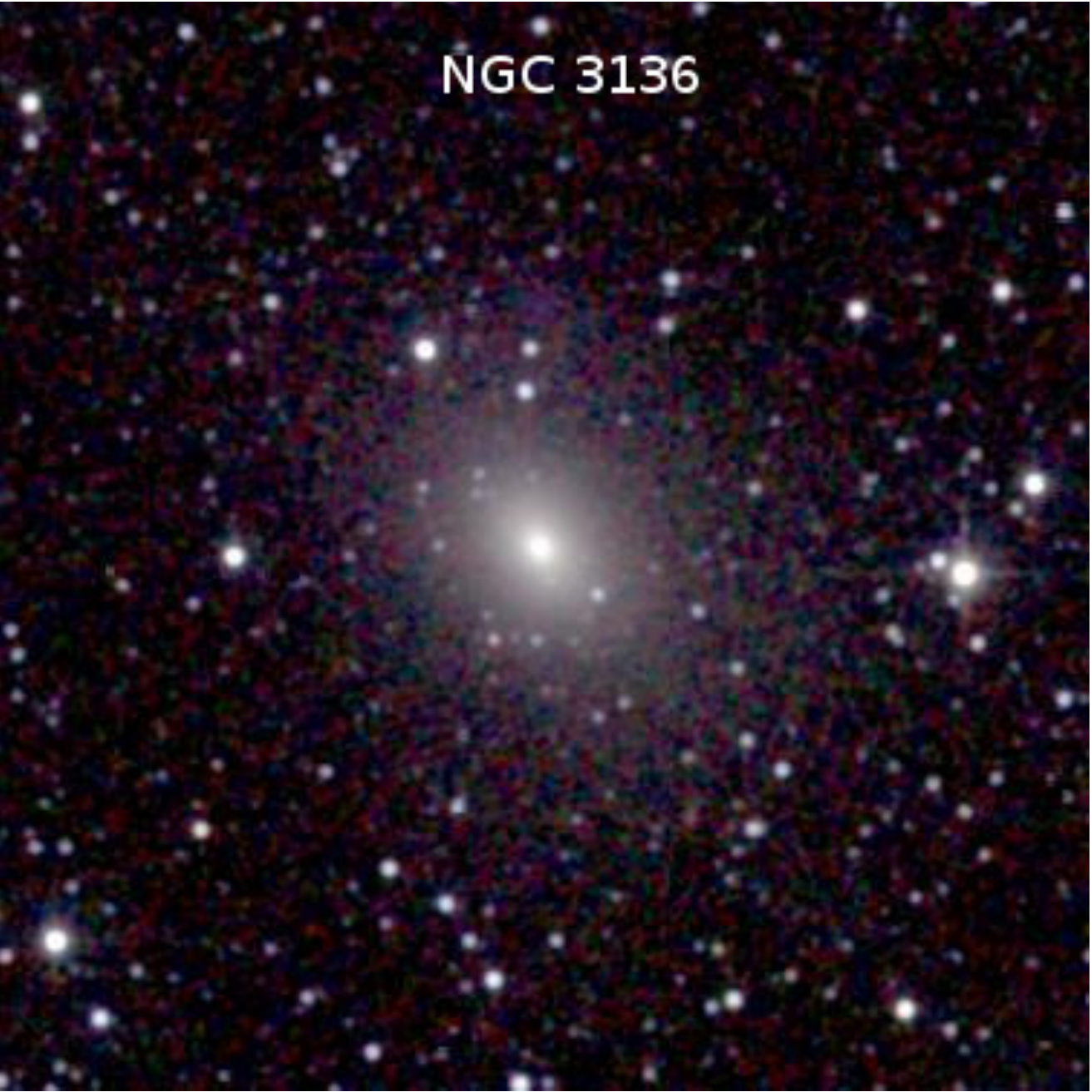}
\includegraphics[scale=0.255]{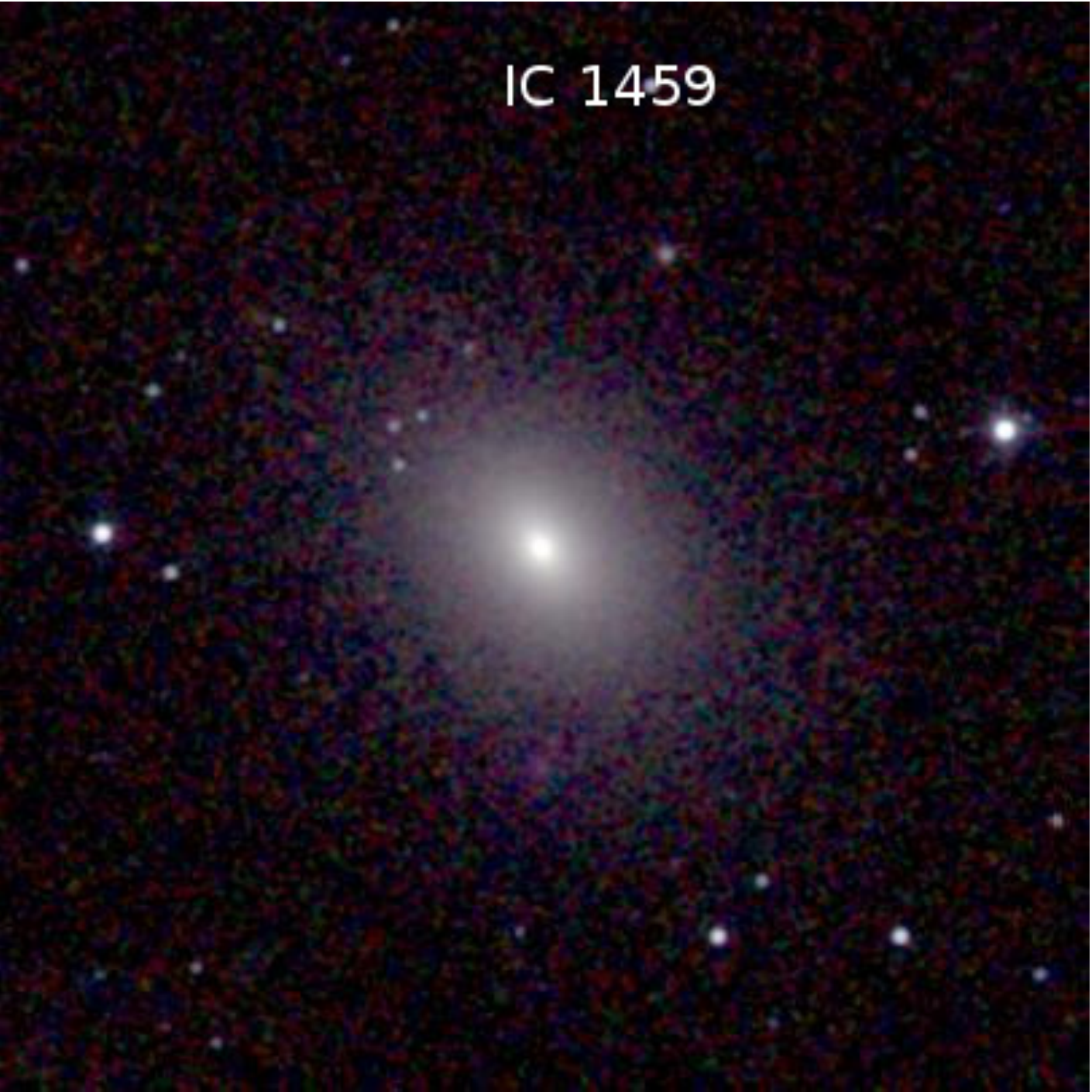}
\hspace{-0.15cm}
\vspace{0.3cm}
\includegraphics[scale=0.18]{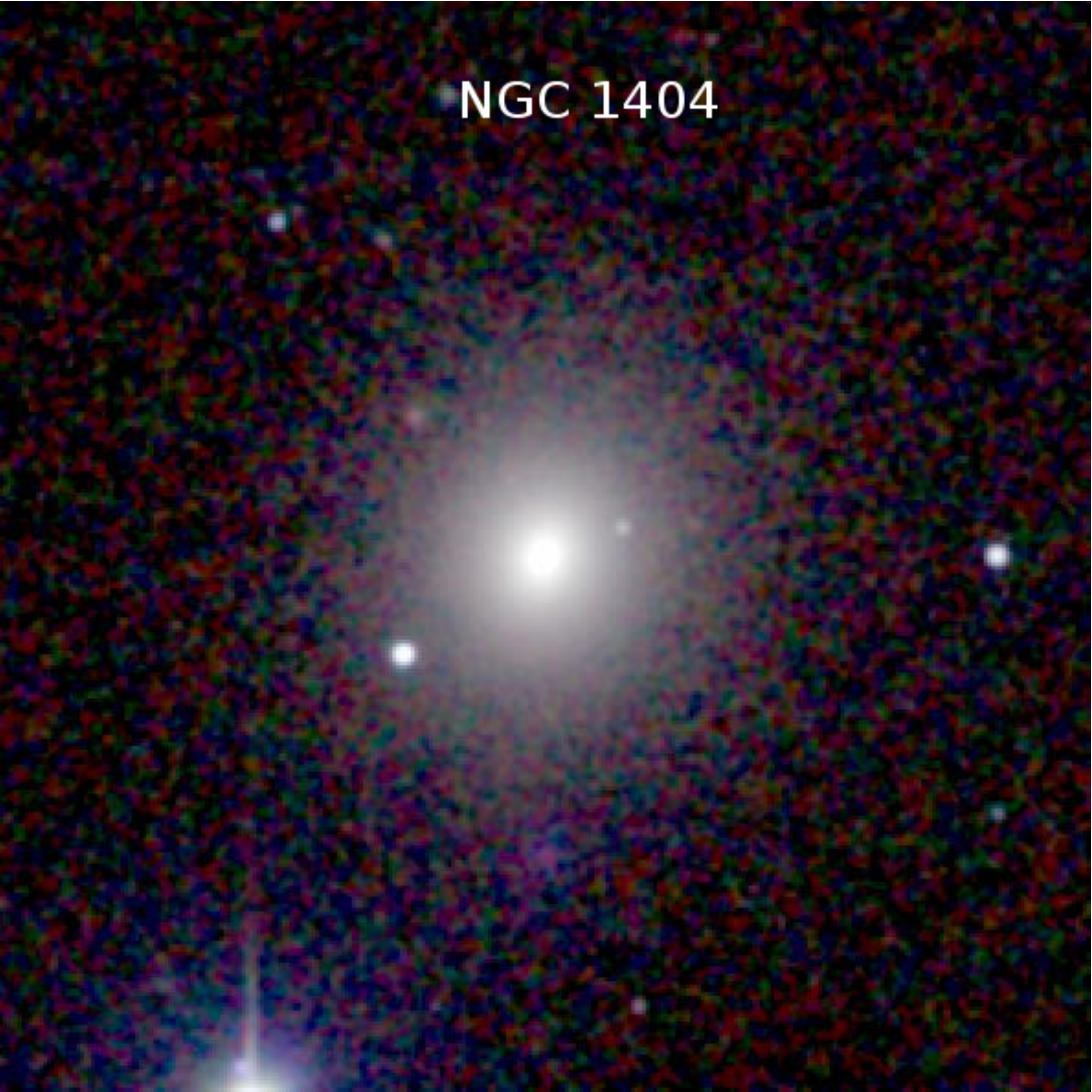}
\includegraphics[scale=0.235]{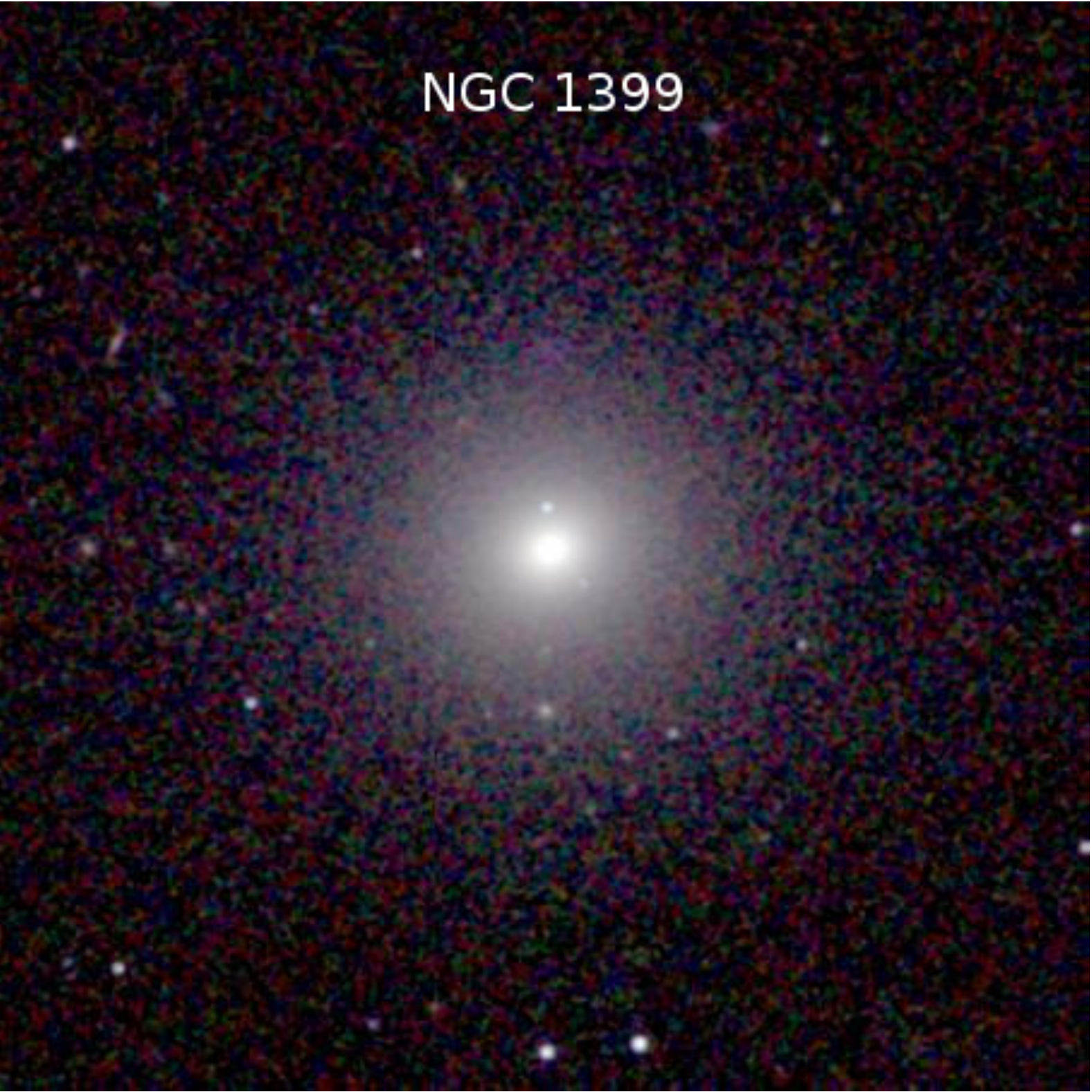}

\end{center}
\caption{The sample of galaxies studied in this work (\textit{JHK}-2MASS).  \label{jhk_images}
}
\end{figure*}

The detection of LLAGNs is favoured in nearby galaxies because of their low luminosity. Besides, one is more likely to find LINERs in ETGs with high central stellar velocity dispersion \citep{1986AJ.....91.1062P,2010A&A...519A..40A}. Thereby, we selected a sample of 10 ETGs from the Southern hemisphere, shown in table \ref{sample}\footnote{The NASA/IPAC Extragalactic Database (NED) is operated by the Jet Propulsion Laboratory, California Institute of Technology, under contract with the National Aeronautics and Space Administration.} and in Figs. \ref{jhk_images}, \ref{HST_images_I} and \ref{HST_images_not_I}, with distances of up to 31 Mpc and stellar velocity dispersions $>$ 200 km s$^{-1}$. 

The sample was basically selected from a survey of H$\alpha$ and [N II] emission in southern galaxies by \citet{1986AJ.....91.1062P}. We selected nearby (d $\leq$ 30 Mpc) galaxies with broad [N II] (FWHM $>$ 400 km s$^{-1}$) emission. Two galaxies (NGC 1399 and NGC 1404) were selected for not having emission lines detected by \citet{1986AJ.....91.1062P}. The stellar velocity dispersion values were obtained from Hyperleda\footnote{http://leda.univ-lyon1.fr} \citep{2003A&A...412...45P} and turned out to be all larger than 200 km s$^{-1}$; this is probably related to the broad [N II] emission.

In this sample, one galaxy has a previously known LINER-like AGN (IC 1459 - see section \ref{case_IC1459} for a review). Moreover, emission lines were detected in seven additional galaxies, although they were not related to AGNs \citep{1986AJ.....91.1062P,1987ApJ...318..531G}. In the other two (NGC 1399 and NGC 1404), emission lines were not previously reported in the literature. Since this sample is statistically incomplete, one is not able to extract statistical information about massive ETGs (e.g. detection rates of AGNs in these objects), as the choice of the sample may be biased. 

\renewcommand{\thefigure}{\arabic{figure}\alph{subfigure}}
\setcounter{subfigure}{1}

\begin{figure*}
\begin{center}
\includegraphics[scale=0.80]{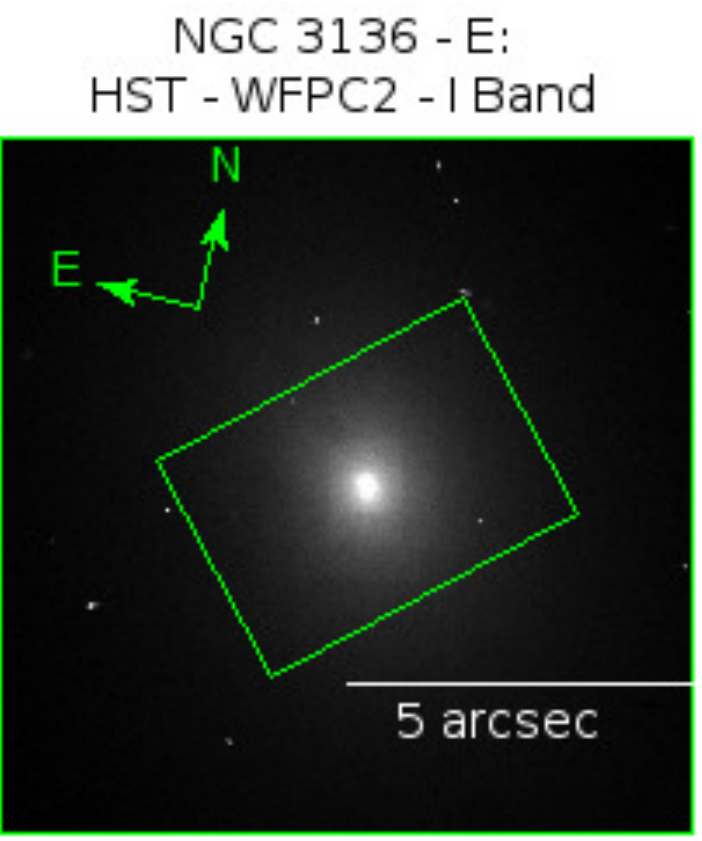}
\hspace{-0.15cm}
\vspace{0.3cm}
\includegraphics[scale=0.80]{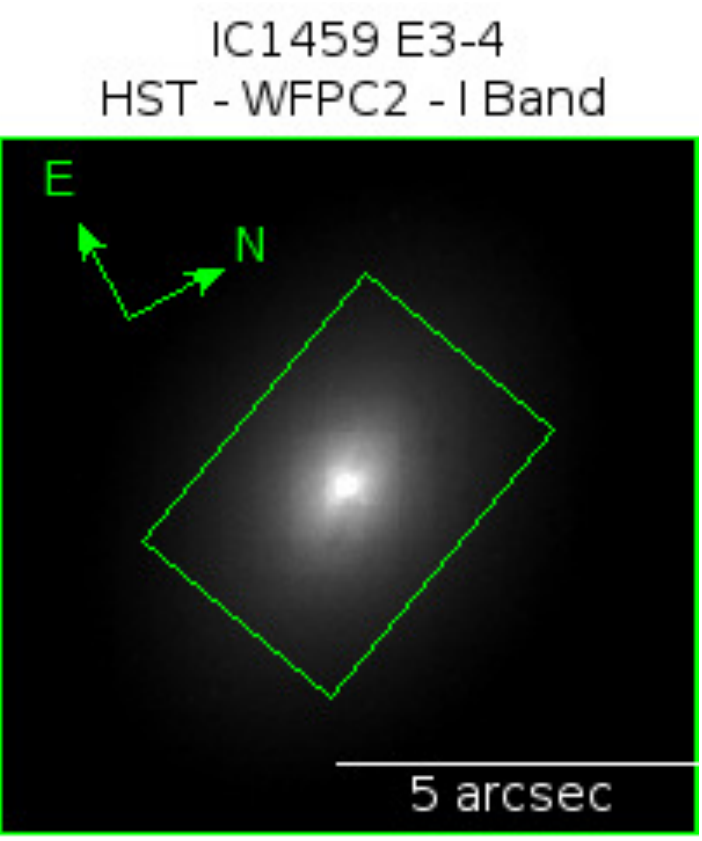}
\includegraphics[scale=0.80]{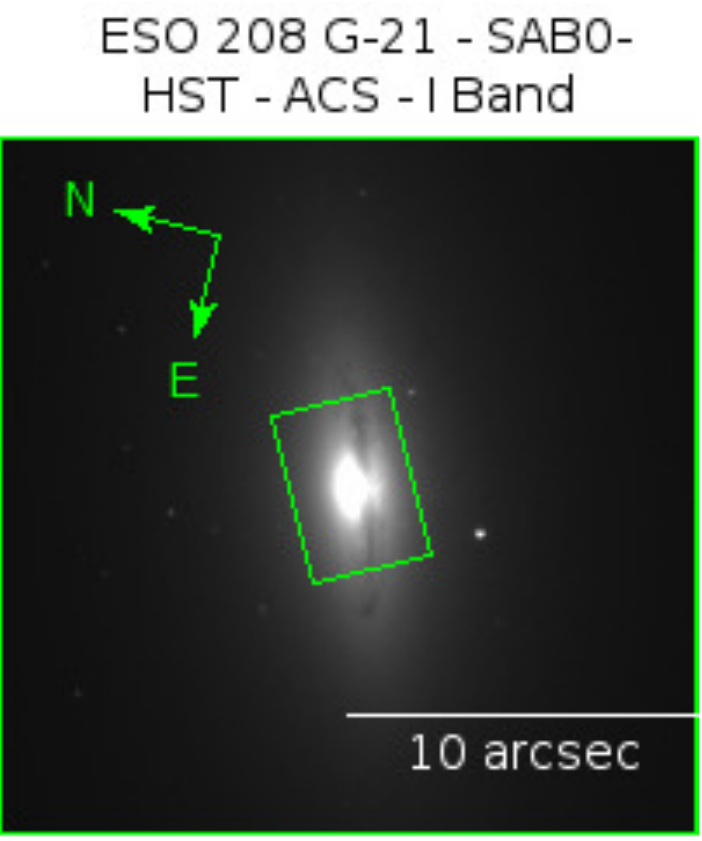}
\hspace{-0.15cm}
\vspace{0.3cm}
\includegraphics[scale=0.80]{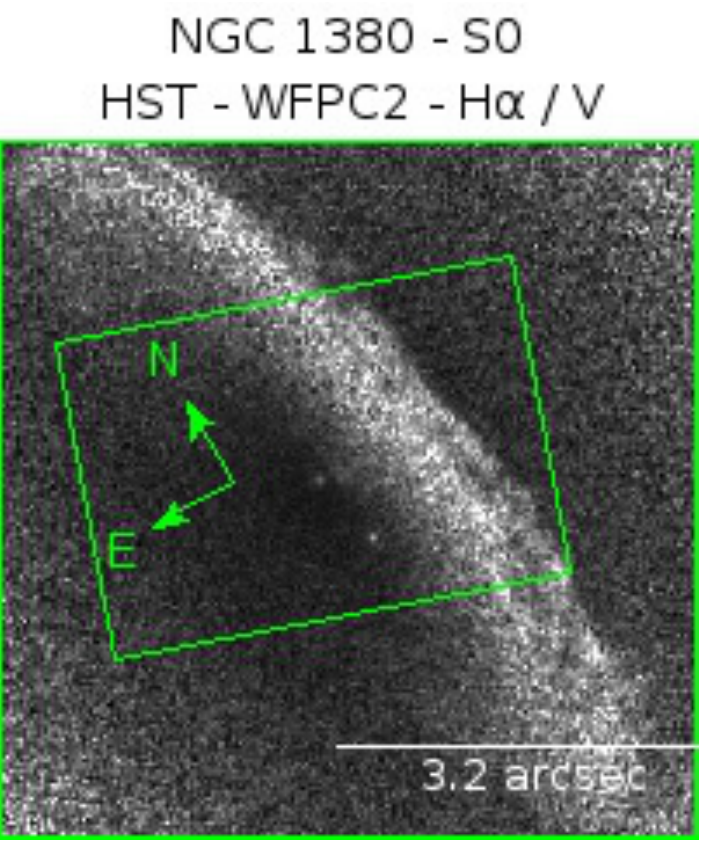}
\includegraphics[scale=0.80]{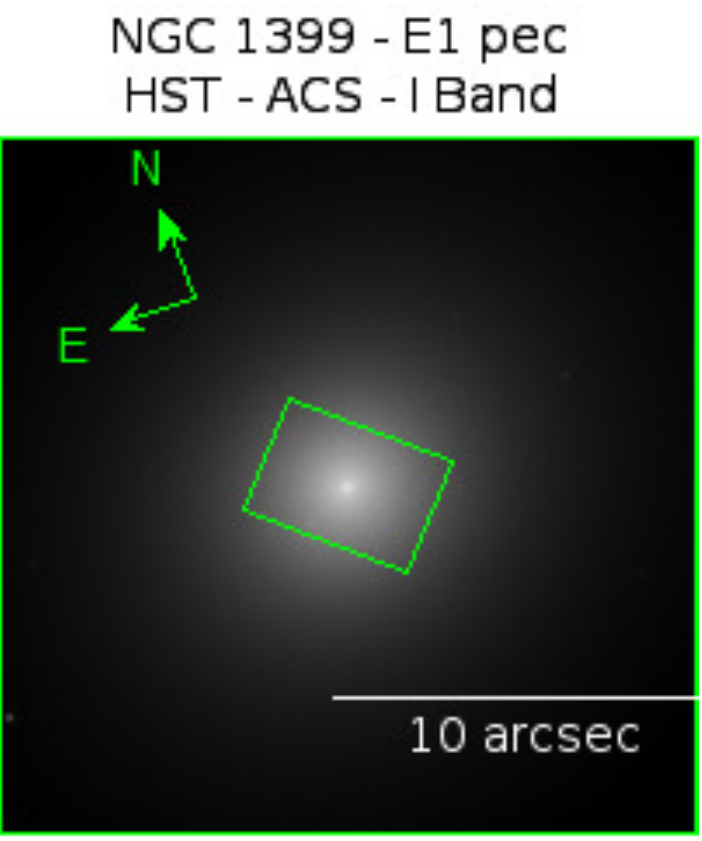}
\hspace{-0.15cm}
\vspace{0.3cm}
\includegraphics[scale=0.80]{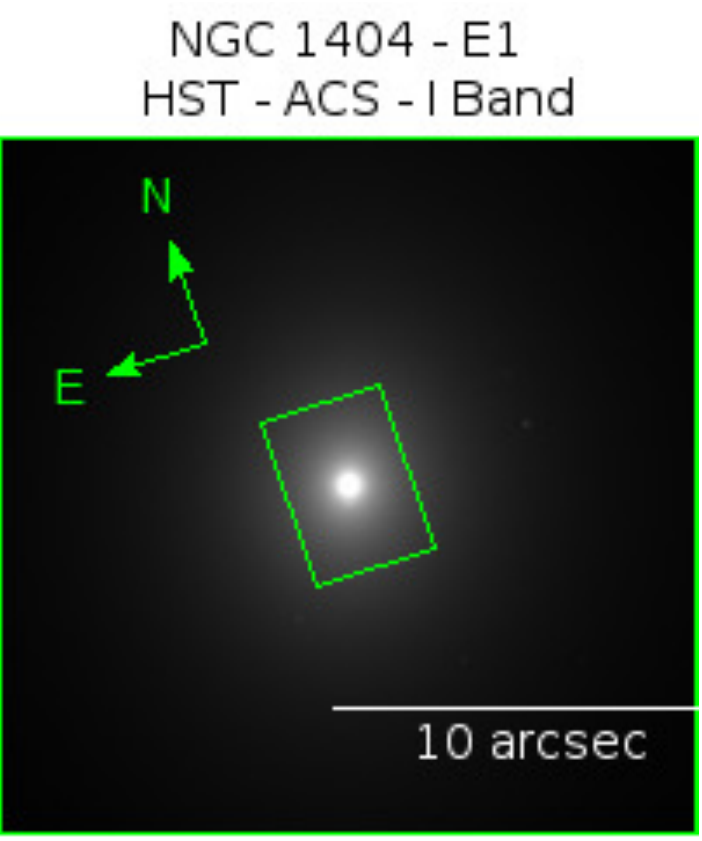}
\end{center}
\caption{Images from the \textit{HST} data archive. The rectangle indicates the FOV of the IFU observation, whose spatial dimensions are 3.5 arcsec x 5 arcsec.\label{HST_images_I}
}
\end{figure*}

\addtocounter{figure}{-1}
\addtocounter{subfigure}{1}

\begin{figure*}
\begin{center}
\includegraphics[scale=0.80]{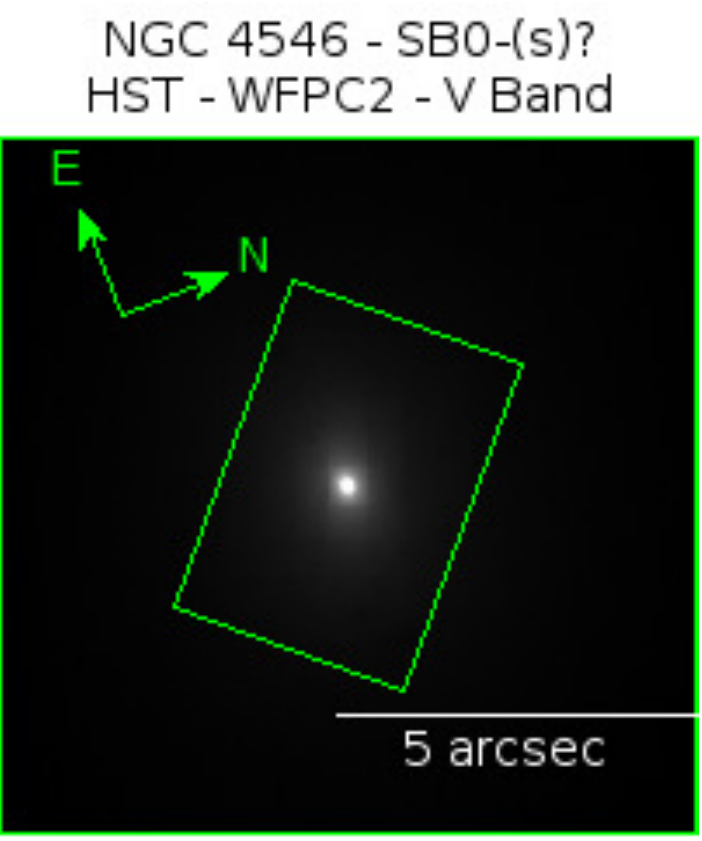}
\hspace{-0.15cm}
\vspace{0.3cm}
\includegraphics[scale=0.80]{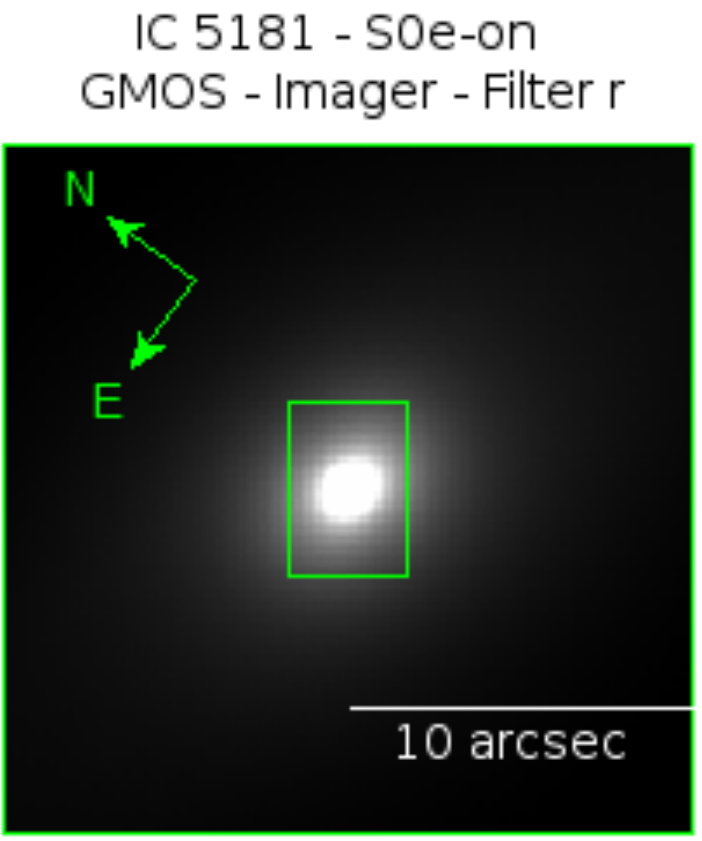}
\includegraphics[scale=0.80]{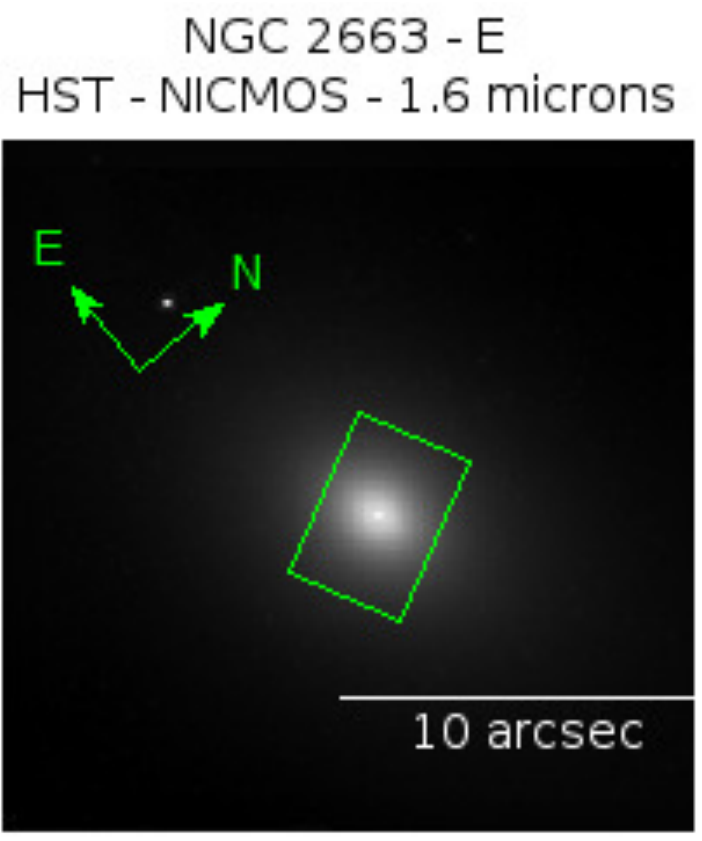}
\hspace{-0.15cm}
\vspace{0.3cm}
\includegraphics[scale=0.80]{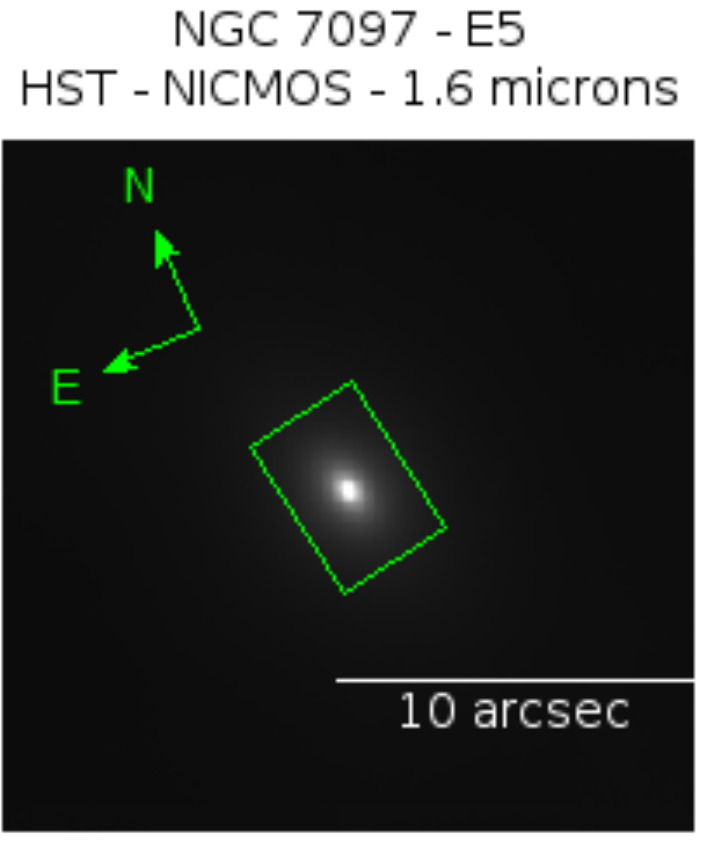}
\end{center}
\caption{Images from the \textit{HST} data archive, except IC 5181, which was obtained with the GMOS imager located at the Gemini South Telescope. \label{HST_images_not_I}
}
\end{figure*}

\renewcommand{\thefigure}{\arabic{figure}}

\begin{table*}
\scriptsize
\begin{center}
\begin{tabular}{lccccccccc} 
\hline
Galaxy name &  $V_{hel}$ & Morphology & $T$ & $R_e$ & $M_K$& $n_s$ & $\epsilon$ & $d$ & EL \\
&(km s$^{-1})$&&&arcsec&&&&(Mpc) & \\
&(1)&(2)&(3)&(4)&(5)& (6) &(7) &(8) &(9) \\
\hline
NGC 1399 &  1425& E1 pec&-5.0&40&-24.95& 4.9& 0.00&18$\pm$2 &No \\
IC 1459 & 1802&E3-4&-5.0&34&-25.23&4.9 & 0.21&27$\pm$3& Yes\\
NGC 2663&  2102&E &-4.6&55&-25.24& 5.2&0.27 &27$\pm$3&Yes \\
NGC 7097&  2616&E5&-5.0&18&-23.70& 4.0&0.34 &31$\pm$1&Yes \\
NGC 3136&  1713&E:&-5.0&37&-24.59& 4.9&0.26 &24$\pm$2&Yes\\
NGC 1404&  1947&E1&-5.0&24&-24.49& 3.7&0.10 &19$\pm$2&No\\
IC 5181& 1987&S0e-on&-2.0&11&-23.76& 4.6&0.68 &25$\pm$11&Yes\\
NGC 4546&  1050&SB0-(s)?&-3.0&27&-23.71&3.6 &0.54 &17$\pm$4&Yes$^*$\\
NGC 1380&  1877&S0&-2.0&40&-24.32& 3.4&0.56 &18$\pm$3&Yes\\
ESO 208 G-21&  1085&SAB0-&-3.0&26&-23.18&3.6 &0.34 &17$\pm$3&Yes\\
\hline
\end{tabular}
\caption{Overview of the observed sample. (1) - Heliocentric radial velocity (km s$^{-1}$), (2) - Morphology, (3) - Hubble type, (4) - effective radius  in arcsec, (5) \textit{K} band absolute magnitude, (6) - S\'ersic index in \textit{B} band, (7) ellipticity, (8) - distance (Mpc) and (9) emission lines previously known in literature. References for collumns: (1) - NED; (2), (3) and (4) - \citet{1991trcb.book.....D}; (5) - 2MASS \citep{2006AJ....131.1163S}; (6) - \citet{1998A&AS..128..299P}; (7) - NED; (8) - NED; (9) - in seven cases, the emission lines were detected by \citet{1986AJ.....91.1062P}. In NGC 4546, emission lines were first reported by \citet{1987ApJ...318..531G}.} \label{sample}

\end{center}
\end{table*}

\section{Observations and data reduction} \label{sec:observation_data_reduction}

The 10 galaxies from the sample were observed with the Gemini South Telescope (programmes GS2008A-Q-51 and GS2008-B-Q-21). We used the Gemini Multi-Object Spectrograph (GMOS) in integral field unit (IFU) mode \citep{2002PASP..114..892A,2004PASP..116..425H} with one slit. This configuration uses 750 micro-lenses located at the focal plane of the telescope, 500 of which are used to observe the object and 250 are dedicated to the sky, both separated by 1 arcmin. The micro-lenses divide the image of the object in 0.2 arcsec slices and are coupled to an array of optical fibres. These fibres are arranged linearly on the nominal location of the spectrograph slit (pseudo-slit). The 750 spectra are horizontally aligned in 3 CCDs. The final product is a data cube, with two spatial dimensions covering a FOV of 3.5 arcsec x 5 arcsec and one spectral dimension. With data cubes, one is able to create images in a specific wavelength range or to extract spectra from different spatial regions. The spatial pixels of a data cube are called spaxels. In programme GS2008-B-Q-21, we used the B600-G5323 grating, with a central wavelength of 5650 \AA. The spectra covered a range of 4228 - 7120 \AA\, with a 1.8 \AA\ resolution, estimated with the O I$\lambda$5577 sky line. In programme GS2008A-Q-51, we used the R831-G5322 grating, with a central wavelength of 5800 \AA. The spectral range was 4736 - 6806 \AA, with a 1.3 \AA\ resolution, also estimated with the sky line mentioned above. The observations are summarized in table \ref{sampledescription}.

Flat-Field exposures (lamp and twilight), bias and CuAr lamp spectra were obtained for calibration and correction of the data cubes. Spectrophotometric calibrations were performed using the LTT 9239 star for the GS2008-B-Q-21 programme and EG 21 for the GS2008A-Q-51 programme. The seeing values of the observations were measured in the acquisition images obtained with the GMOS imager in filter \textit{r} (SDSS system) and are shown in table \ref{sampledescription}. 

\begin{table*}
\scriptsize
\begin{center}
\begin{tabular}{lccccccc} 
\hline
Galaxy name & Seeing & Obs time & Programme & Grating & Resolution & Spectral range & Spatial scale \\
&(arcsec)&&&& (\AA) & (\AA) & (pc/arcsec) \\
 & (1) &(2)&(3)&(4)&(5)&(6)&(7) \\
\hline
NGC 1399 & 1.14& Aug 4 2008 & GS-2008B-Q-21 & B600 - G5323 & 1.8 & 4228 - 7121&88  \\
IC 1459 & 0.70 &Aug 3 2008 & GS-2008B-Q-21 & B600 - G5323 & 1.8 & 4228 - 7121&130\\
NGC 2663& 0.84& Feb 3 2008 & GS-2008A-Q-51 & R831 - G5322 & 1.3 & 4735 - 6806&130 \\
NGC 7097& 1.05& Jul 31 2008 & GS-2008B-Q-21 & B600 - G5323 & 1.8 & 4228 - 7121&150 \\
NGC 3136& 0.69& Feb 6 2008 & GS-2008A-Q-51 & R831 - G5322 & 1.3 & 4735 - 6857&116\\
NGC 1404& 1.02& Aug 5 2008 & GS-2008B-Q-21 & B600 - G5323 & 1.8 & 4231 - 7120&92\\
IC 5181& 0.79 &Aug 3 2008 & GS-2008B-Q-21 & B600 - G5323 & 1.8 & 4234 - 7105&122\\
NGC 4546& 0.71& Feb 17 2008 & GS-2008A-Q-51 & R831 - G5322 & 1.3 & 4736 - 6806&82\\
NGC 1380& 0.76& Aug 7 2008 & GS-2008B-Q-21 & B600 - G5323 & 1.8 & 4228 - 7121&88\\
ESO 208 G-21& 0.99& Feb 3 2008 & GS-2008A-Q-51 & R831 - G5322 & 1.3 & 4734 - 6806&82\\
\hline
\end{tabular}
\caption{Observing parameters. (1) - Gaussian FWHM of the seeing of the observations, in arcsec, (2) - observation date, (3) Gemini programme for the observations, (4) - grating, (5) - spectral resolution (FWHM), in \AA, (6) - spectral range in \AA, (7) - spatial scale in pc/arcsec.  \label{sampledescription}
}
\end{center}
\end{table*}

Data reduction was performed using the standard IRAF package for the Gemini telescope. The data were bias and background subtracted. Bad pixel maps were applied to each CCD. Then, the position of the fibres were identified with the lamp flat-field. Each spectrum of the lamp flat-field was extracted, divided by the average spectra and multiplied by the relative collapsed spectra of the twilight flat-field exposure. This produced a master flat-field that could account for the different responses between each fibre and also remove high-frequency patterns in spectral dimension. Cosmic rays were removed with the LACOS algorithm \citep{2001PASP..113.1420V}. Finally, the objects' spectra were extracted, wavelength calibrated and sky subtracted, resulting in a FITS file containing 500 spectra, each associated with a different spatial position of the object. 

Spectrophotometric stars were reduced in the same way mentioned above. After sky subtraction, a spectrum was obtained from the sum of the spectra of every fibre. This representative spectrum was used to obtain the sensibility function of the CCDs, which was applied to all 500 spectra of the objects. After this, the data cubes were built with a spatial sampling of 0.05 arcsec, resulting in an FOV with 66 $\times$ 98 spaxels.

Besides the basic data reduction, the data cubes also possess instrumental artefacts that may hamper the analysis. In Appendix \ref{tratamento_dados_complementares}, complementary data cube treatment procedures are presented, in particular to remove low and high frequency noises contained in both spatial and spectral dimensions, apart from the correction of the differential atmospheric refraction effect. In addition, each data cube image was deconvolved using the Richardson Lucy method (\citealt{1972JOSA...62...55R,1974AJ.....79..745L}) with six iterations and a Gaussian point spread function (PSF) with an FWHM = seeing of each observation, shown in table \ref{sampledescription}. The procedures were carried out in the order presented in Appendix \ref{tratamento_dados_complementares}. Nevertheless, one should always remain aware when removing the low and high frequency noises before the deconvolution process because deconvolution may mask these noises instead of removing them. 

\section{Results} \label{sec:results}

\subsection{PCA Tomography: a brief introduction} \label{pca_tomography}

In astronomy, it is usual to collect observational data by means of direct images or through spectroscopy. Recent technology allows the simultaneous acquisition of images and spectra with the IFUs. This technique results in data cubes, which have information on both spatial and spectral dimensions of an object. With tens of millions of pixels to be analysed, the application of an efficient method to extract relevant information from a data cube is necessary. An effective option is PCA Tomography \citep{1997ApJ...475..173H,2009MNRAS.395...64S}. The use of this method in optical \citep{2011ApJ...734L..10R,2011MNRAS.413..149S,2010IAUS..267..123M} and near-infrared \citep{2011MNRAS.416..493R,2014MNRAS.438.2597M} data cubes has highlighted the power of the technique. 

PCA is a statistical tool of multidimensional data analysis that is able to optimally detect correlations between $m$ parameters in a sample of $n$ objects. This consists in obtaining a set of eigenvectors that will define a new basis for the object sample. Each eigenvector of this basis is defined by a linear combination of the $m$ parameters, whose weight gives a direct measure of the correlations between the parameters in the sample. The construction of this set of eigenvectors is done in such a way that statistically relevant information of the sample is compressed in a number $l$ $<<$ $m$ of eigenvectors. 

The base transformation associated with PCA is given by

\begin{equation}
	T_{\beta k} = I_{\beta \lambda}\cdot E_{\lambda k}, \label{transf_PCA}
\end{equation}
where $T_{\beta k}$ is the data set in the new coordinate system, $I_{\beta \lambda}$ is the same set in the original basis and $E_{\lambda k}$ is the eigenvector's set that defines the new basis. Regarding the indices, $\beta$ is related to the $\beta^{th}$ object of the sample, $k$ is the k$^{th}$ eigenvector of the basis and $\lambda$ is the $\lambda^{th}$ parameter associated with the objects of the sample. One may demonstrate that the basis that minimizes the loss of information in the $l$ eigenvector is defined by the eigenvector set that results from the covariance matrix $C_{cov}$ of the parameters of the sample (see \citealt{fukunaga} or \citealt{1997ApJ...475..173H}), defined as 

\begin{equation}
	C_{cov} = \frac{\left[I_{\beta \lambda}\right]^T \cdot I_{\beta \lambda}}{n - 1}. \label{cov_matrix}
\end{equation}
The eigenvalue $\Lambda_k$, related to the eigenvector $E_k$, is a direct measurement of the variance contribution that this eigenvector possesses in the data set. This means that the first eigenvector corresponds to the greatest variance of the data set, the second eigenvector accounts for the second most important variance of this set and so on. Thus, one is able to extract useful information from this sample by analysing only the first \textit{l} eigenvectors. One way to define which is the last eigenvector that still has useful information is the scree test, which consists in building a plot of the variance fraction of the eigenvectors as a function of the eigenvector numbers. In this plot, one defines a straight line that fits the variance fractions related to the highest eigenvectors (i.e. those with the lowest respective eigenvalues that contain only noise). The eigenvectors that have variance fractions clearly above this line should contain important signal for the sample analysis (see \citealt{2009MNRAS.395...64S} for more details).   

PCA Tomography consists in applying the principal component analysis to data cubes. A data cube structure is given by $I_{ij\lambda}$, where $i$ and $j$ correspond to the $x$ and $y$ coordinates, respectively, of the spatial dimension of the data cube and $\lambda$ corresponds to the spectral dimension of the data cube. This $I_{ij\lambda}$ structure is, then, re-organized in a two-dimensional matrix $I_{\beta \lambda}$, where $\beta$ is the index related to each spaxel of the data cube. The covariance matrix of the data cube is calculated with equation \ref{cov_matrix}. The eigenvectors $E_{\lambda k}$ of this covariance matrix (also called eigenspectra in this case) show the correlations between the wavelengths while the tomograms $T_{\beta k}$, calculated with the equation \ref{transf_PCA}, reveal where the correlations detected in the eigenspectra $E_{\lambda k}$ occur on spatial dimension. 

Among the most useful procedures of PCA Tomography are feature enhancement and feature suppression. In feature enhancement procedures, one may select eigenvectors related to a physical process `A' (e.g. eigenvectors whose correlations show typical AGN emission lines) and build a new data cube selecting the feature factor $\Gamma_k$ = 1 for the eigenvectors associated with `A' and $\Gamma_k$ = 0 for the other eigenvectors. To suppress the features, the procedure is similar; however, one sets $\Gamma_k$ = 0 for eigenvectors related to a characteristic that must be removed from the original data cube (e.g. instrumental defects) and $\Gamma_k$ = 1 for all other eigenvectors. This may be done as

\begin{equation}
	I^\prime_{\beta \lambda} = T_{\beta k}\cdot \left[(E_{\lambda k})_\Gamma \right]^T, \label{feat_enh}
\end{equation}
where $I^\prime_{\beta \lambda}$ is the two-dimensional matrix of the rebuilt data cube and  $(E_{\lambda k})_\Gamma$ is associated with the matrix $E_{\lambda k}$, where each eigenvector $E_k$ is multiplied by its corresponding feature factor $\Gamma_k$. After the achievement of $I^\prime_{\beta \lambda}$, the data are reorganized in a data cube structure $I^\prime_{ij\lambda}$ for subsequent analysis. 

The eigenvectors are orthogonal to each other. This means that distinct physical phenomena (central AGN; circumnuclear ionized gas disc or stellar disc), by not having correlated variance, may become associated with distinct eigenvectors. There is no guarantee that this will occur. For instance, if a stellar disc and gas disc are co-planar but in counter-rotation, they may be related to a single eigenvector. This means that the results (eigenspectra and tomograms) require careful interpretation. This may, however, reveal unexpected information-rich phenomena, such as demonstrated in NGC 7097 by \citet{2011ApJ...734L..10R}.

\subsection{The case of IC 1459 - a previously known LINER} \label{case_IC1459}

The elliptical galaxy IC 1459 was included in this sample as a test object to be analysed with PCA Tomography. It is a previously known LINER that has already been studied in some detail \citep{1986AJ.....91.1062P,2010A&A...519A..40A}. This galaxy is a member of a loose group containing several spiral galaxies. \citet{1985MNRAS.217...87S} detected a dust lane located south-west from the nucleus. In a \textit{V - I} image obtained with the \textit{HST}, \citet{1997ApJ...481..710C} confirmed the presence of dust in the central region of the galaxy, with a disc-like structure. In the same region, they found a blue point-like source. \citet{1988ApJ...327L..55F} detected a stellar core that counter-rotated with the external region of IC 1459 and was aligned with the semi-major axis (P.A. $\sim -146^o$; \citealt{2000AJ....120.1221V})\footnote{The position angles (P.A.) in this work vary from $-180^o$ to $180^o$. The P.A. has its origin in the north and raises eastward until the kinematic component in redshift. All the measurements extracted from the literature are shown following this definition.}. IC 1459 contains a gas disc corotating with the external region of the galaxy \citep{1988ApJ...327L..55F,2000AJ....120.1221V,2002ApJ...578..787C}, with a P.A. $\sim 34^o$ measured in a H$\alpha$+[N II] emission map obtained with the \textit{HST} \citep{2000AJ....120.1221V}. The line ratios resemble those of the emission of LINERs \citep{1986AJ.....91.1062P,2010A&A...519A..40A}. The X-ray properties \citep{2003ApJ...588..175F,2009A&A...506.1107G}, a compact radio source \citep{1994MNRAS.269..928S} and the blue point-like source revealed by \citet{1997ApJ...481..710C} indicate that, in the nuclear region, this LINER must be photoionized by an AGN.

\begin{figure*}
\includegraphics[scale=0.4]{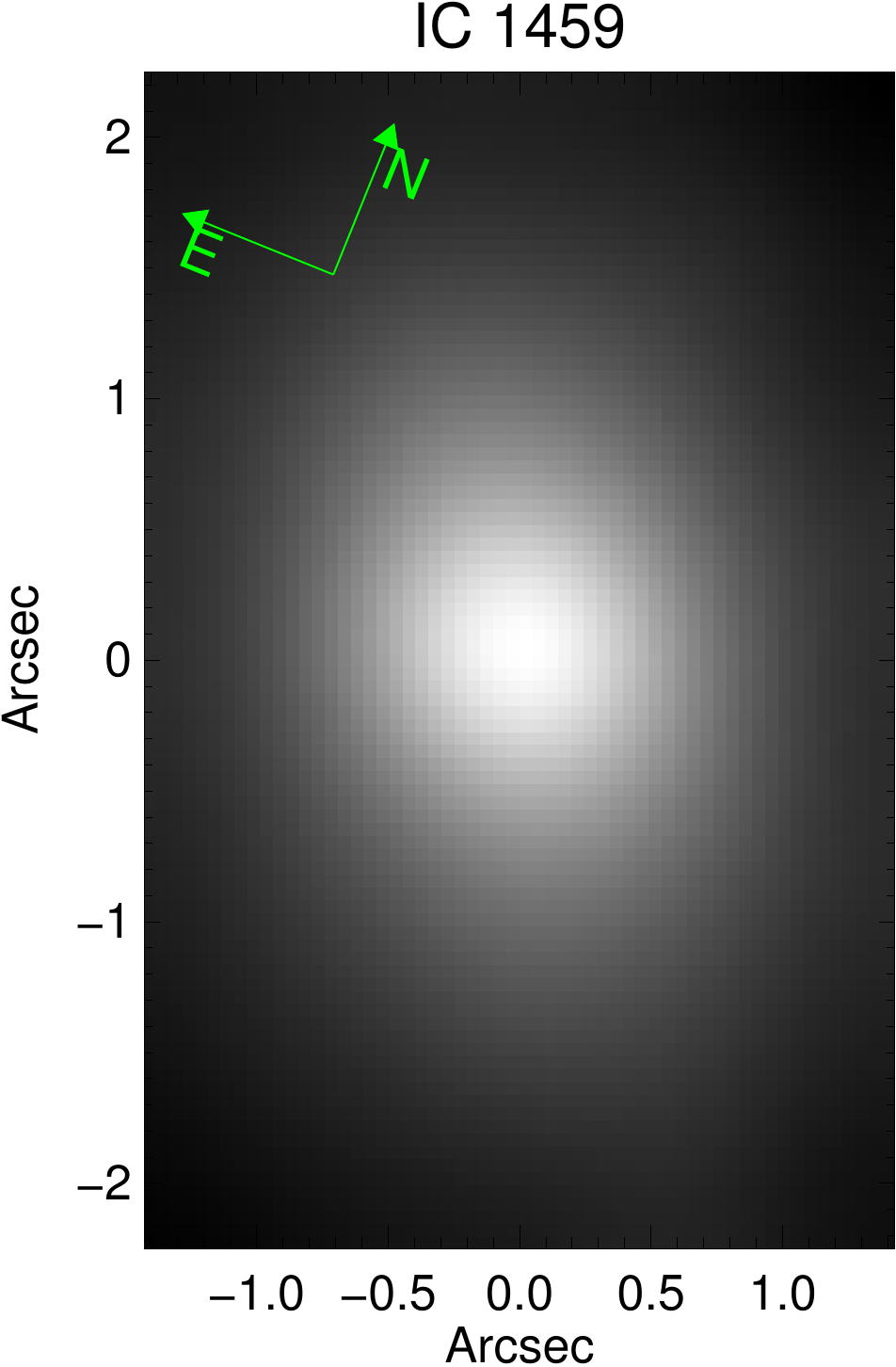}
\vspace{1cm}
\hspace{1cm}
\includegraphics[scale=0.4]{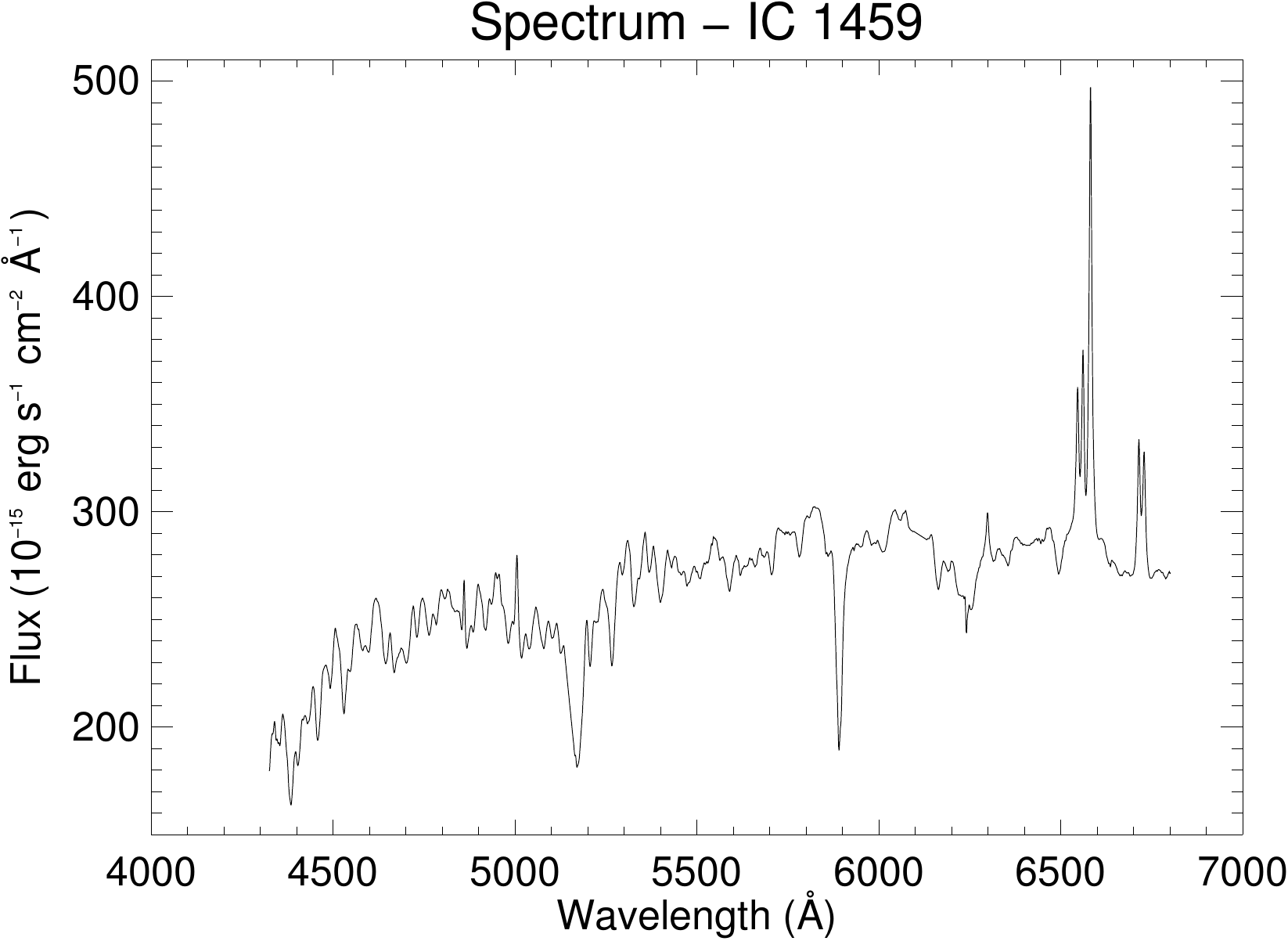}
\caption{Collapsed image and the sum of all spectra from the data cube of IC 1459. \label{cube_results_IC1459}} 
\end{figure*}

\begin{figure*}
\includegraphics[scale=0.4]{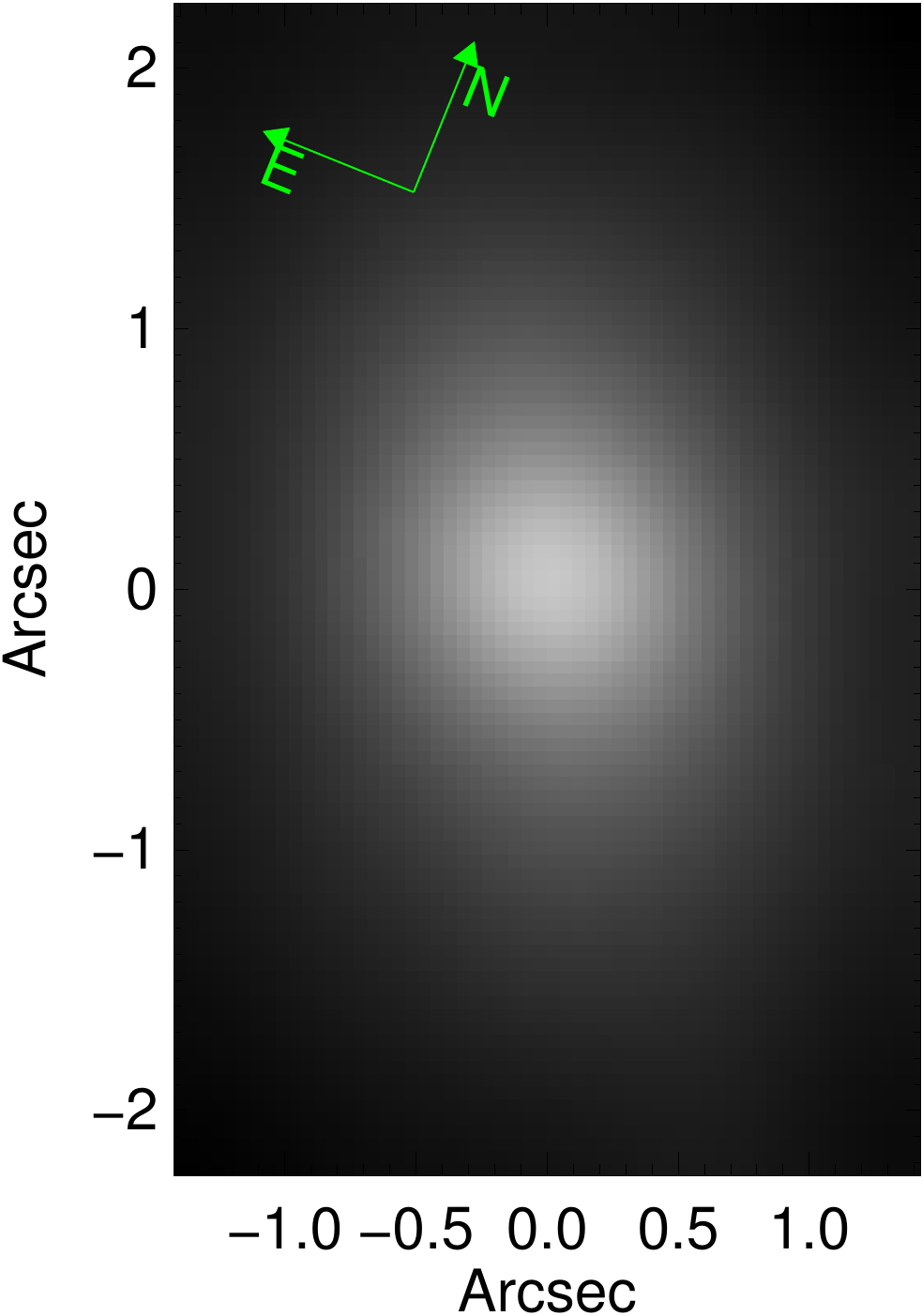}
\vspace{1cm}
\hspace{3cm}
\includegraphics[scale=0.4]{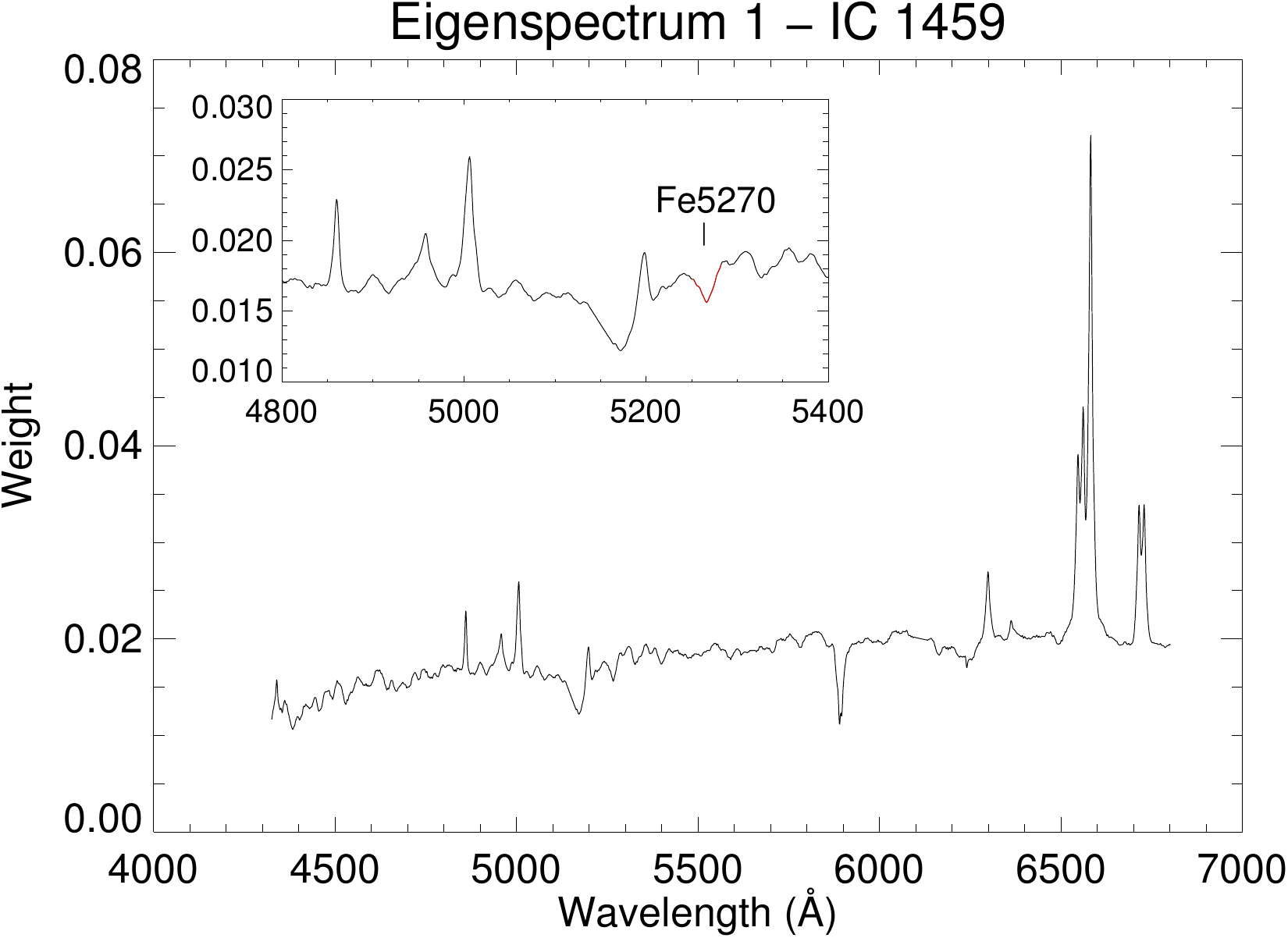}
\includegraphics[scale=0.4]{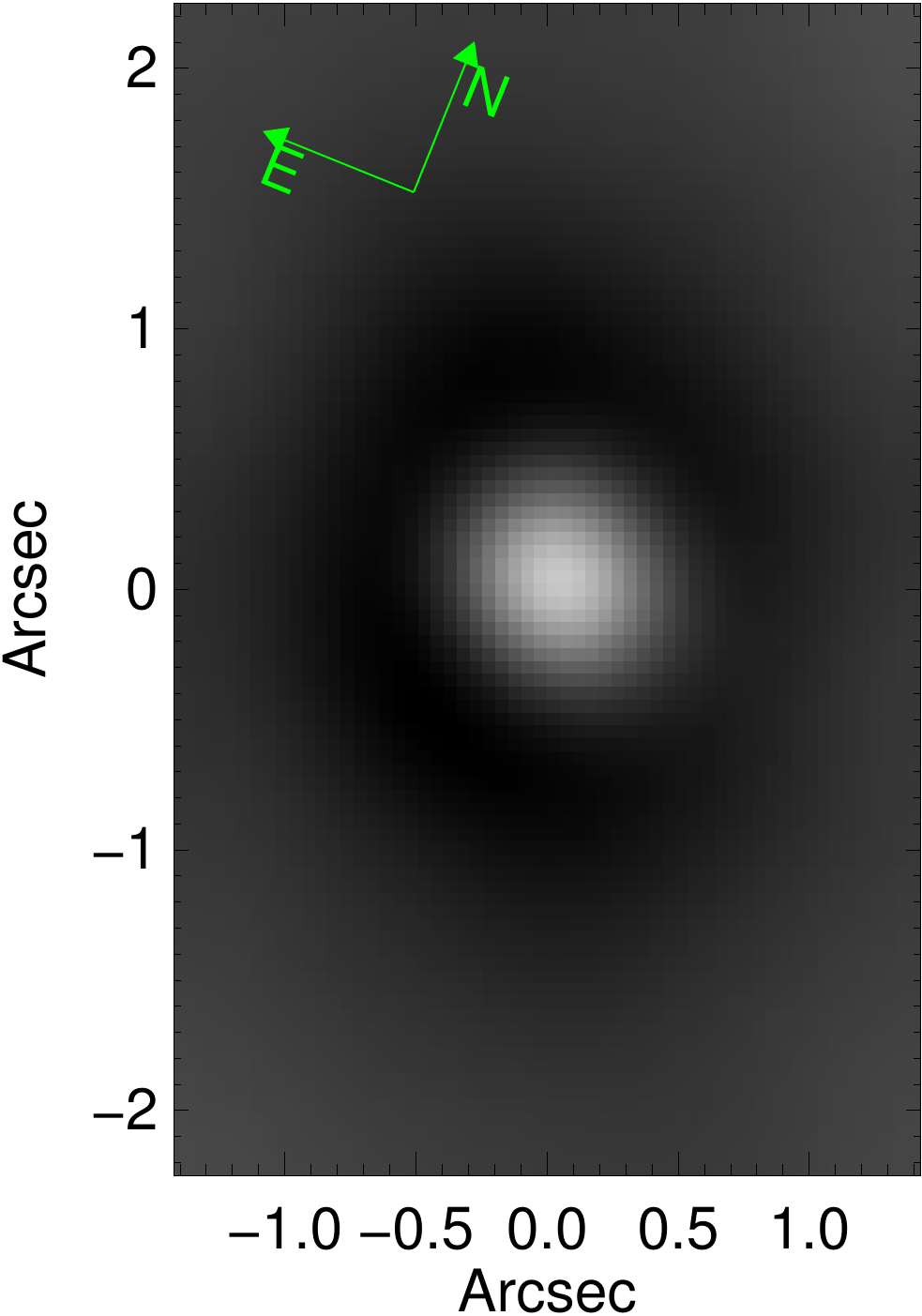}
\vspace{1cm}
\hspace{3cm}
\includegraphics[scale=0.4]{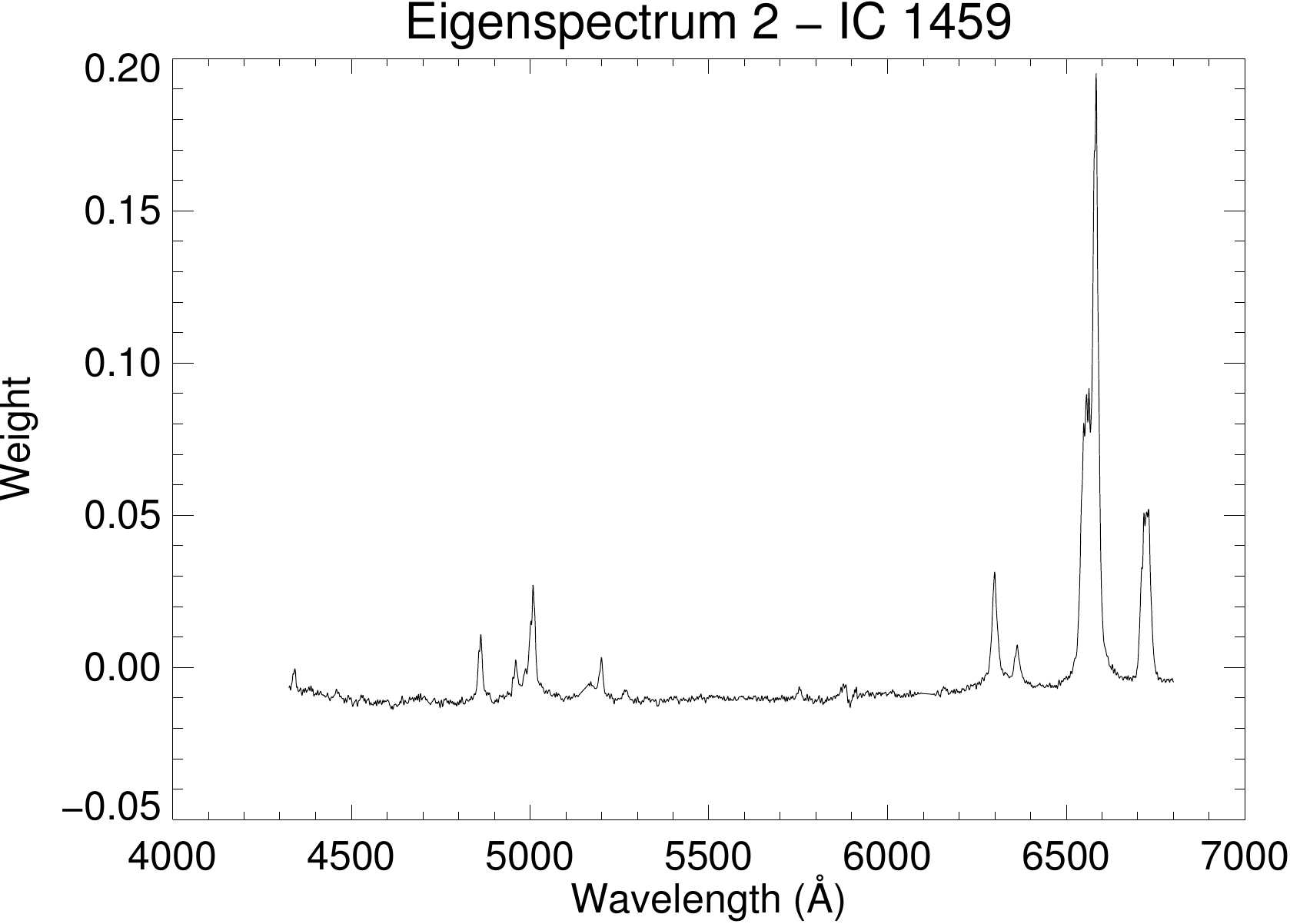}
\includegraphics[scale=0.4]{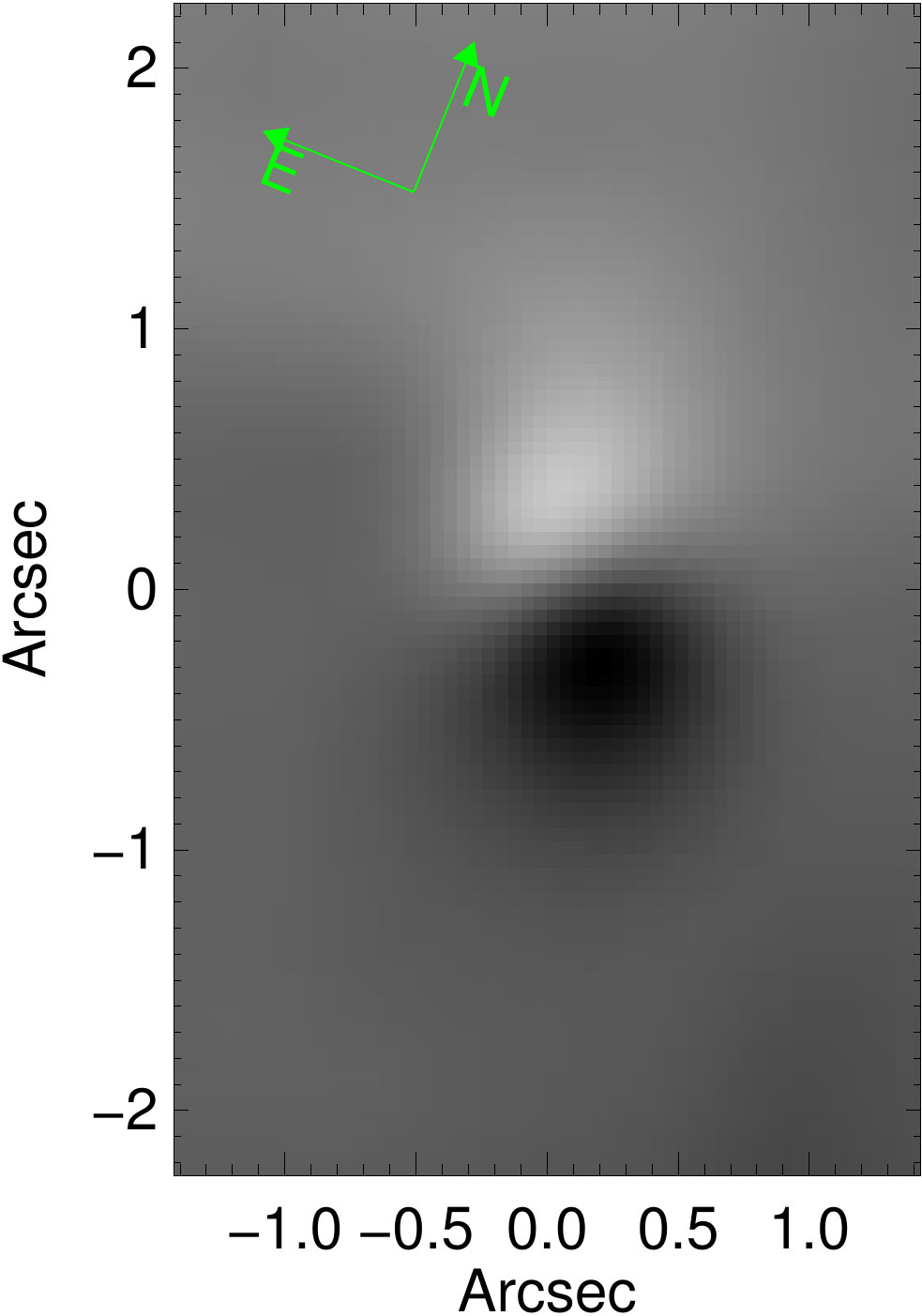}
\vspace{1cm}
\hspace{3cm}
\includegraphics[scale=0.4]{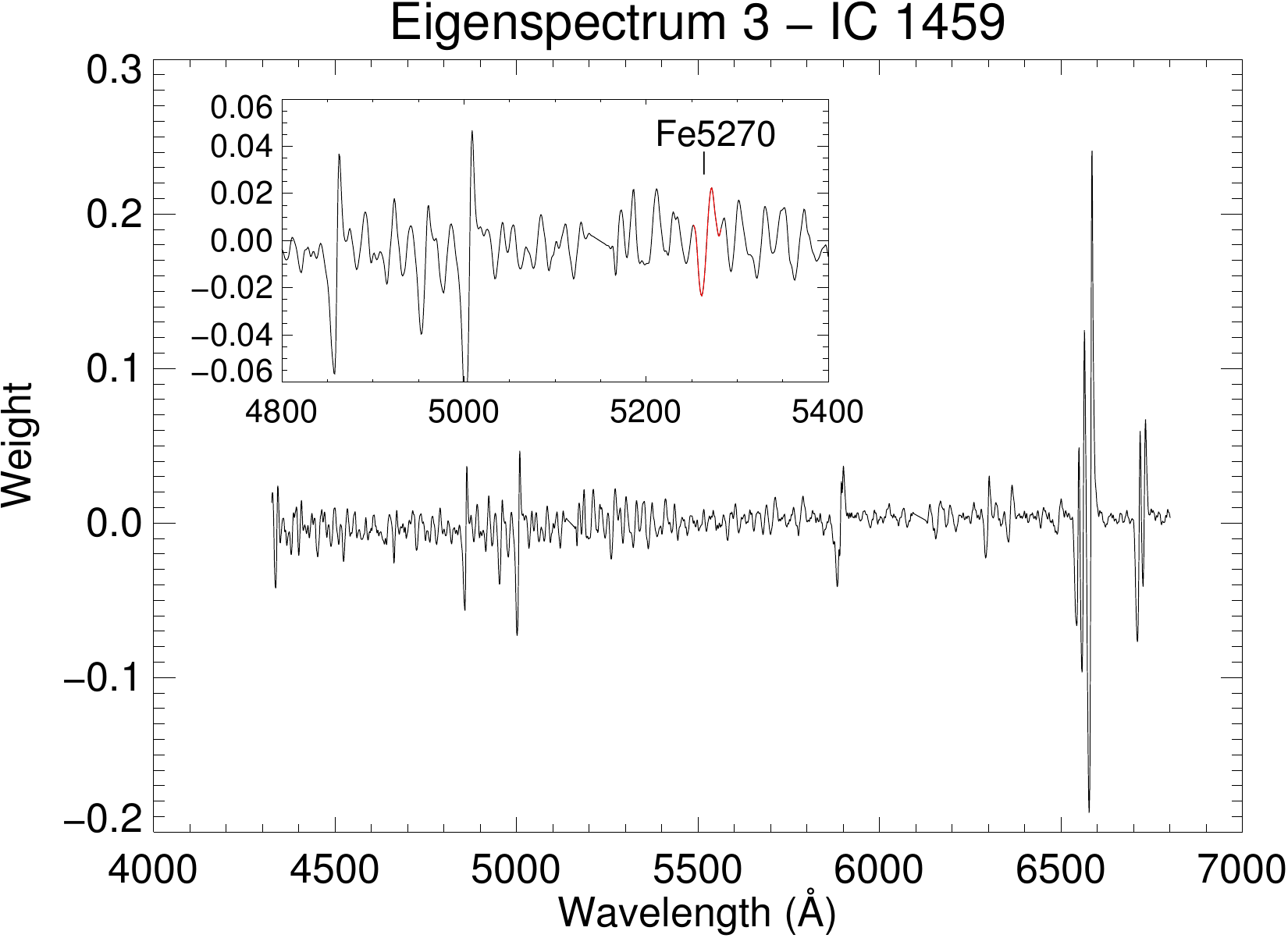}
\caption{Eigenvectors and tomograms 1-3 from the data cube of IC 1459. In both the first and third eigenvectors, the Fe5270 stellar line is highlighted in red. Note, in eigenvector 3, that the red wing of the Fe5270 is correlated with the red wing of the H$\beta$ emission line. Since the Fe5270 is an absorption line, this correlation indicates that both stellar and gas components have opposite kinematics, leading us to the conclusion that these structures are counter-rotating.}  \label{results_IC1459}
\end{figure*}

\begin{figure}
		\hspace{-0.5cm}
		\includegraphics[scale=0.50]{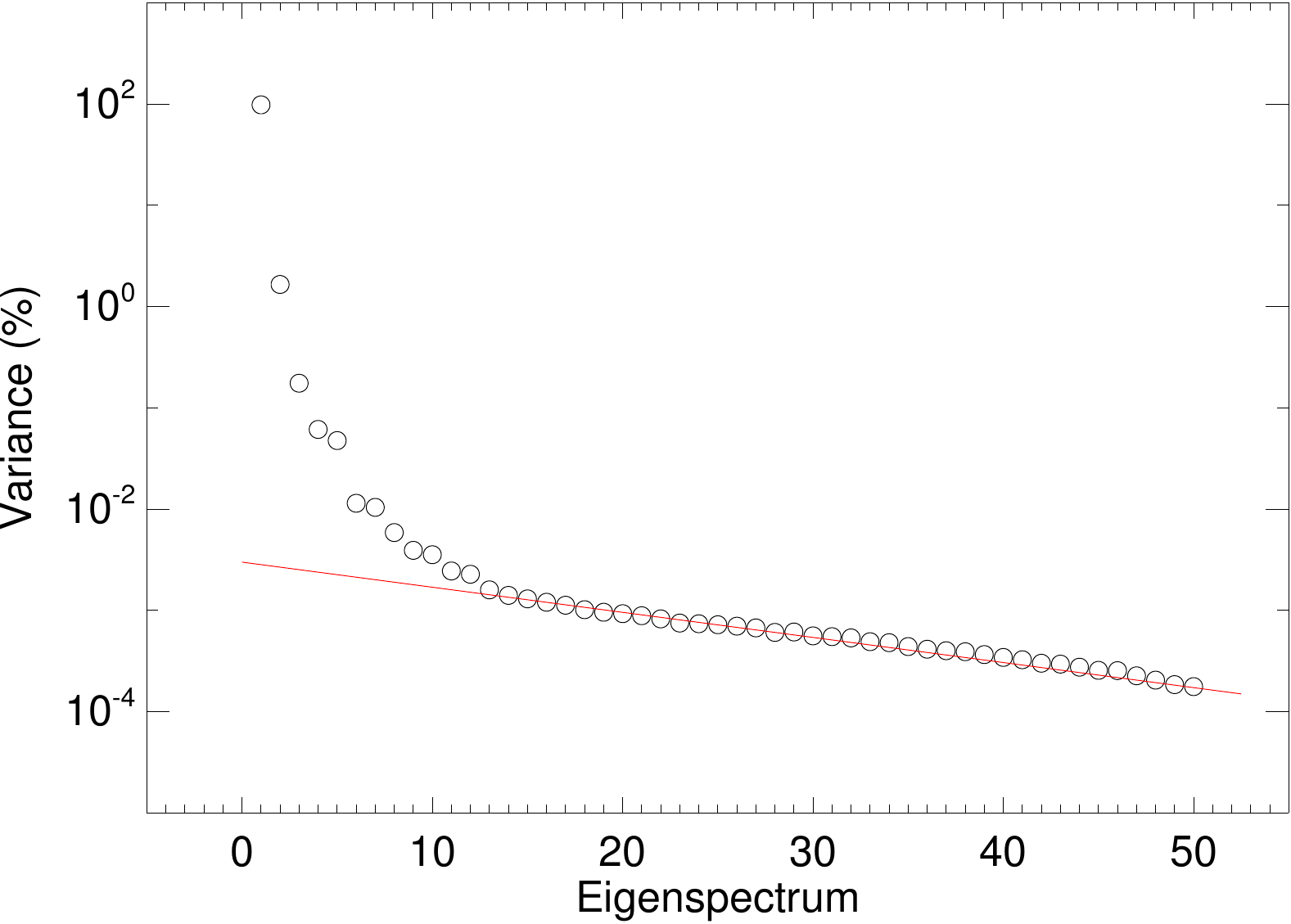}
	\caption{Scree test resulted from PCA Tomography applied to the data cube of IC 1459. According to this test, there is useful information until eigenvector 12. Beyond this eigenvector, all the results correspond to noises of the data cube. }
	\label{fig:scree_test_IC1459}
\end{figure}

We applied PCA Tomography \citep{2009MNRAS.395...64S} to the data cube of IC 1459 from 4325 to 6800 \AA. The collapsed image, as well as the sum of all spectra from the data cube of this galaxy, is shown in Fig. \ref{cube_results_IC1459}. The goals are to verify if PCA Tomography is able to recover the physical phenomena that are occurring in the central region of this galaxy, described by the literature, and to assess the limitations of this technique. The first three eigenvectors are shown in Fig. \ref{results_IC1459}. Their corresponding eigenvalues are presented in table \ref{eigenvalues_whole_pca}. The scree test (see \citealt{2009MNRAS.395...64S} and section \ref{pca_tomography} for details) related to these eigenvalues is shown in Fig. \ref{fig:scree_test_IC1459}.

The first eigenspectrum revealed correlations between the stellar continuum and the absorption lines. Emission lines are also spatially correlated with the stellar features and its tomogram has revealed the central stellar structure of IC 1459. One can see that the eigenspectrum and the tomogram 1 are very similar to, respectively, the average spectrum and the image from the data cube. This highlights the great redundancy contained in the data cubes. In the case of IC 1459, eigenvector 1 accounts for $\sim$ 98\% of the variance (see table \ref{eigenvalues_whole_pca}). The same trend is seen in the other galaxies of the sample, always with variances $>$ 98\%. 

In the second eigenvector, whose variance is $\sim$ 1.65\%, correlations between the wavelengths associated with the H$\beta$, [O III]$\lambda \lambda$4959, 5007, [N I]$\lambda \lambda$5198, 5200, [O I]$\lambda \lambda$6300, 6363, H$\alpha$+[N II]$\lambda \lambda$6548, 6583 and [S II]$\lambda \lambda$6716, 6731 emission lines were detected. These correlations indicate an eigenspectrum that resembles a lot a typical LINER spectrum. In particular, the relative correlations between [O III] and H$\beta$ lines, on one hand, and between [O I], [S II] and [N II] with H$\alpha$, on the other, are surprisingly similar to a LINER spectrum. 

\begin{figure*}
\includegraphics[scale=0.3]{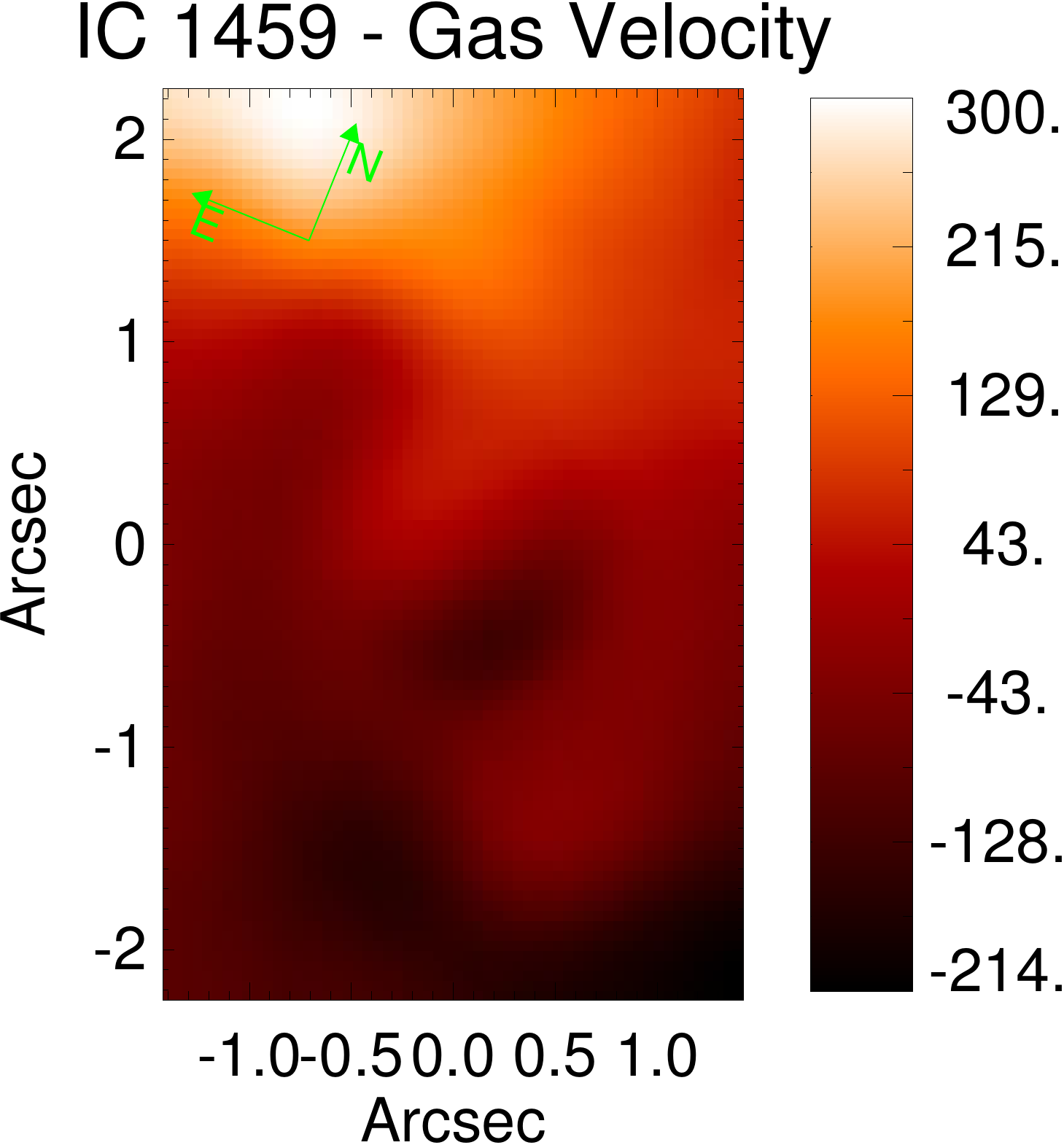}
\includegraphics[scale=0.3]{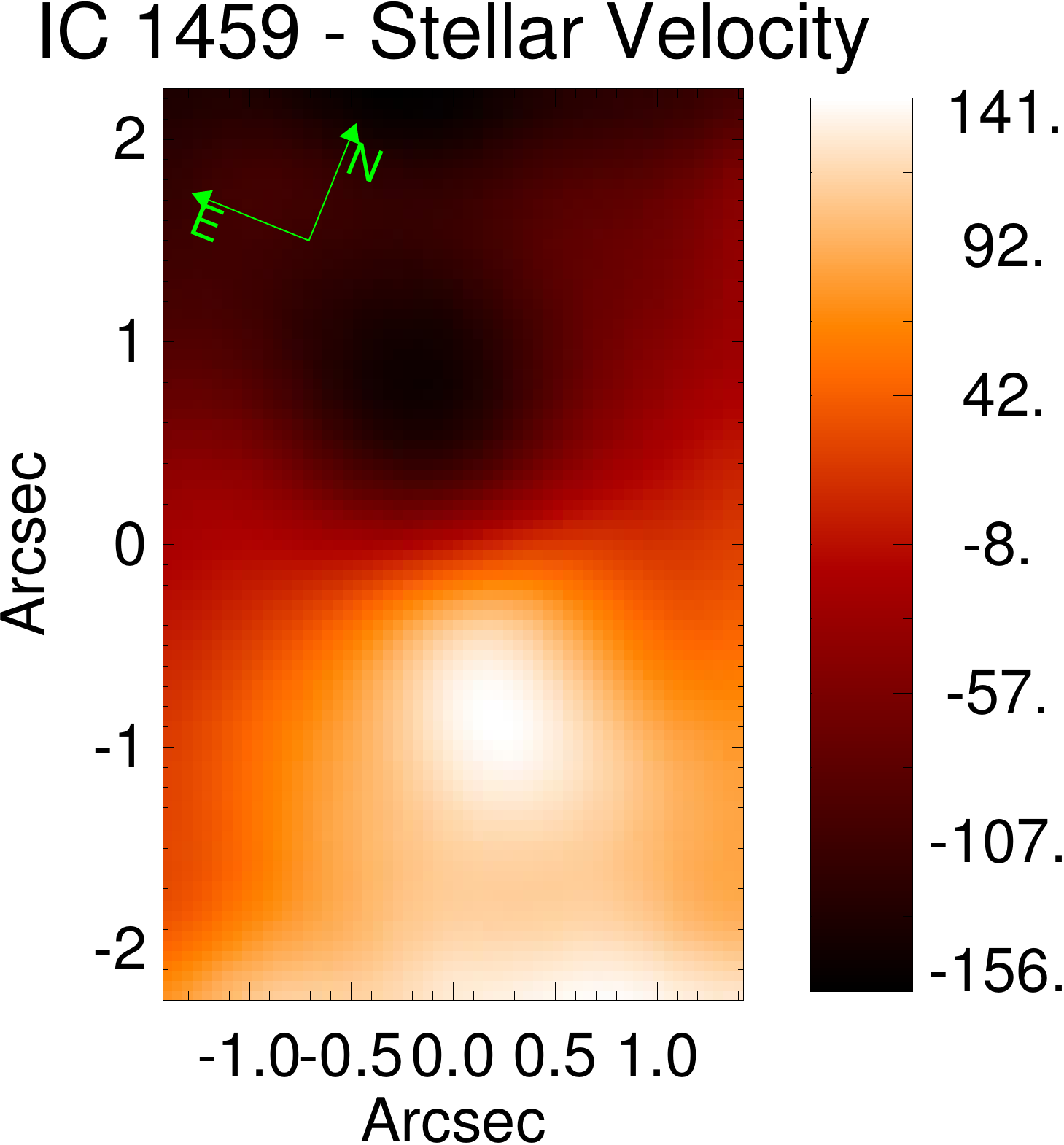}
\vspace{1.0cm}
\includegraphics[scale=0.3]{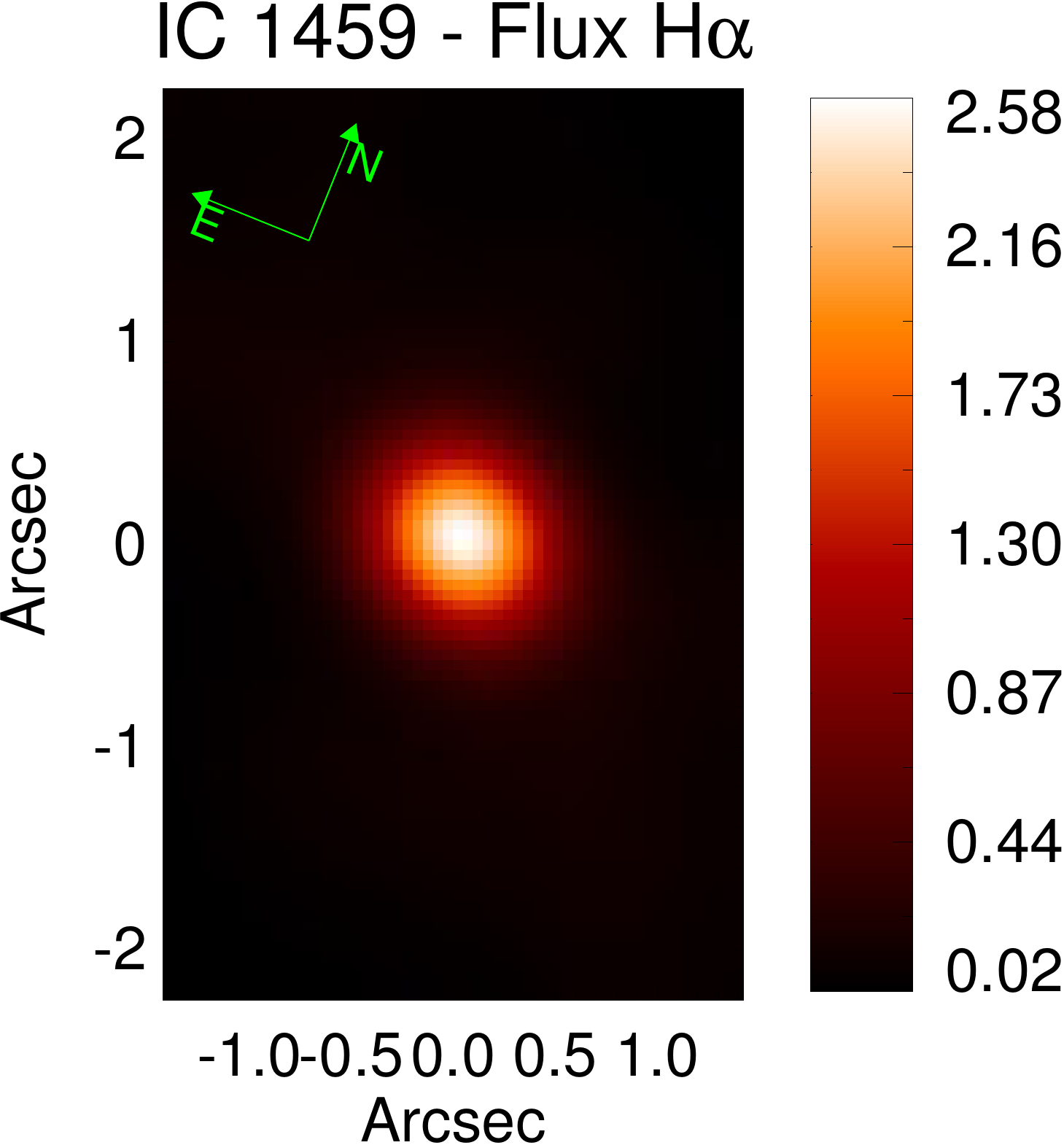}
\caption{Left: gas velocity map of IC 1459, in km s$^{-1}$. Centre: stellar velocity map of IC 1459, in km s$^{-1}$. Right: flux map of the H$\alpha$ emission line, in erg s$^{-1}$. The stellar velocity map was extracted with the ppxf procedure \citep{2004PASP..116..138C} applied in each spectrum of the data cube, while both gas velocity and flux maps were extracted by fitting a Gaussian function to the narrow component of the H$\alpha$ emission line from each spectrum of the data cube. In this figure, it is clear that the gaseous and stellar components are counter-rotating. One may also notice that the bulk of the narrow component of H$\alpha$ intensity is located on the AGN. \label{results_2_IC1459}}
\end{figure*}

Tomogram 2 shows strong positive correlation at the centre of the FOV. This marks the location of the AGN in this case at the centre of the galaxy. In Fig. \ref{results_2_IC1459}, the flux map of the H$\alpha$ line, extracted by fitting a Gaussian function to this line in each spectrum of the data cube after the subtraction of the stellar component (see Paper III for more details), shows that the bulk of this emission is concentrated on the AGN.

The third eigenvector reveals anti correlations between the red and blue wings of all emission lines. In the tomogram, these correlations are seen in distinct regions and are relatively symmetric to the AGN. We associated this structure with a disc of ionized gas. This interpretation is quite similar to that proposed for NGC 4736 \citep{2009MNRAS.395...64S}, NGC 7097 \citep{2011ApJ...734L..10R} and M 81 \citep{2011MNRAS.413..149S}. Moreover, \citet{1997ApJ...475..173H} applied PCA technique to a simulated rigid rotating cloud and their results are equivalent to what is seen in Fig. \ref{results_IC1459} (compare to their Figs. 2 and 3). In IC 1459, the ionized gas disc has been already mentioned by other researchers \citep{1988ApJ...327L..55F,2000AJ....120.1221V,2002ApJ...578..787C}. Fig. \ref{results_2_IC1459} shows the velocity map of the H$\alpha$ line, obtained with the Gaussian fitting procedure mentioned in the last paragraph. 

It is interesting to note that in the third eigenvector, the emission lines are not the only ones showing kinematic features of a disc; the atmospheric stellar lines also reveal this characteristic. However, in this case, the stellar disc and the gas disc are counter-rotating. In eigenvector 3, for instance, shown in Fig. \ref{results_IC1459}, the red wing of the Fe5270 line is correlated with the red wing of the H$\beta$ emission line. However, the Fe5270 is in absorption, indicating, therefore, an opposite kinematics when compared to the one highlighted by the H$\beta$ line. Thus, one may conclude that the stellar and the gas components are counter-rotating. In Fig. \ref{results_2_IC1459}, we show the stellar velocity map, obtained with the penalized pixel fitting procedure (ppxf;  \citealt{2004PASP..116..138C}). The comparison between the gas and stellar velocity maps confirms the counter-rotation between both components in the most central region of IC 1459.

It is worth mentioning that this third tomogram, as any other kinematic tomogram, does not give a hint on the real values of the velocities of both stellar and gas component. Tomograms are the projections of the data cube on to a given eigenspectrum. So, the highlighted regions in a kinematic tomogram are a combination of the intensity (once the variation in the total stellar continuum is accounted for) and the shift of the spectra to the red or to the blue. One should notice that, in Fig \ref{results_2_IC1459}, high velocity values (both stellar and gas) are seen in regions where the H$\alpha$ flux is weaker, whereas in regions where the velocity values are smaller, the H$\alpha$ flux is more conspicuous. A deeper analysis of the velocity maps, as well as the flux maps, will be held in Paper III 

Although the scree test revealed that, in this galaxy, signal dominates noise until eigenvector 12, we were not able to interpret the eigenvectors in terms of physical meaning beyond the third one. Indeed, only one galaxy of the sample had its eigenvectors 4 and 5 interpreted in terms of physical meaning, which is NGC 7097. These results and conclusions were presented by \citet{2011ApJ...734L..10R}.

These results from IC 1459 have shown that PCA Tomography applied to this data cube was able to recover the LINER and the gas and stellar core rotations. This also shows that the PCA Tomography technique is capable of detecting kinematic features, related both to the gas and to the stellar spectra. However, the fact that the same eigenvector detected both gas and stellar kinematics may result in a more complex tomogram to analyse, since both structures are superposed along the spaxels. So, applying PCA Tomography to different wavelength ranges, where one of the ranges is dominated by absorption lines and the other one is dominated by emission lines, may be an effective way to study the kinematics of both components separately. This will be presented in details in Section \ref{tom_disc}.


\subsection{The AGN: the second eigenvectors} \label{agn_eigen2}

\begin{figure*}
\begin{center}
\includegraphics[width=70mm,height=55mm]{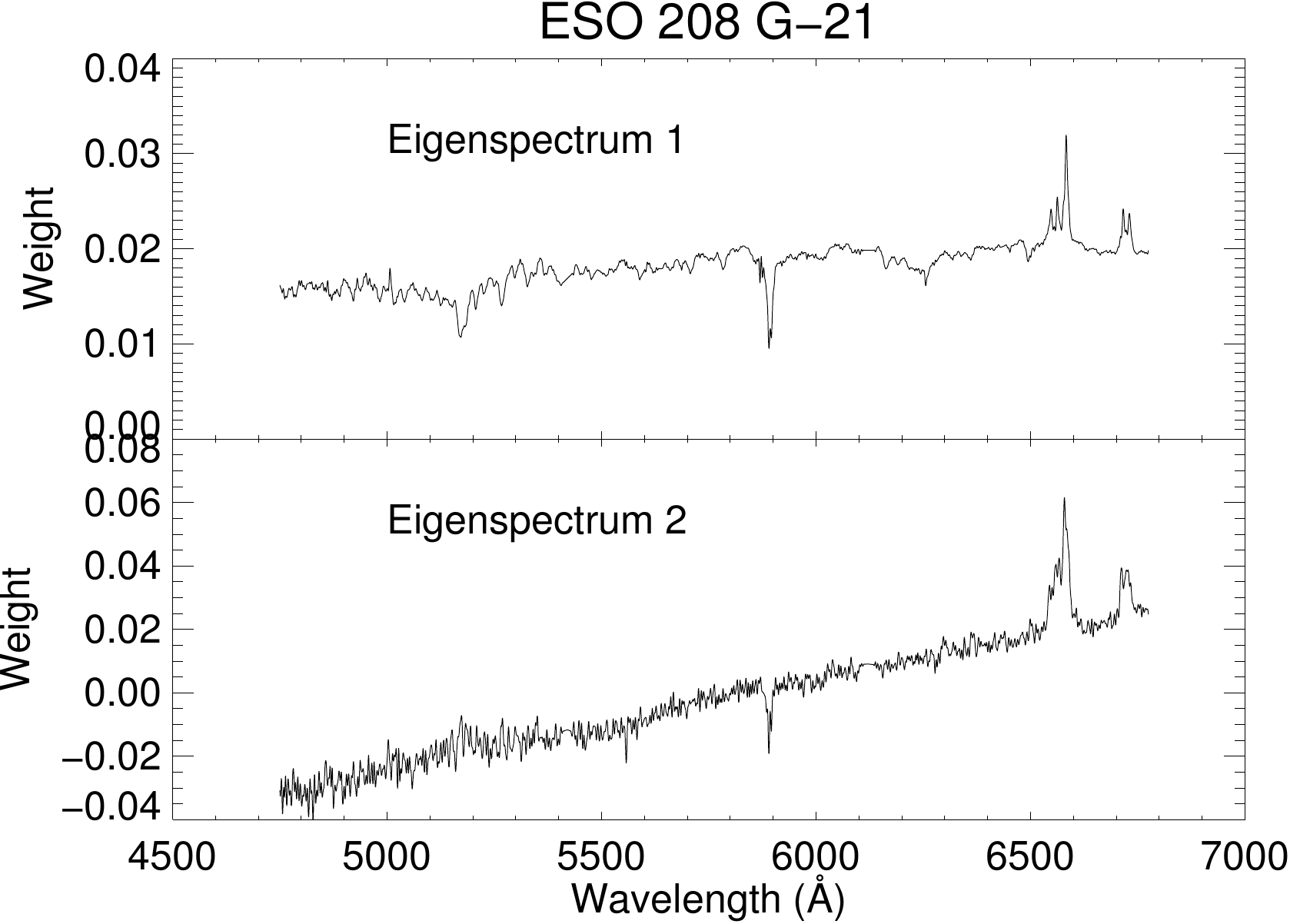}
\includegraphics[width=70mm,height=55mm]{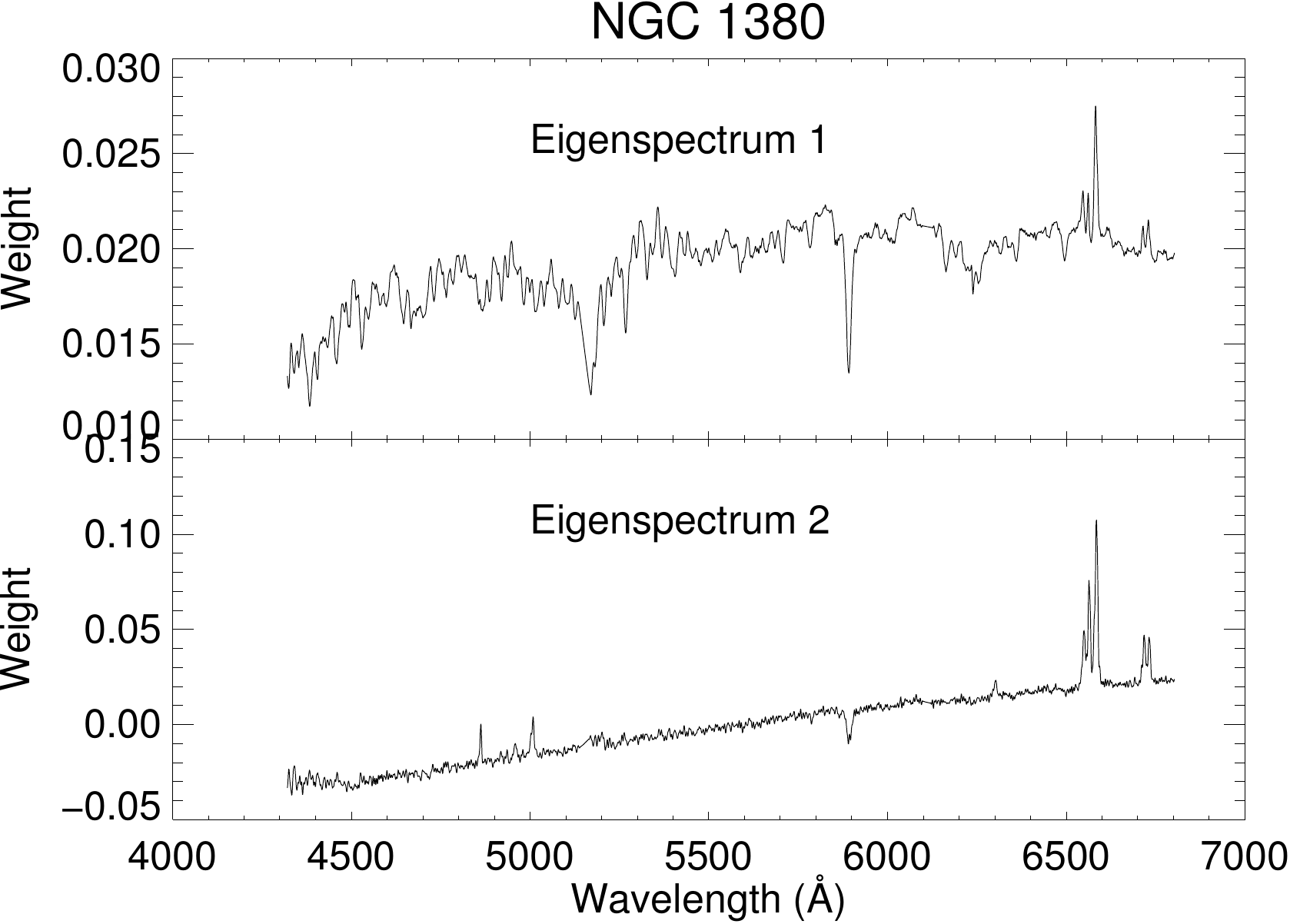}
\includegraphics[width=70mm,height=55mm]{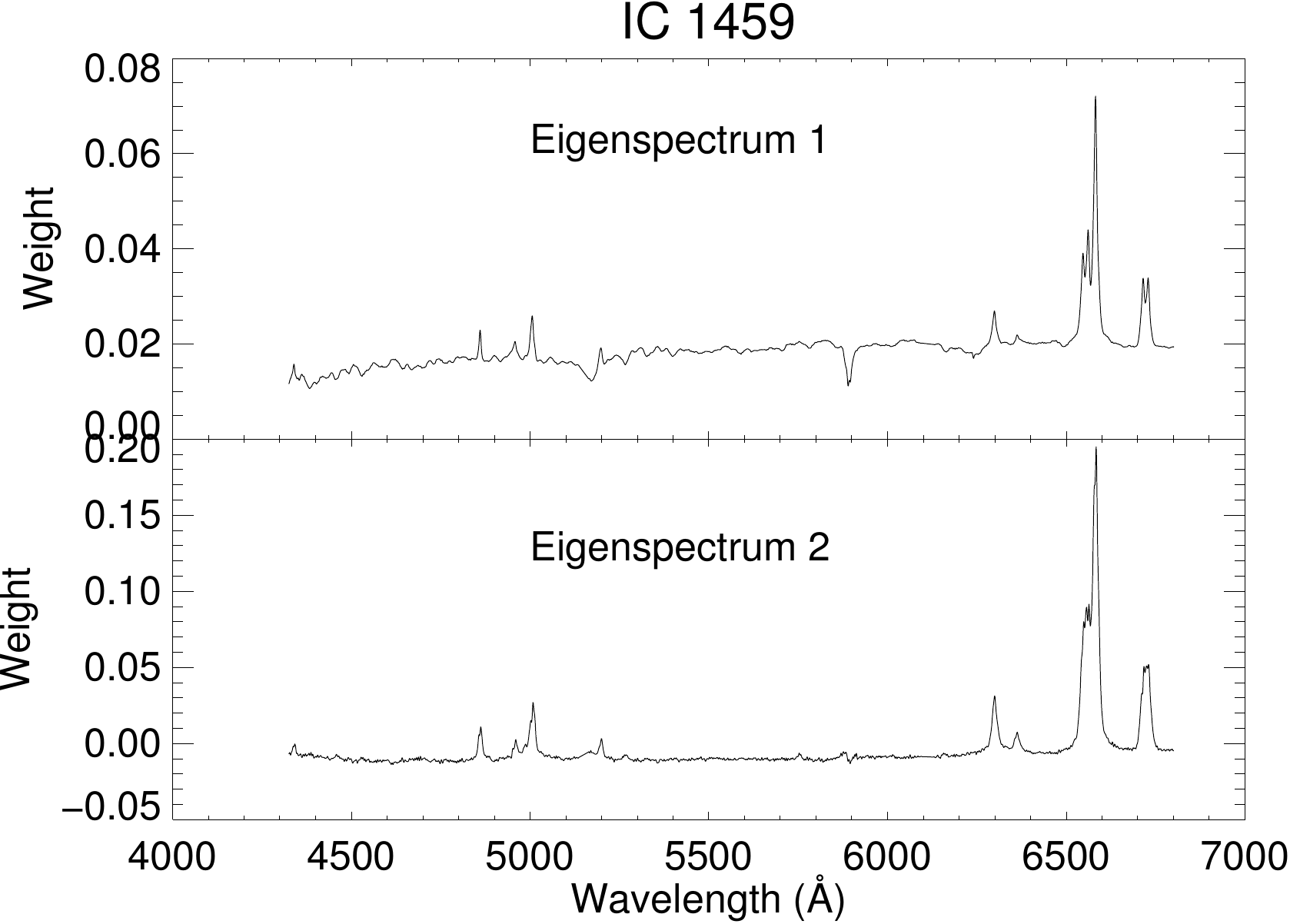}
\includegraphics[width=70mm,height=55mm]{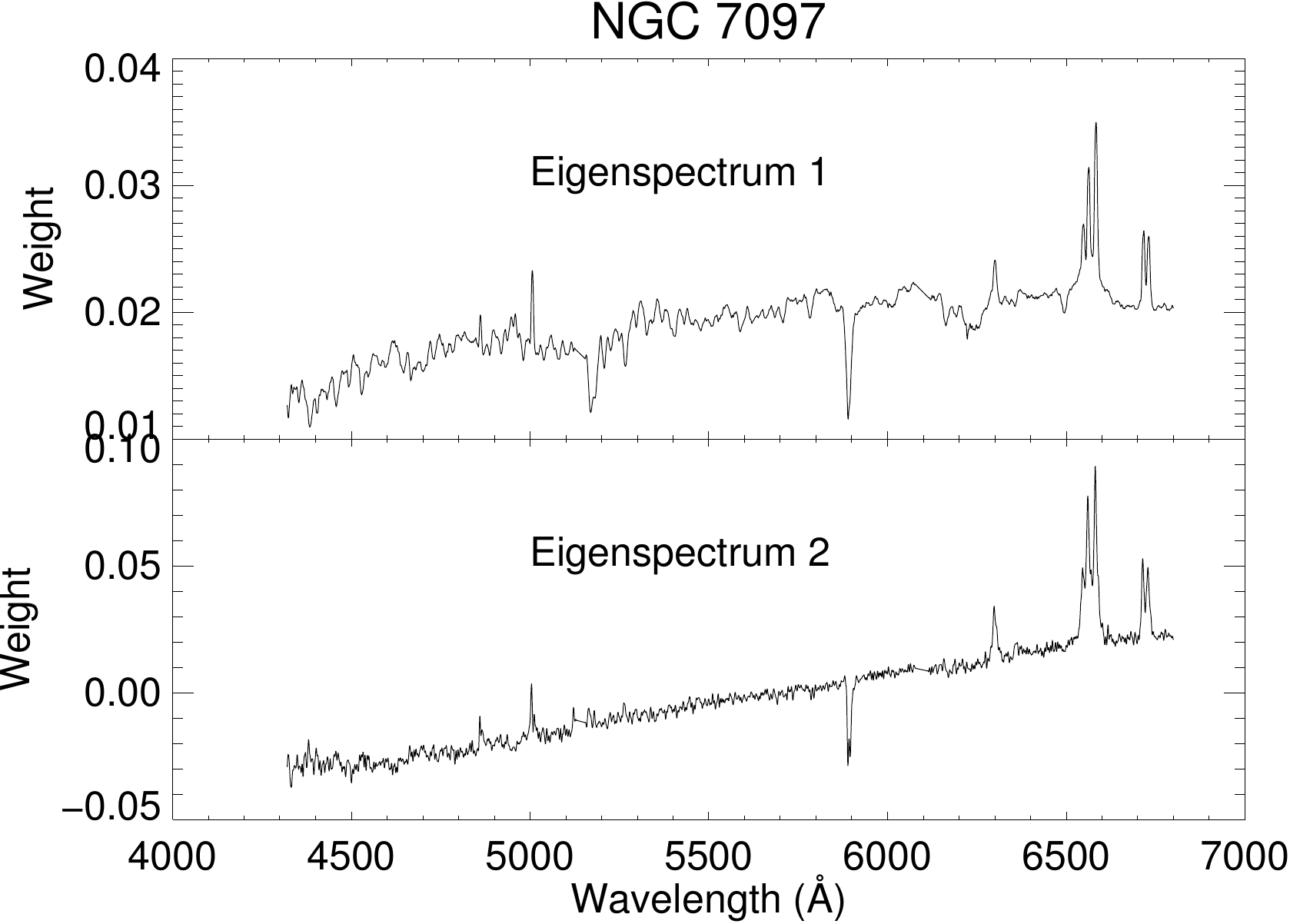}
\includegraphics[width=70mm,height=55mm]{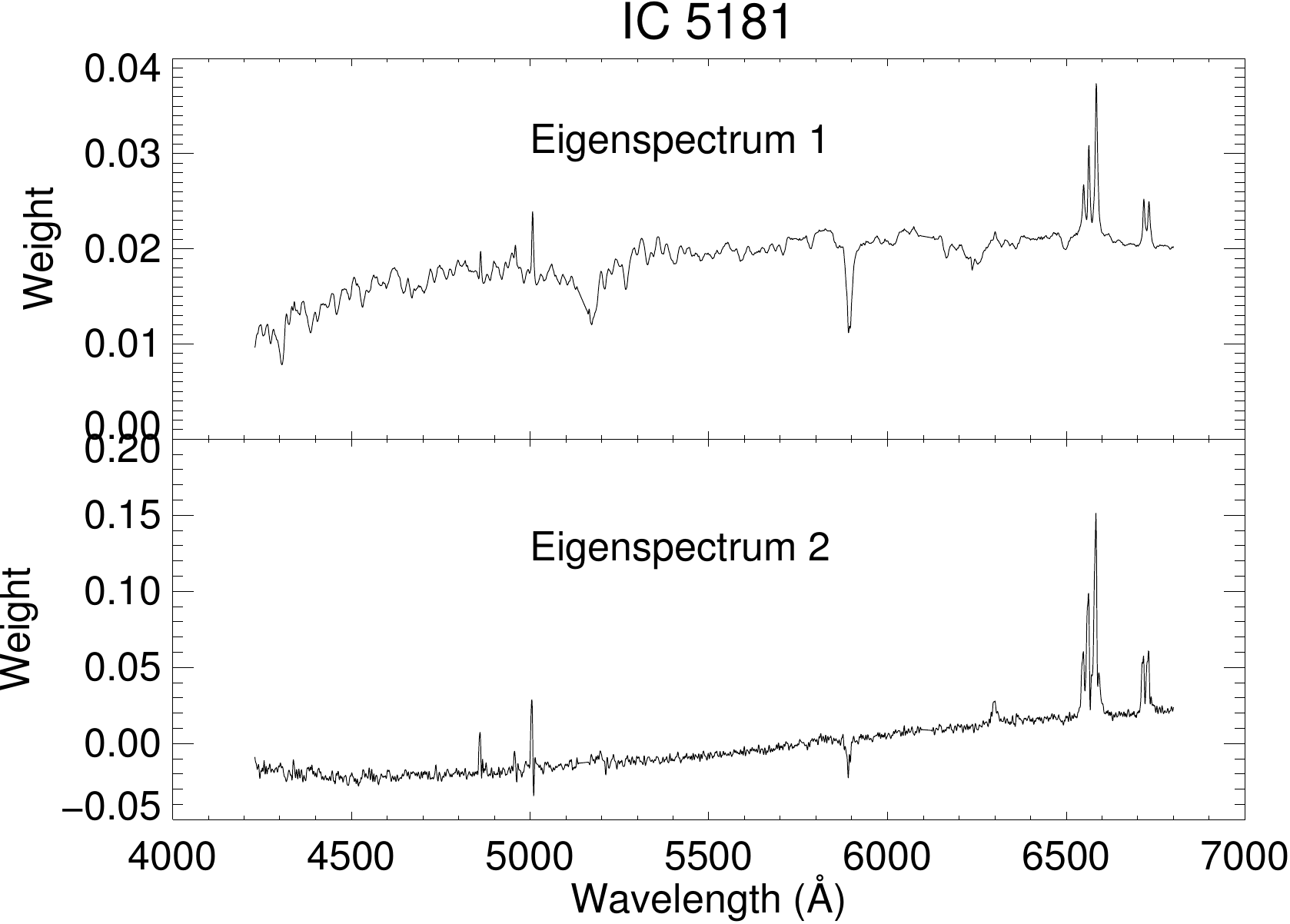}
\includegraphics[width=70mm,height=55mm]{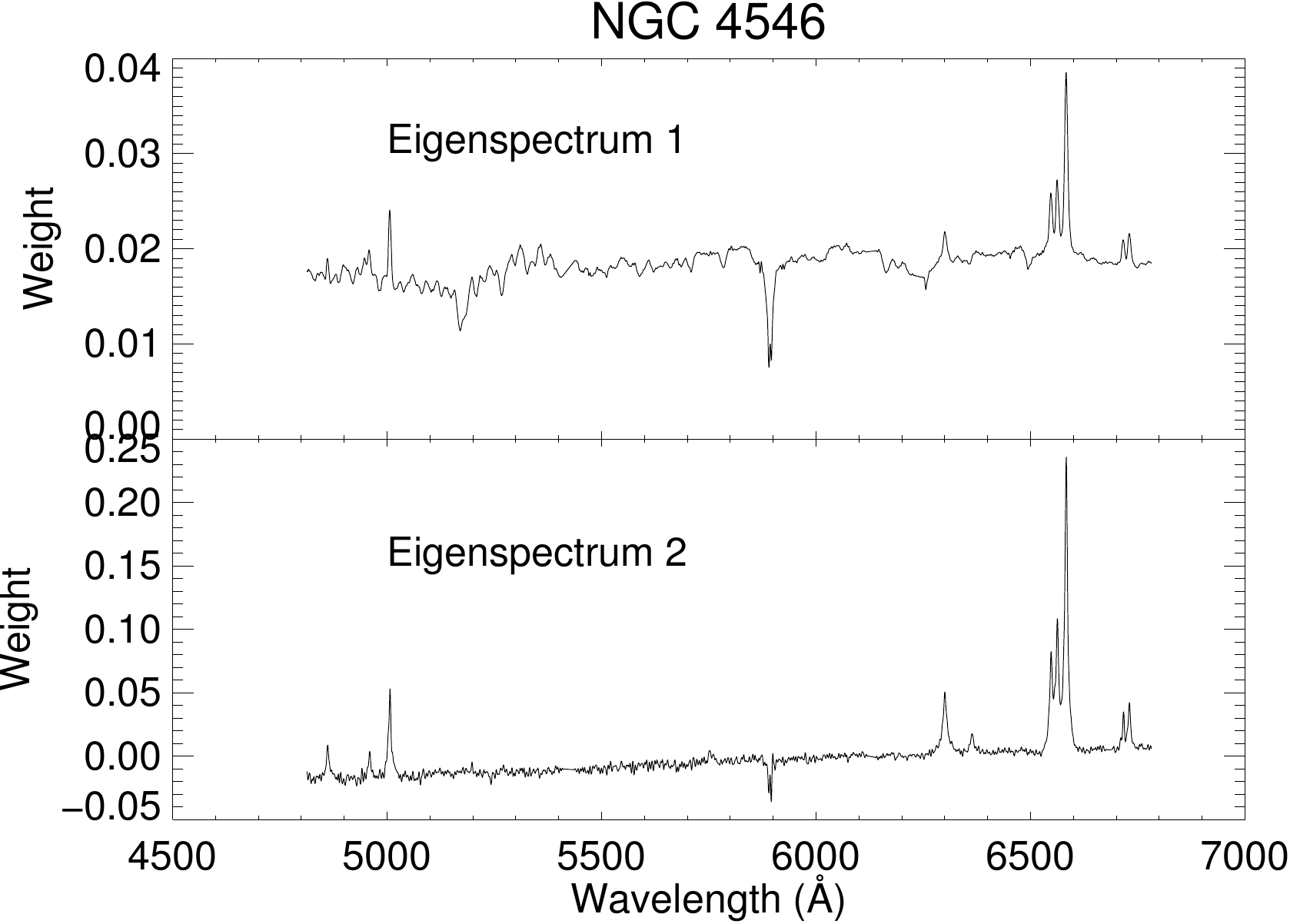}
\includegraphics[width=70mm,height=55mm]{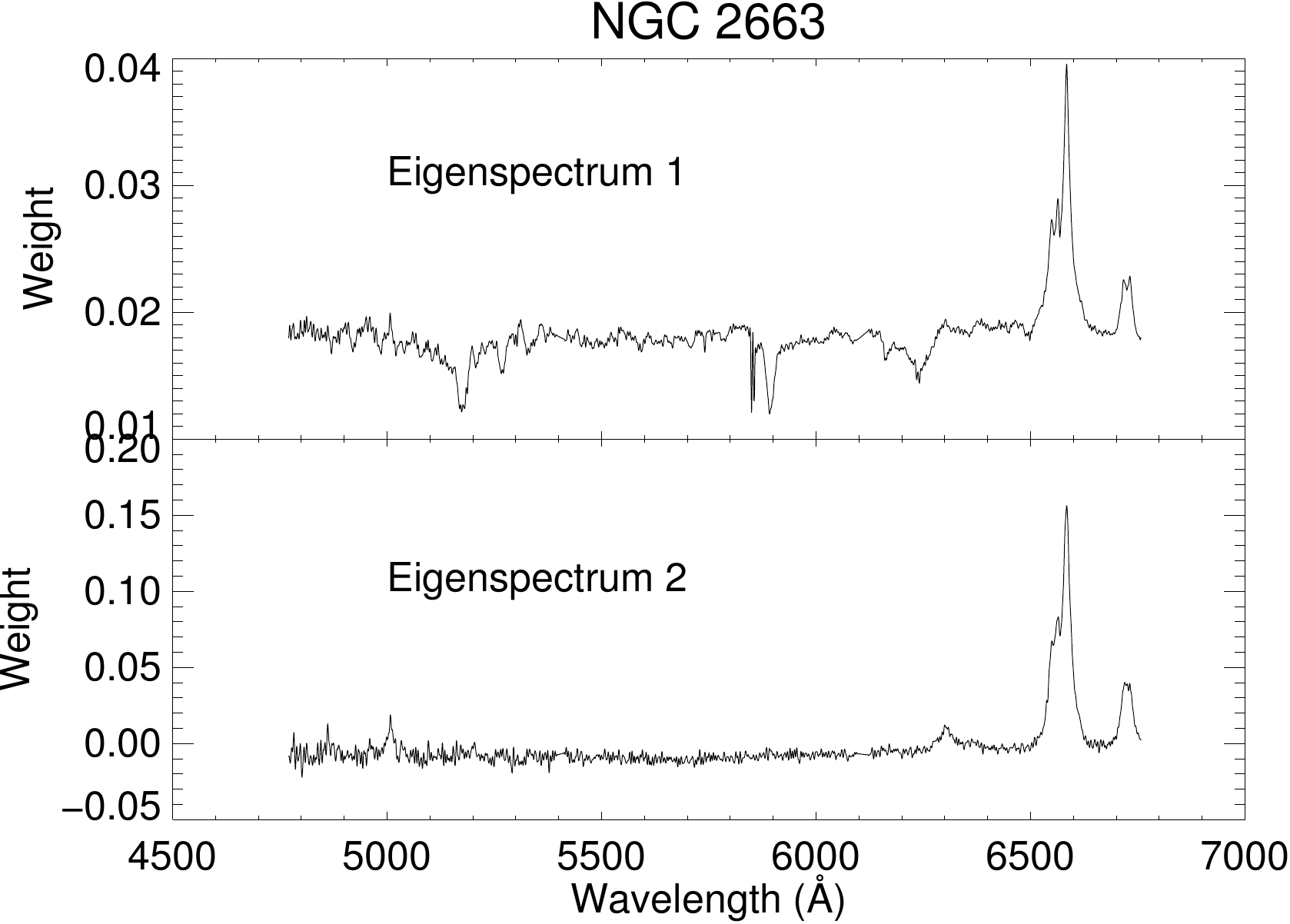}
\includegraphics[width=70mm,height=55mm]{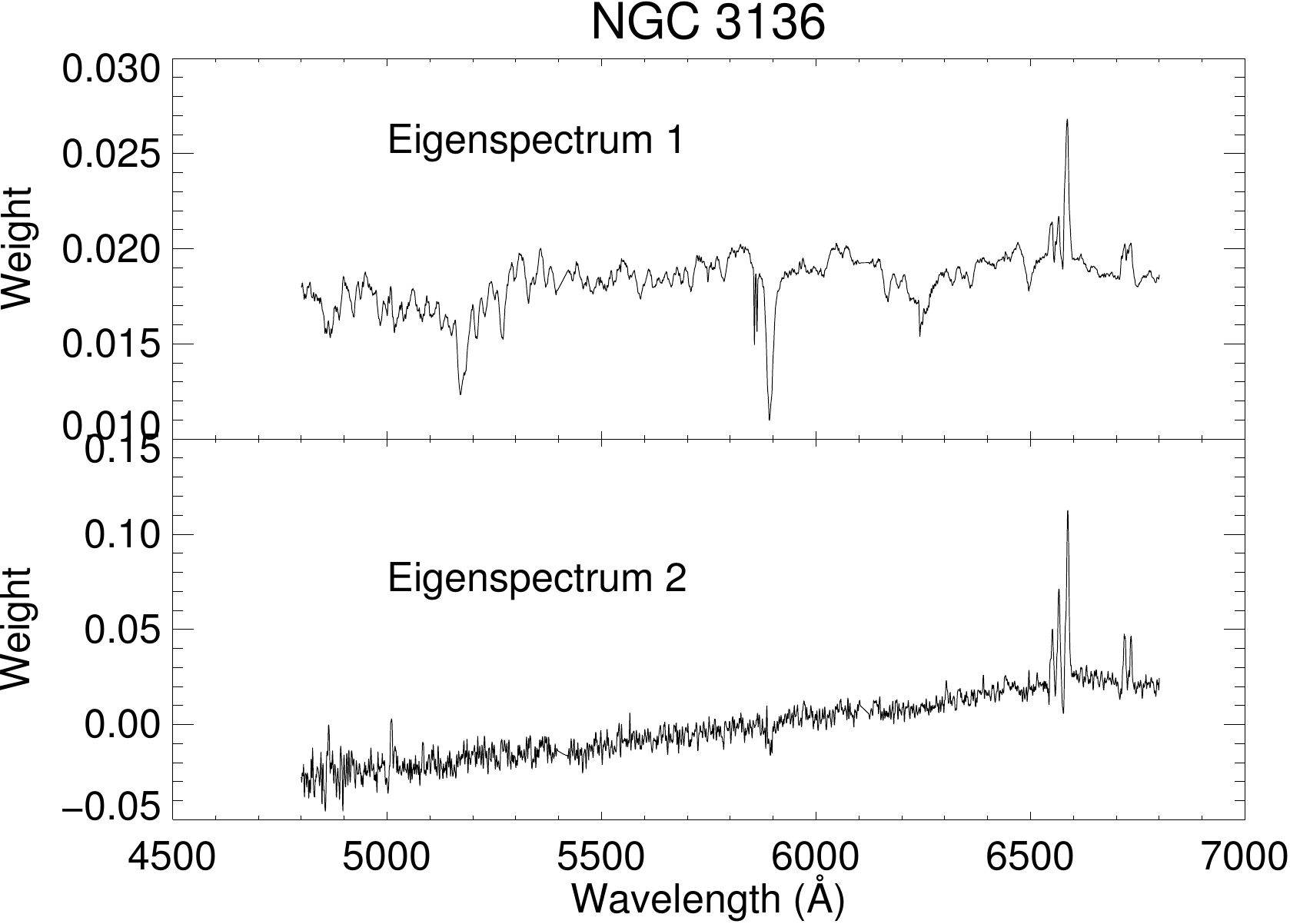}
\caption{First and second eigenspectra resulting from PCA Tomography applied to eight data cubes of the sample, covering the spectral range from 4325 - 6800 \AA. The first eigenspectra are, typically, similar to the average spectra of the data cubes. In the second eigenspectra, the gas component is efficiently isolated from the star light, allowing a more accurate analysis of the galactic nuclei emission lines.  \label{autoespectros_AGN}
}
\end{center}
\end{figure*}

\begin{figure*}
\begin{center}
\includegraphics[width=70mm,height=55mm]{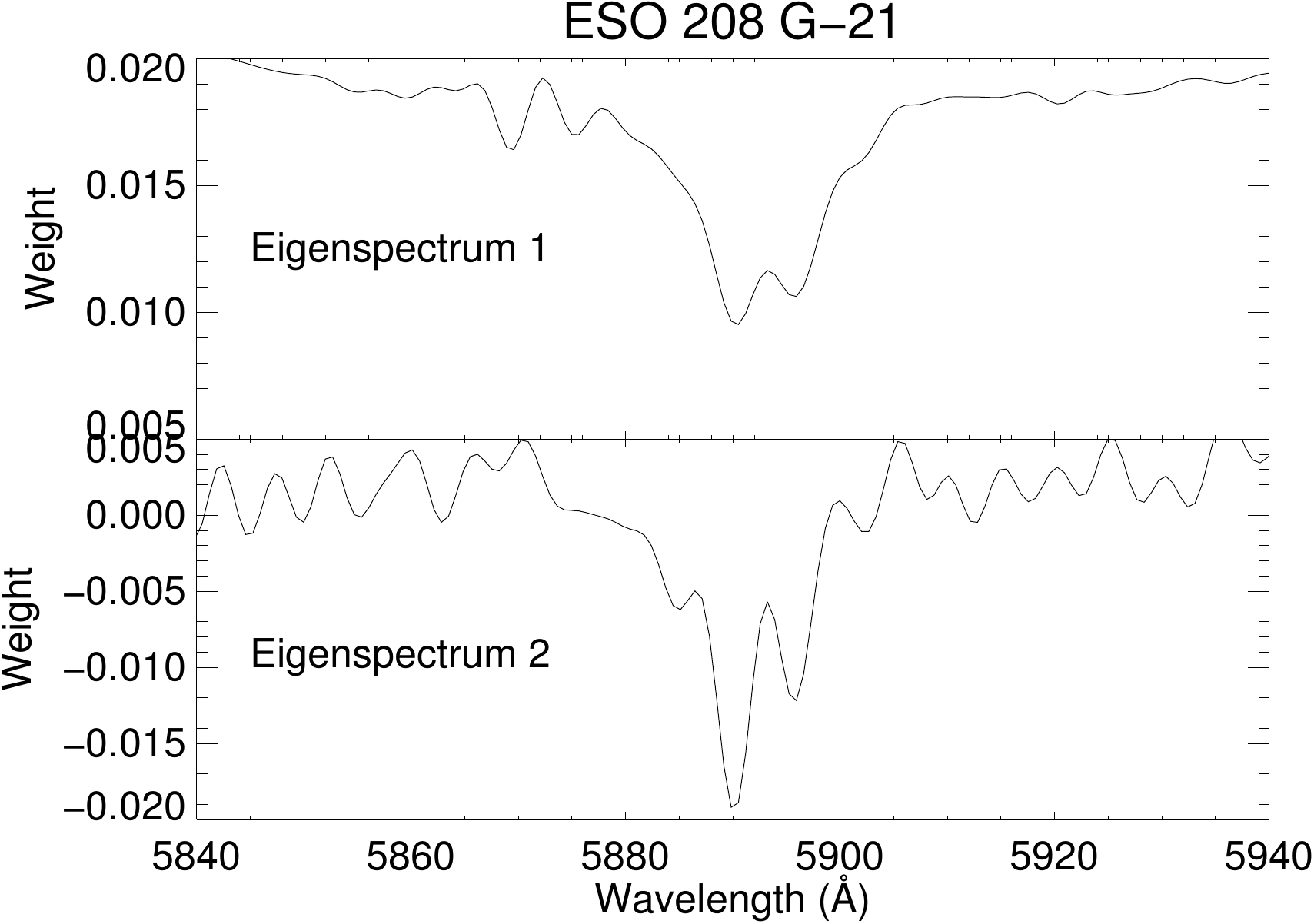}
\includegraphics[width=70mm,height=55mm]{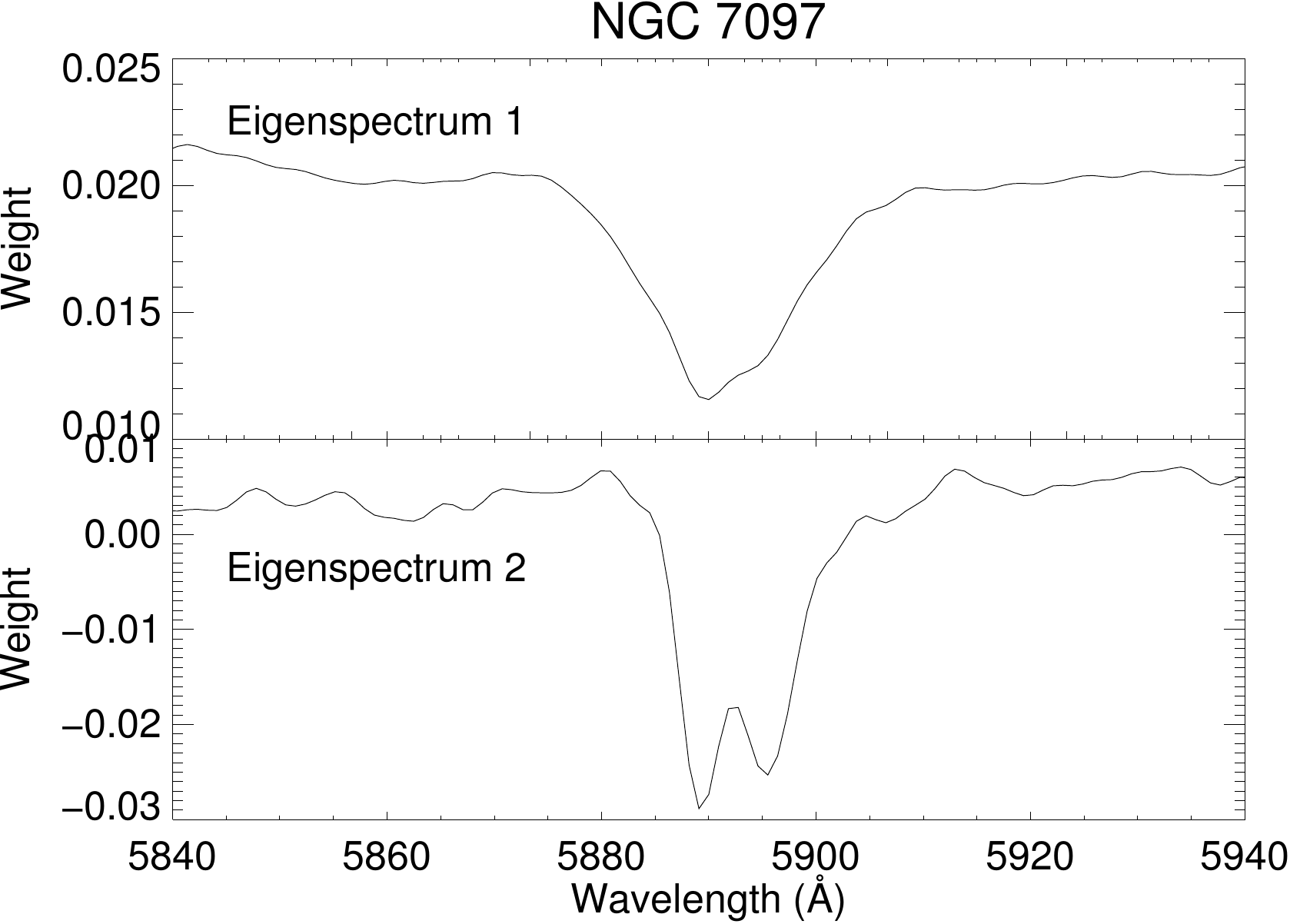}
\caption{Same as in Fig. \ref{autoespectros_AGN}, but focusing on Na D lines. Both ESO 208 G-21 and NGC 7097 have the reddest eigenspectrum 2 among the sample. Note that in the first eigenspectra, the Na D lines are blended together since, in this case, the stellar component plays an important role. However in eigenspectra 2, both 5891 and 5896\AA\ lines are seen, which indicates that the interstellar component is dominant in this case. Thus, the reddening in eigenspectra 2 in both galaxies and also in IC 5181, NGC 1380, NGC 3136 and NGC 4546 may be mainly caused by dust associated with neutral gas (Na D).   \label{autoespectros_AGN_on_NaD}
}
\end{center}
\end{figure*}

\begin{figure}
\includegraphics[scale=0.4]{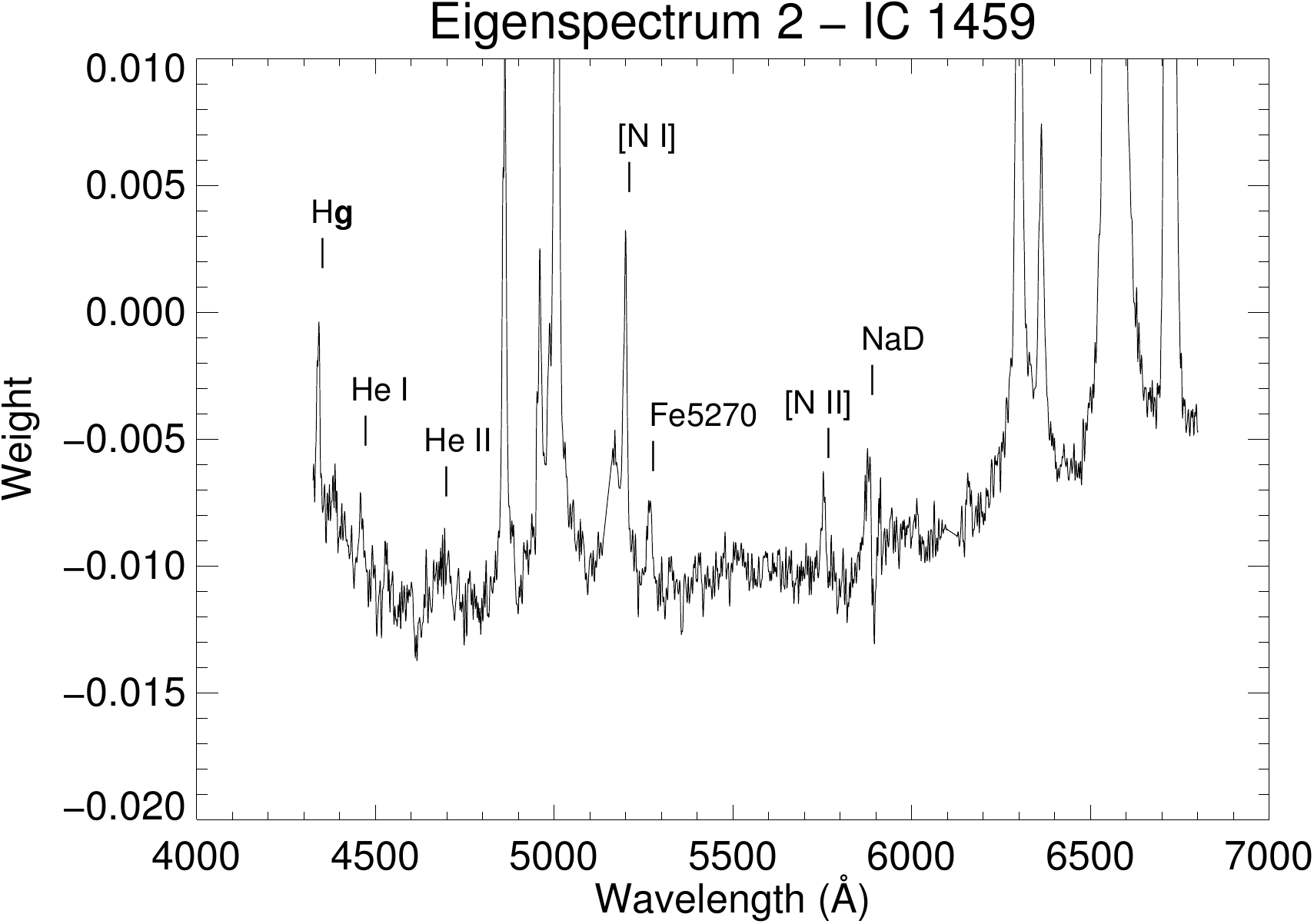}
\caption{Eigenspectrum 2 of IC 1459, focusing on the weak lines. Note the correlation between the emission lines (H$\gamma$, He I, He II, [N I] and [N II]) and the absorption lines (Fe5270 and Na D - stellar component). This result may be caused by a featureless continuum, since this component decreases the equivalent width of the stellar lines in the nuclear region. } \label{eigen_2_ic1459}
\end{figure}

PCA Tomography was applied to the other nine data cubes in the spectral range between 4325 and 6800 \AA. In all eight galaxies where emission lines were previously known, the eigenvectors 2 and their respective tomograms revealed features typical of an AGN, as shown in Figs. \ref{autoespectros_AGN}, \ref{autoespectros_AGN_on_NaD}, \ref{eigen_2_ic1459}, \ref{fig_gas_disc_1}, \ref{fig_gas_disc_2} and \ref{fig_gas_disc_3}. In these objects, the correlation between [N II], H$\alpha$ and [S II] lines resembles LINERs spectra. Only in ESO 208 G-21, the H$\beta$ and [O III] lines are not intense enough to overcome the noise. In four objects, the [O I] lines are intense, whereas in the other four, these lines are quite weak, but detectable. The tomograms showed that the projections of the data cubes of the galaxies of the sample into the eigenvectors 2  resulted in point-like sources. In Figs. \ref{fig_gas_disc_1}, \ref{fig_gas_disc_2} and \ref{fig_gas_disc_3}, only the tips of the weights of the tomograms related to the second eigenvectors of the galaxies of the sample are shown in green. In other words, we show only the tips of the PSFs related to the AGN of the galaxies, since the projection of the data cubes on to their respective eigenspectra 2 produced higher weights in the position of the AGNs. The eigenvalues of the first three eigenvectors are presented in table \ref{eigenvalues_whole_pca}.

In ESO 208 G-21, NGC 1380, IC 5181, NGC 4546 and NGC 7097, the red component of the continuum is correlated with the emission lines in eigenspectrum 2. It is an indicative that the second eigenvector applies a ``correction'' to eigenvector 1, which has red spectral characteristics. Three possible explanations are proposed: 1 - the AGN has a featureless continuum that is intrinsically red, i.e. redder than the nuclear stellar populations in this spectral range; 2 - the featureless continuum is intrinsically blue, but reddened in its line of sight (e.g. dust extinction); 3 - the nuclear stellar population is redder than the circumnuclear stellar population. If an extinction is caused by the interstellar medium (ISM), then a correlation between the red component of the featureless continuum and the Na D lines is expected, since Na D lines are partially produced by the ISM. In Fig. \ref{autoespectros_AGN_on_NaD}, we show the eigenspectra 1 and 2 of ESO 208 G-21 and NGC 7097 focusing on the Na D lines. In eigenspectra 1, the Na D lines are blended, since the first eigenspectra in all objects correspond to the stellar component of the galaxies. However, both 5891\AA\ and 5896\AA\ Na I lines are seen in eigenspectra 2 of the two galaxies. Although we have shown only the cases of ESO 208 G-21 and NGC 7097, it also occurs for IC 5181, NGC 1380, NGC 4546 and more weakly in NGC 3136. Since the Na D that is produced by the ISM has a lower velocity dispersion when compared to the stellar populations absorption lines, then we may conclude that the reddening of the featureless continuum in eigenspectra 2 of these five galaxies is mainly caused by dust extinction contained in the ISM. In NGC 2663, no Na D lines are seen in eigenspectrum 2, indicating that the light from its nucleus does not suffer a substantial influence from the ISM. For IC 1459, we propose that the featureless continuum is detected in eigenspectrum 2. In Fig. \ref{eigen_2_ic1459}, we show the second eigenspectrum of IC 1459, but focusing on the weak lines. Note that the emission lines are correlated with the Fe5270 stellar line and with the stellar component of Na D. Since a featureless continuum decreases the equivalent width of the stellar lines on the nuclear region, a correlation between the emission lines and the absorption lines is expected in this case.

In the galaxies IC 1459, IC 5181, NGC 1380 and NGC 7097, whose spectra cover the 4300-4800\AA\ range, a blue upturn is seen in their eigenspectra 2. One should be very careful when interpreting this. One possibility is that it may be caused by the featureless continuum. However, we see the correlation between the emission lines and the stellar lines only in IC 1459. Even if the correlation involving the lines is overcome by the noise, there is no way to assert that the blue upturn is caused by the featureless continuum. Another scenario is that this upturn could be caused by a young stellar population in the nuclear region of the galaxies. In this case, an upturn that is seen close to 5700\AA\ in NGC 2663 could be the result of a blue stellar population. However, only a careful analysis on the stellar population properties would be able to provide a reasonable interpretation for the blue upturns seen in the eigenspectra of these objects. 

\renewcommand{\thefigure}{\arabic{figure}\alph{subfigure}}
\setcounter{subfigure}{1}

\begin{figure*}
\begin{center}
\includegraphics[scale=0.4]{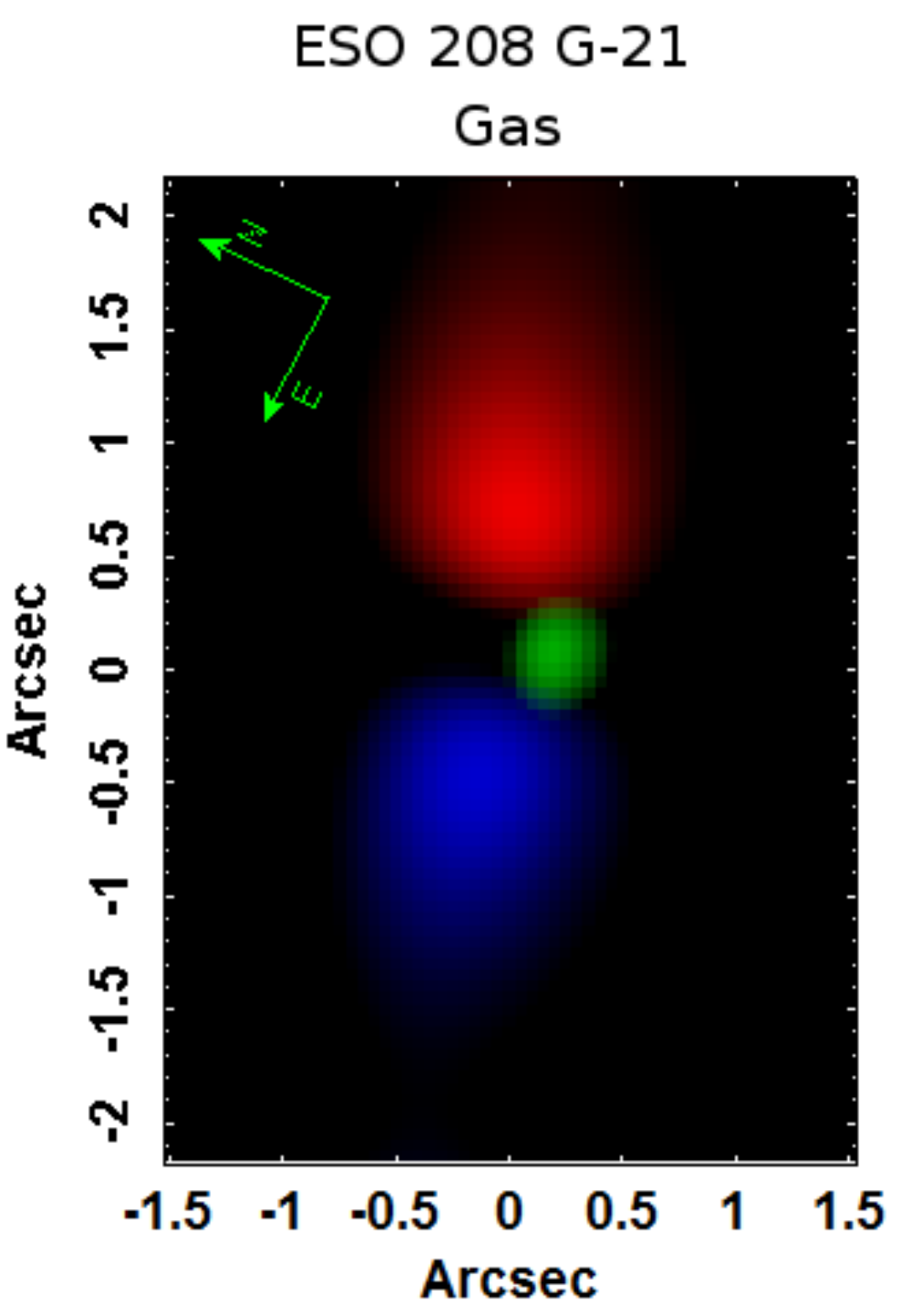}
\includegraphics[width=70mm,height=60mm]{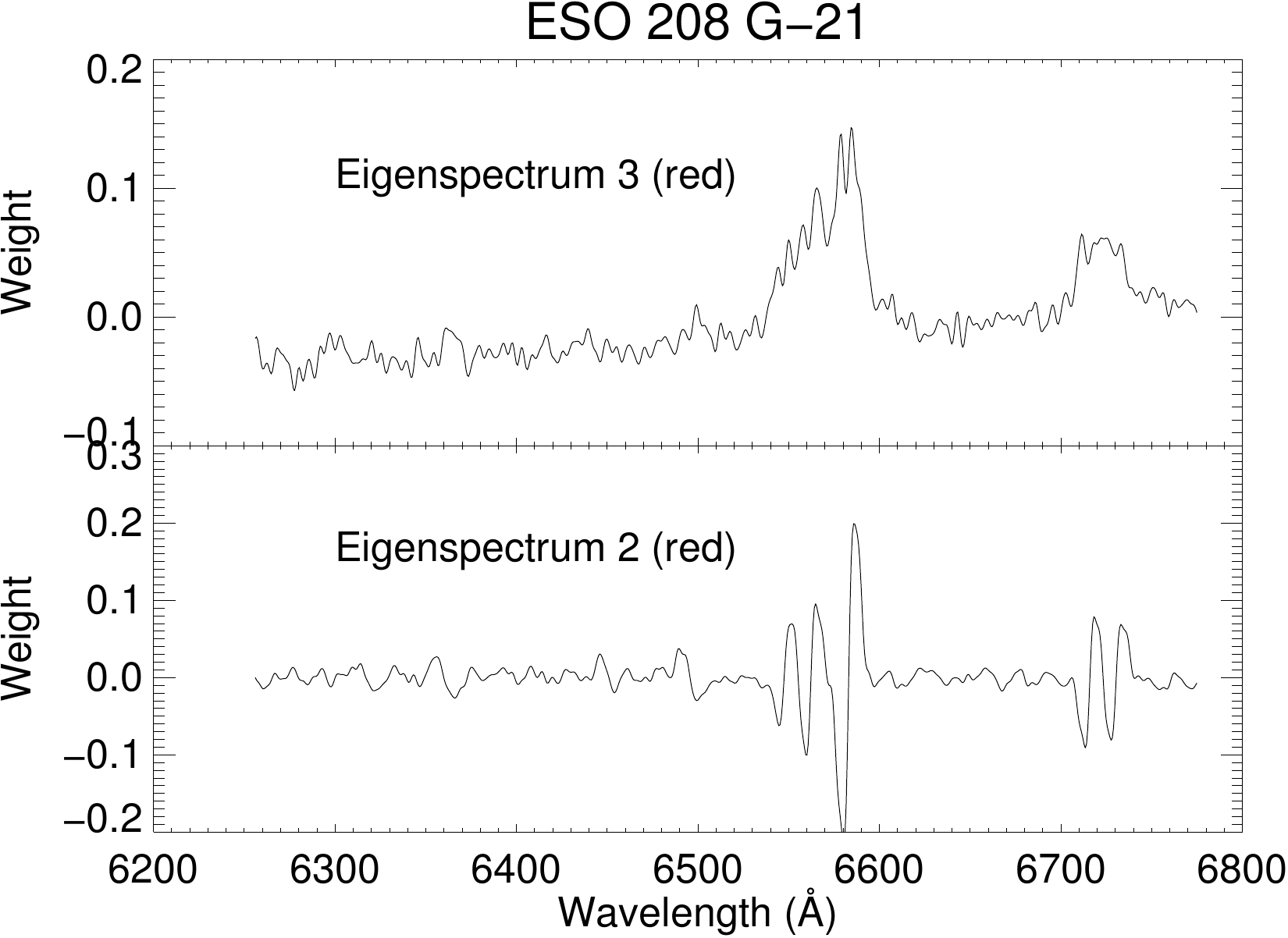}
\vspace{0.5cm}

\includegraphics[scale=0.4]{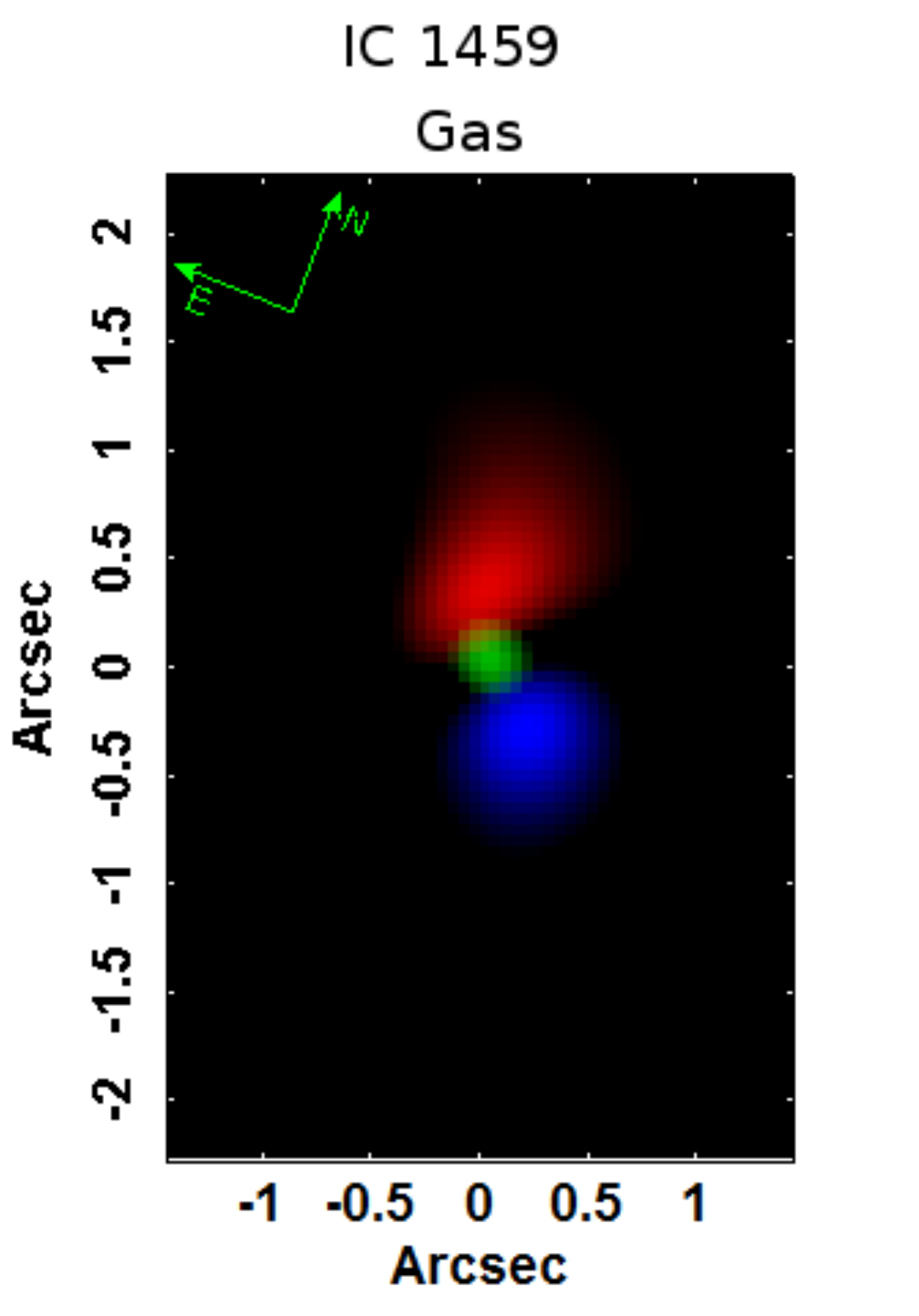}
\includegraphics[width=70mm,height=60mm]{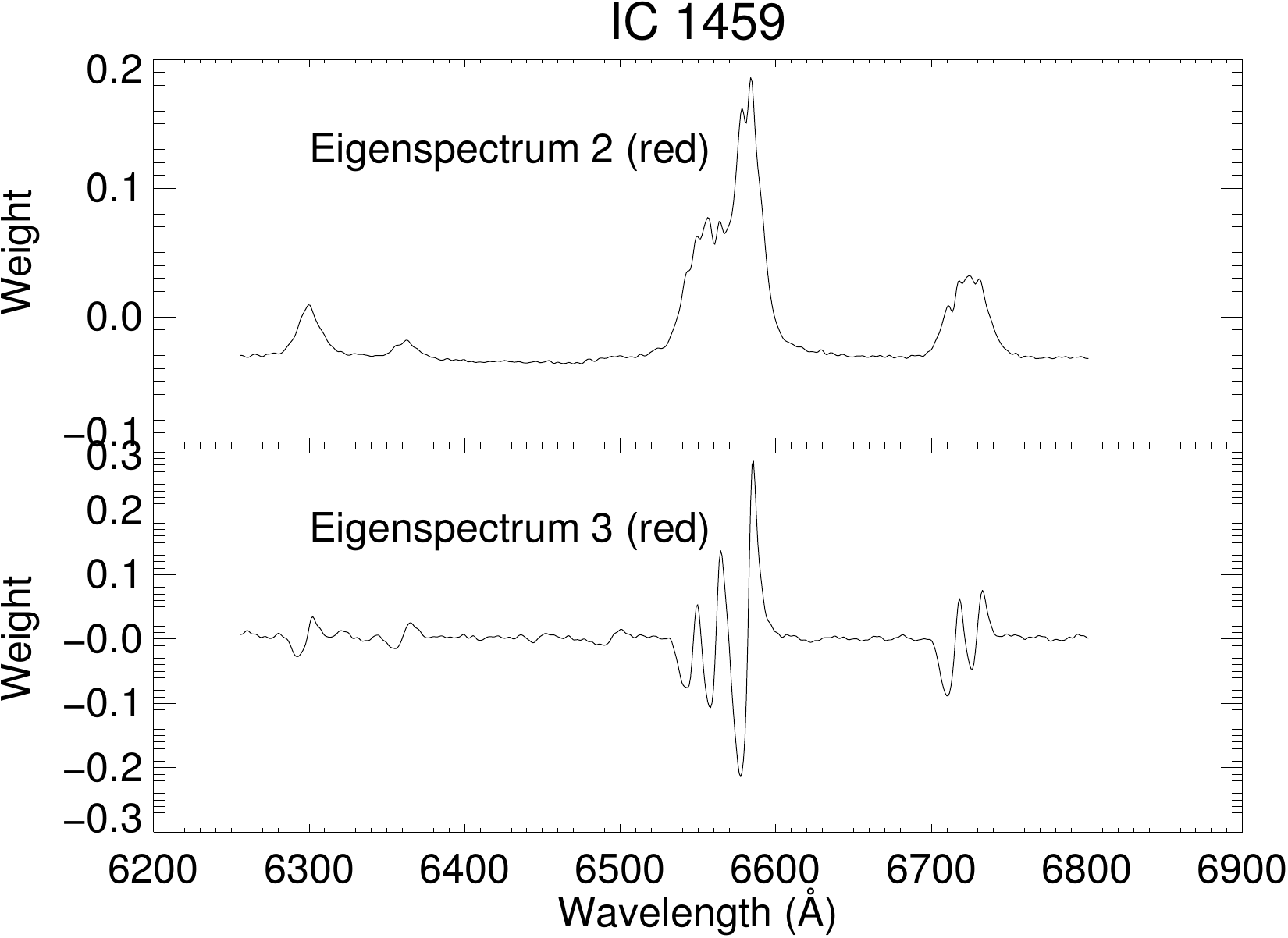}
\vspace{0.5cm}

\includegraphics[scale=0.4]{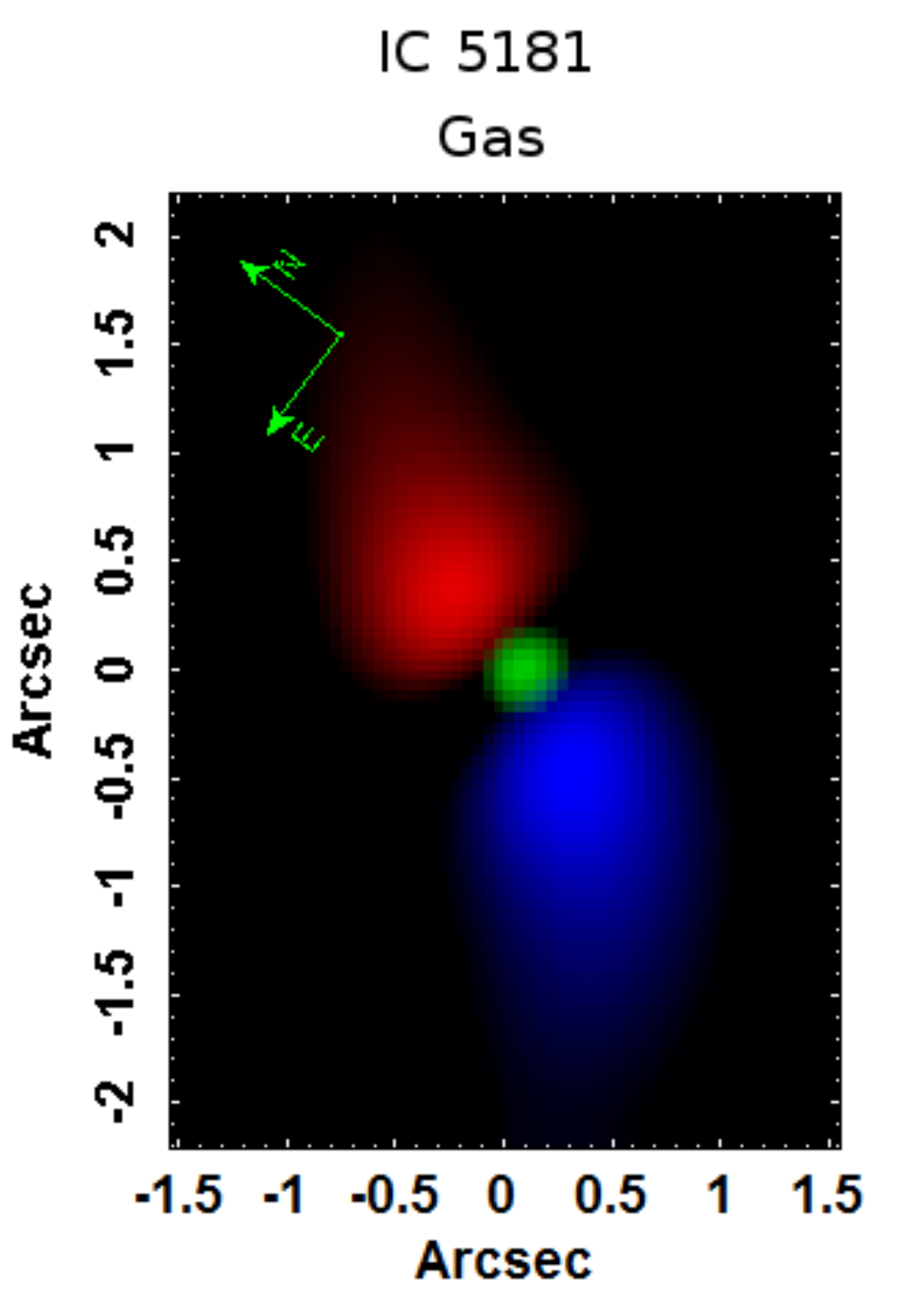}
\includegraphics[width=70mm,height=60mm]{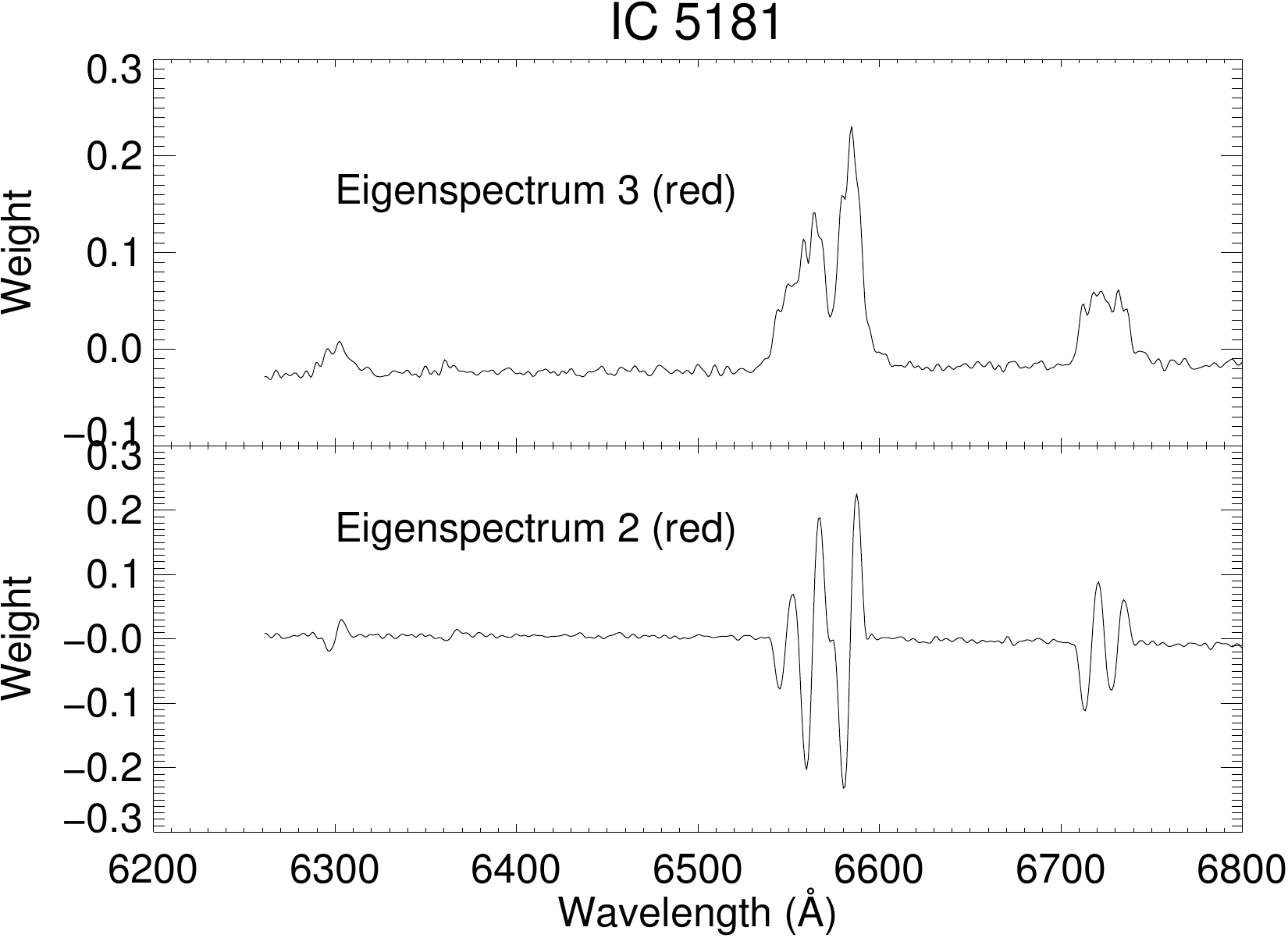}
\vspace{0.5cm}

\caption{PCA Tomography applied to the red region of the data cubes for the galaxies ESO 208 G-21, IC 1459 and IC 5181. In the tomograms shown above, regions blueshifted (redshifted) when compared to the galactic centres are shown in blue (red). Both regions were detected in the same tomogram, where each colour is a pole belonging to the bipolar structure. The tomograms that correspond to the AGNs (Fig. \ref{autoespectros_AGN}) are shown in green. In the eigenspectra, the kinematic signature is characterized by the anti correlation between the red and blue wings of the emission lines. 
\label{fig_gas_disc_1}
}
\end{center}
\end{figure*}

\addtocounter{figure}{-1}
\addtocounter{subfigure}{1}

\begin{figure*}
\begin{center}
\includegraphics[scale=0.4]{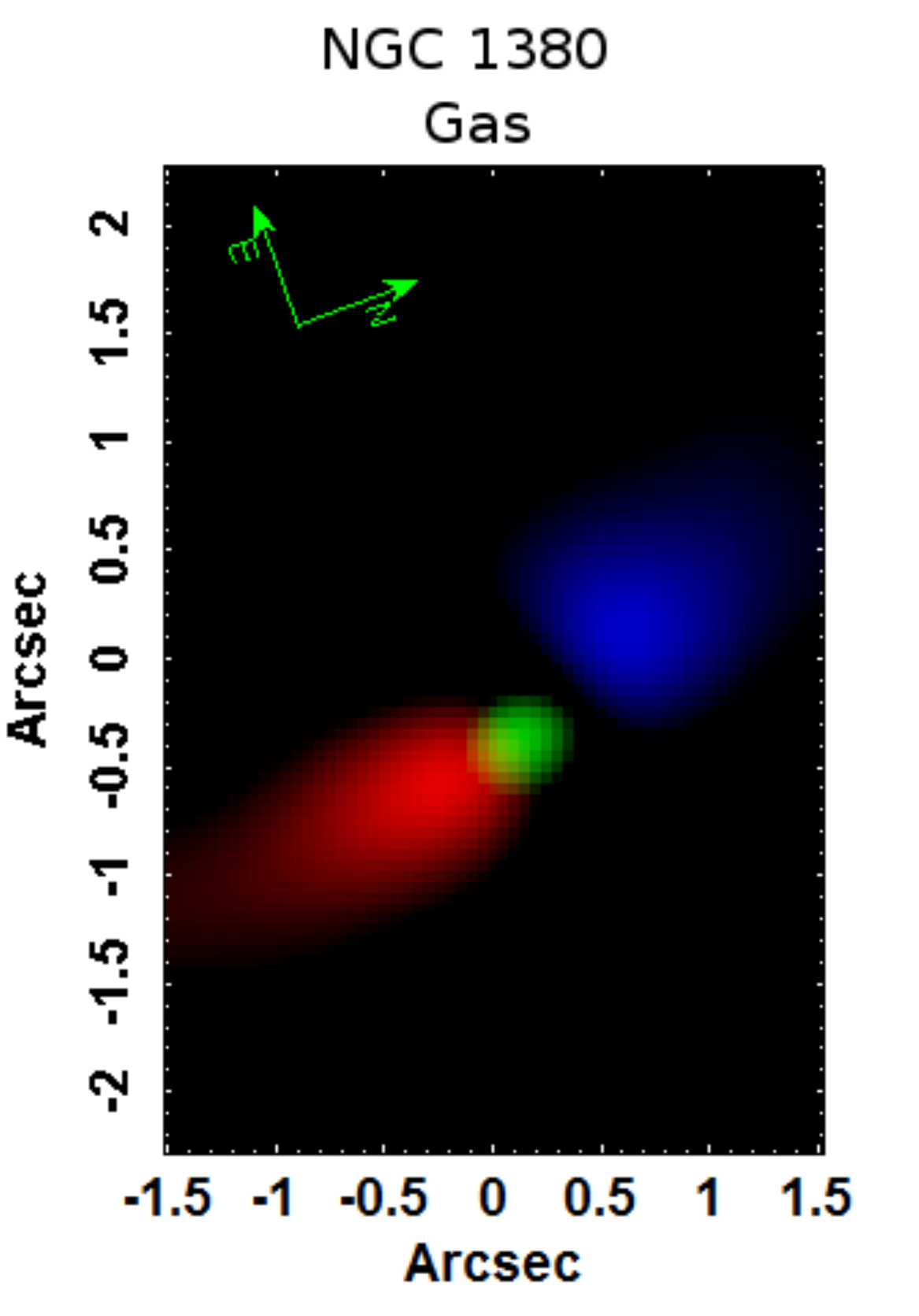}
\includegraphics[width=70mm,height=60mm]{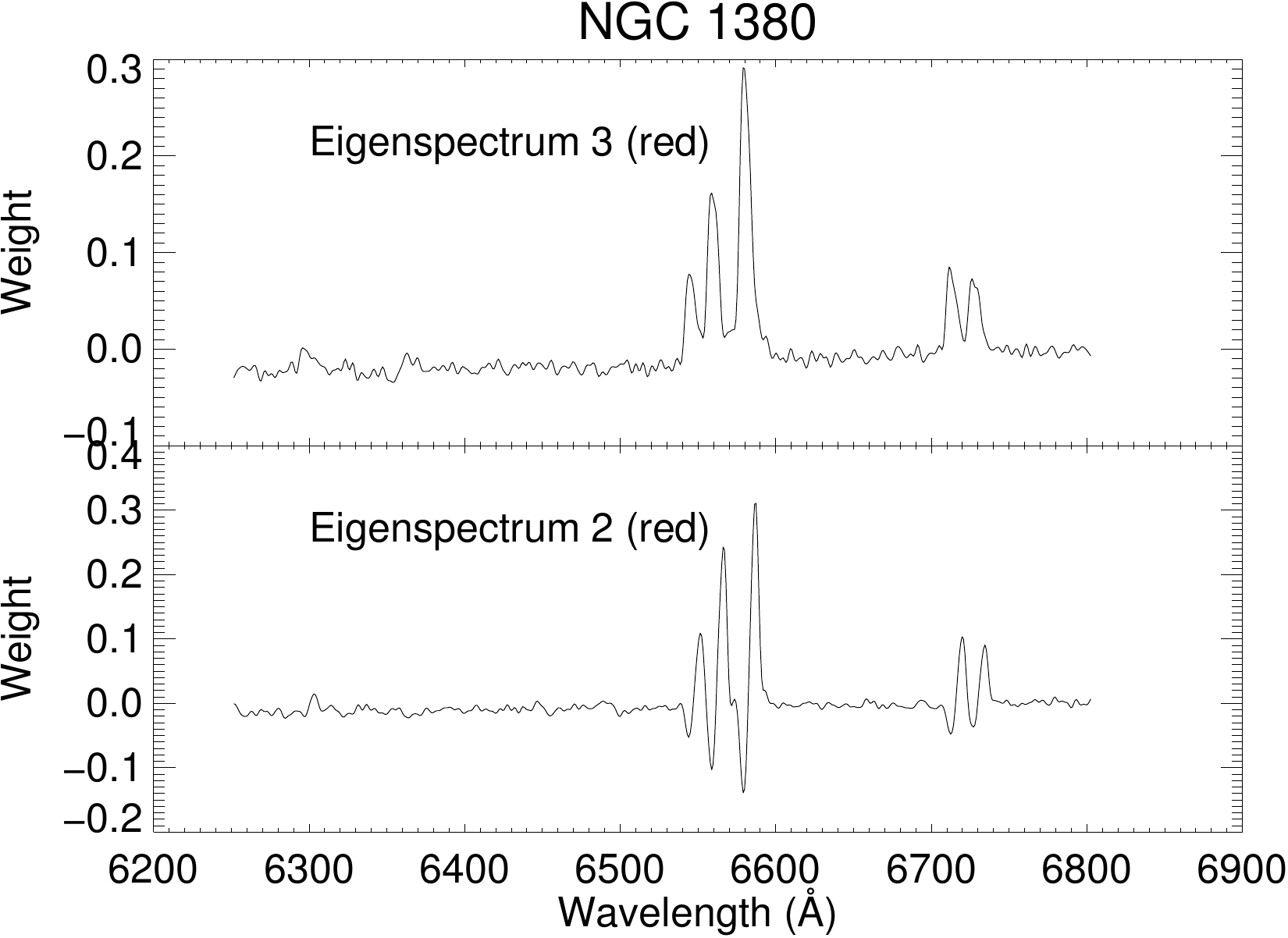}
\vspace{0.5cm}

\includegraphics[scale=0.4]{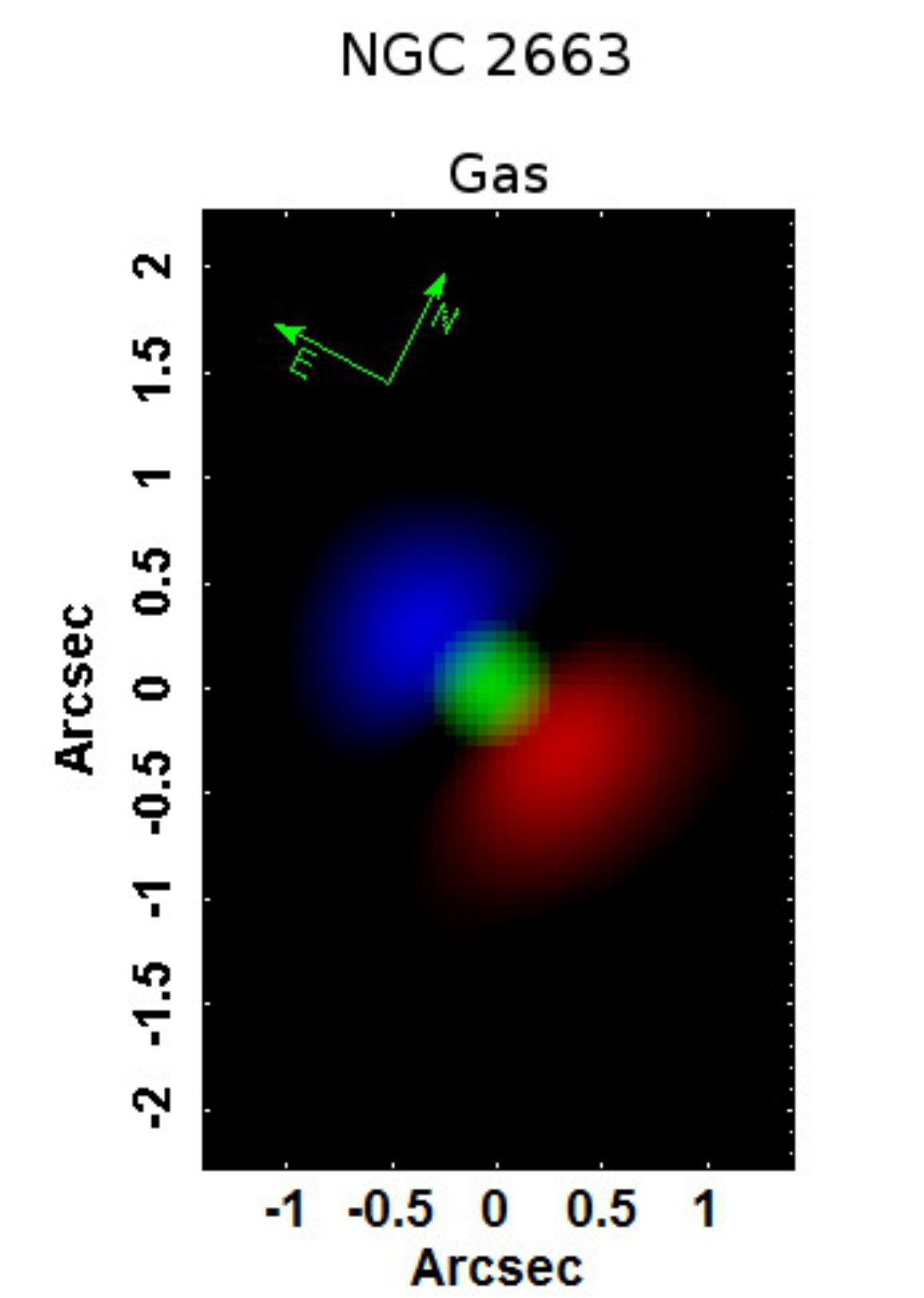}
\includegraphics[width=70mm,height=60mm]{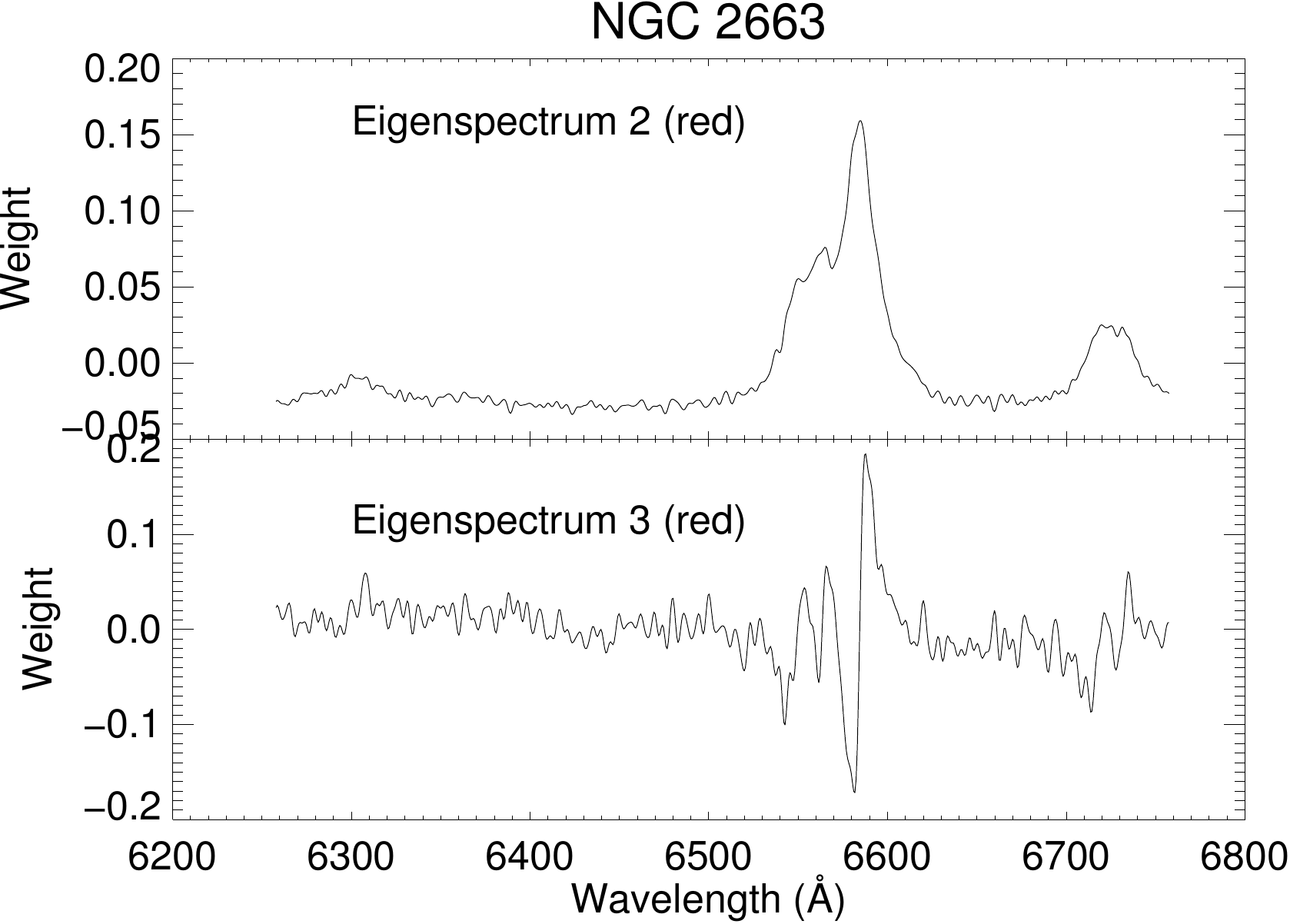}
\vspace{0.5cm}

\includegraphics[scale=0.4]{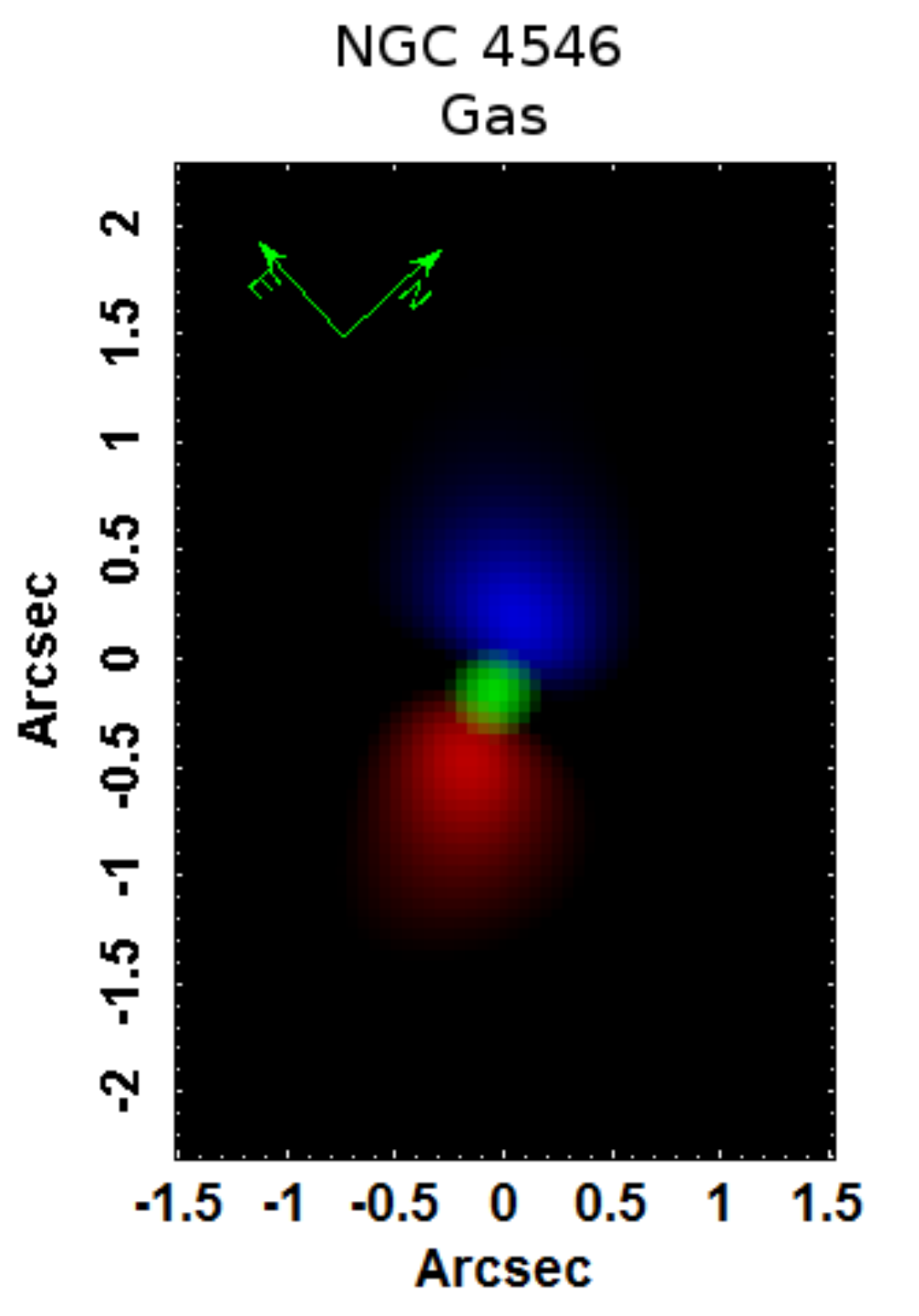}
\includegraphics[width=70mm,height=60mm]{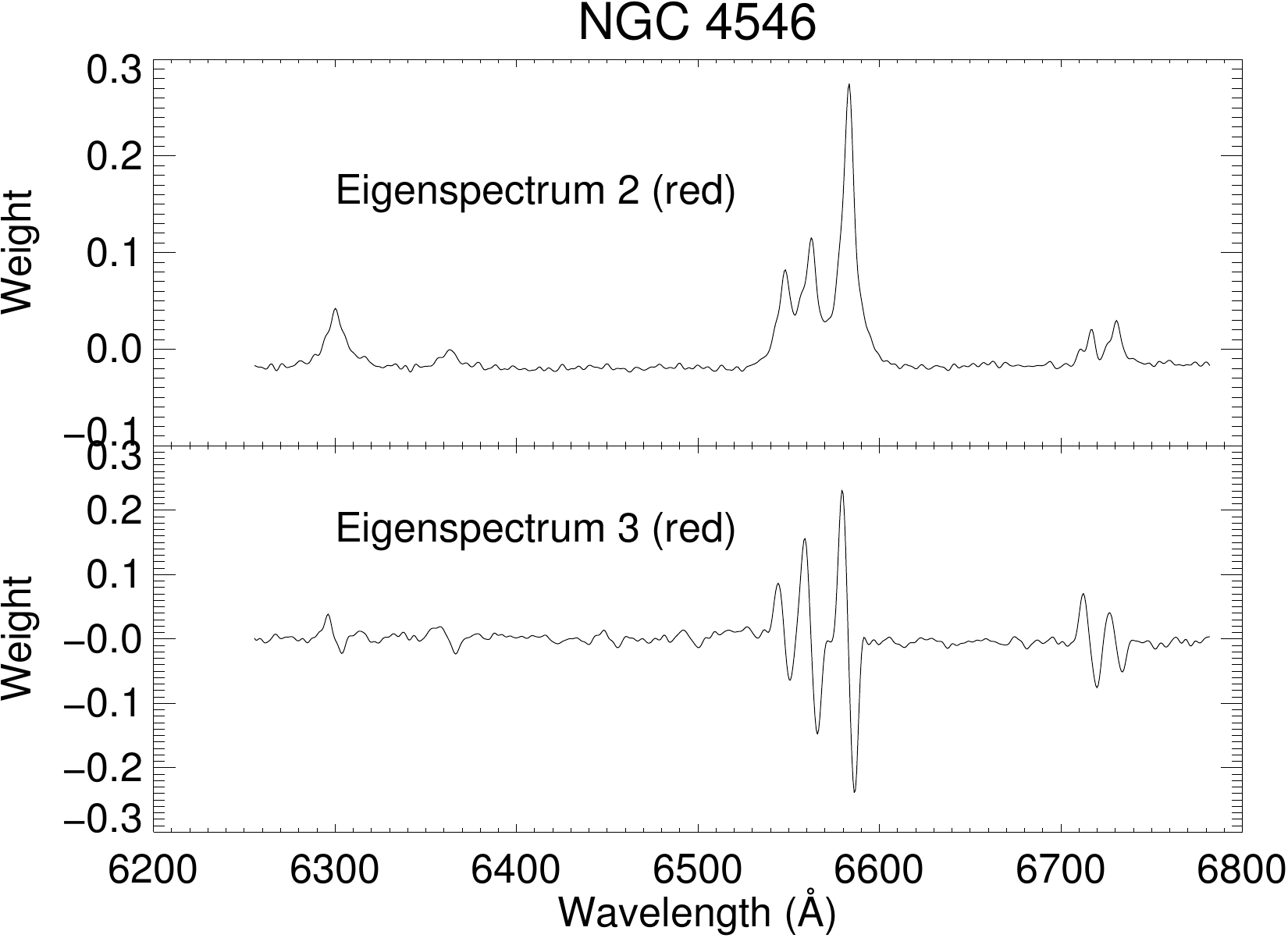}
\vspace{0.5cm}

\caption{The same as in Fig. \ref{fig_gas_disc_1} for the galaxies NGC 1380, NGC 2663 and NGC 4546.  
\label{fig_gas_disc_2}
}
\end{center}
\end{figure*}

\addtocounter{figure}{-1}
\addtocounter{subfigure}{1}

\begin{figure*}
\begin{center}
\includegraphics[scale=0.4]{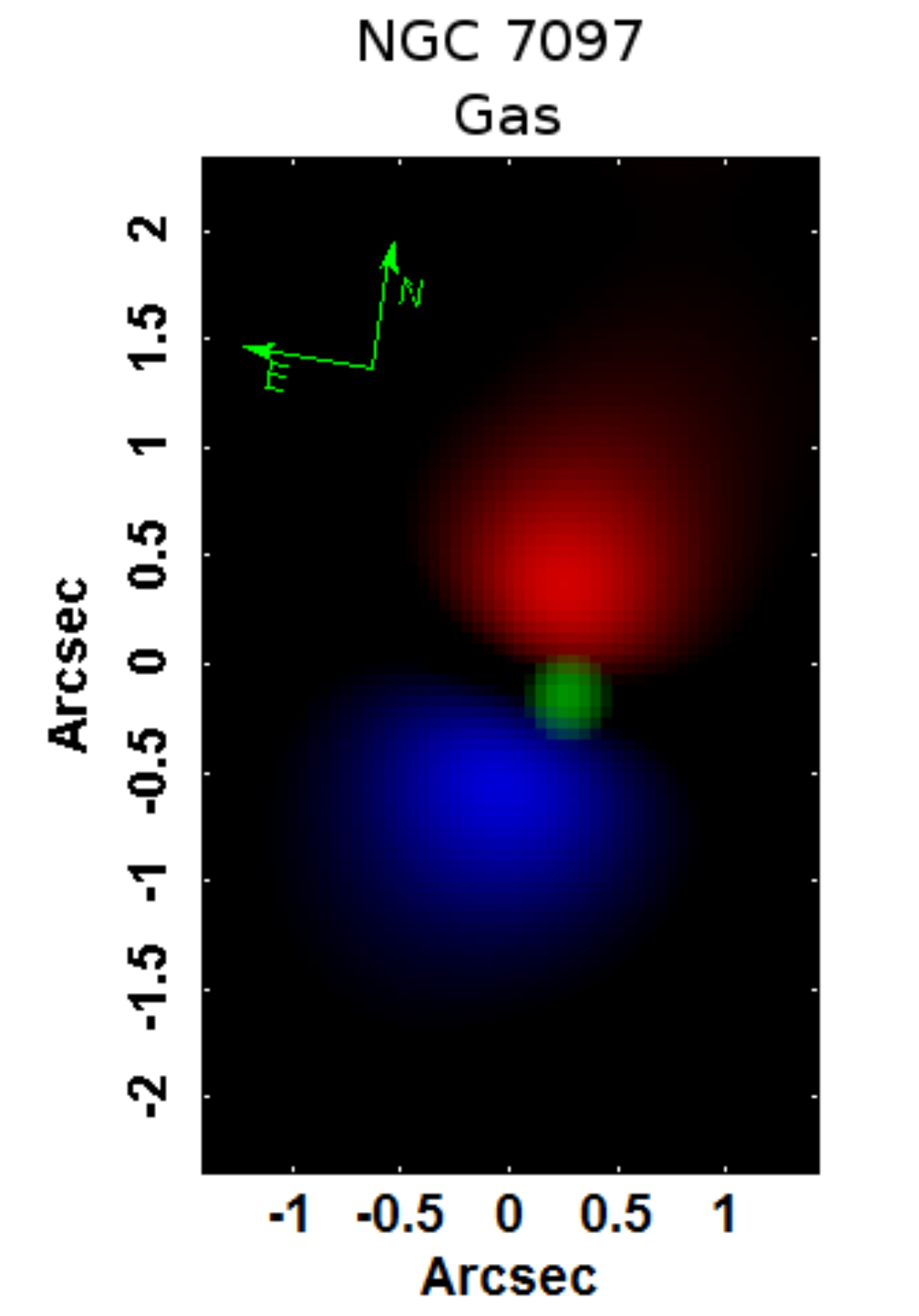}
\includegraphics[width=70mm,height=60mm]{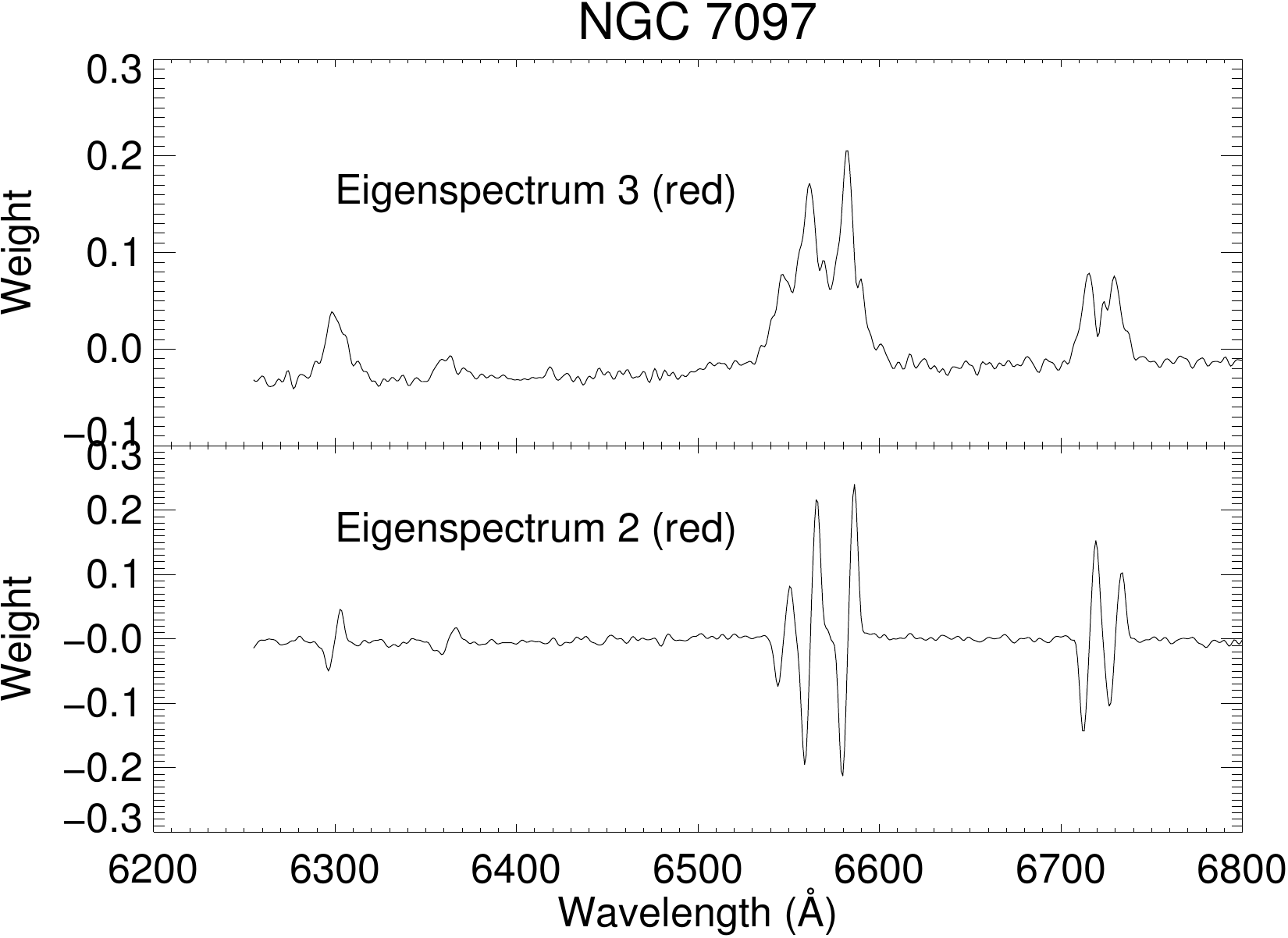}
\vspace{0.5cm}

\caption{The same as in Fig. \ref{fig_gas_disc_1} for the galaxy NGC 7097. 
\label{fig_gas_disc_3}
}
\end{center}
\end{figure*}

\renewcommand{\thefigure}{\arabic{figure}}

\begin{table}
 \scriptsize
 \begin{center}
 \begin{tabular}{@{}lccc}
  \hline
  Galaxy name & $\Lambda_1$ (\%) & $\Lambda_2$ (\%) & $\Lambda_3$ (\%) \\
  \hline
  ESO 208 G-21 & 99.385 & 0.397 & 0.123 \\
  IC 1459 & 97.998 & 1.653 & 0.175\\
  IC 5181 & 99.666 & 0.184 & 0.068\\
  NGC 1380 & 99.505 & 0.361 & 0.050\\
  NGC 2663 & 98.565 & 0.942 & 0.062\\
  NGC 3136 & 99.666 & 0.107 & 0.047\\
  NGC 4546 & 99.654 & 0.197 & 0.066\\
  NGC 7097 & 99.497 & 0.379 & 0.073\\
  \hline
 \end{tabular}
 \caption{Eigenvalues related to eigenspectra 1 - 3 from PCA Tomography applied to eight data cubes of the sample, covering the spectral range from 4325 to 6800 \AA. In all cases, the first eigenvectors correspond closely to the average spectra of the data cubes, which reveals mainly the star light from the central region of the galaxies. Fig. \ref{results_IC1459} shows the eigenspectrum and tomogram 1 associated with galaxy IC 1459. In this table, one can see that the first eigenspectra contain variances $\sim$ 99\%. The eigenvalues that correspond to eigenspectra 2 have variances $<$ 1\%, except in IC 1459. Nevertheless, the second eigenspectra, shown in Fig. \ref{autoespectros_AGN}, are related to the AGNs in eight galaxies of the sample. The third eigenvectors, with variances $<$ 0.2\%, are associated with gas and stellar kinematics of the galaxies. Only information on IC 1459 is shown in Fig. \ref{results_IC1459}. It is interesting to note that, even with a very small variance, PCA Tomography is able to extract important information from the data cube. \label{eigenvalues_whole_pca}
}
 \end{center}
\end{table}

The classification of AGNs in Seyferts, LINERs or TOs in the optical regime is done through emission line ratios. The eigenspectra 2 shown in Fig. \ref{autoespectros_AGN} revealed emission lines related to a point-like source in the centre of these eight galaxies. However, eigenspectra exhibit correlations between wavelengths. It is not adequate to classify AGNs using correlation ratios. Nonetheless, as seen in IC 1459, its eigenspectrum 2 clearly resembles a LINER spectrum. In all eight objects whose eigenvectors 2 are shown in Fig. \ref{autoespectros_AGN}, the [N II]$\lambda$6584/H$\alpha$ correlation ratio $>$ 1.2. 

The point related to the classification of these AGNs following the flux ratio of the emission lines will be addressed in Paper II.

\subsubsection{Non AGN detection in NGC 1399 and NGC 1404} \label{fail_cases}

PCA Tomography applied to the data cubes of NGC 1399 and NGC 1404 did not detect any sign of emission lines. This result does not exclude the possibility of gas emission. In fact, in Paper II, we will show that [N II]+H$\alpha$ lines are unveiled in both objects. A hypothesis is that the variance of these emission lines is similar to the variance caused by the noise. This is highlighted in Paper II, where we will show that the extraction of information from these emission lines is quite complicated. 

\begin{figure*}
\includegraphics[scale=0.4]{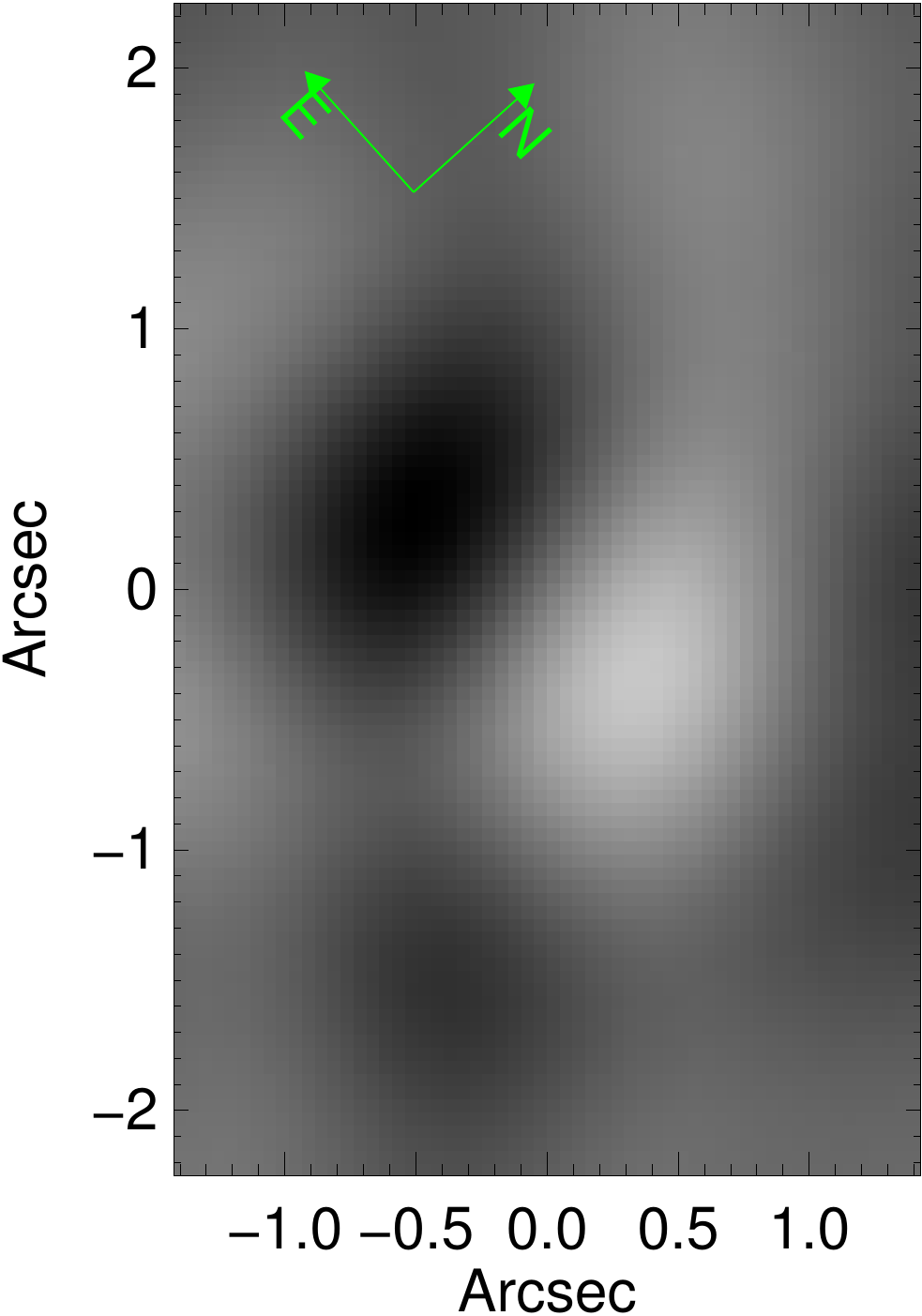}
\vspace{1cm}
\hspace{3cm}
\includegraphics[scale=0.4]{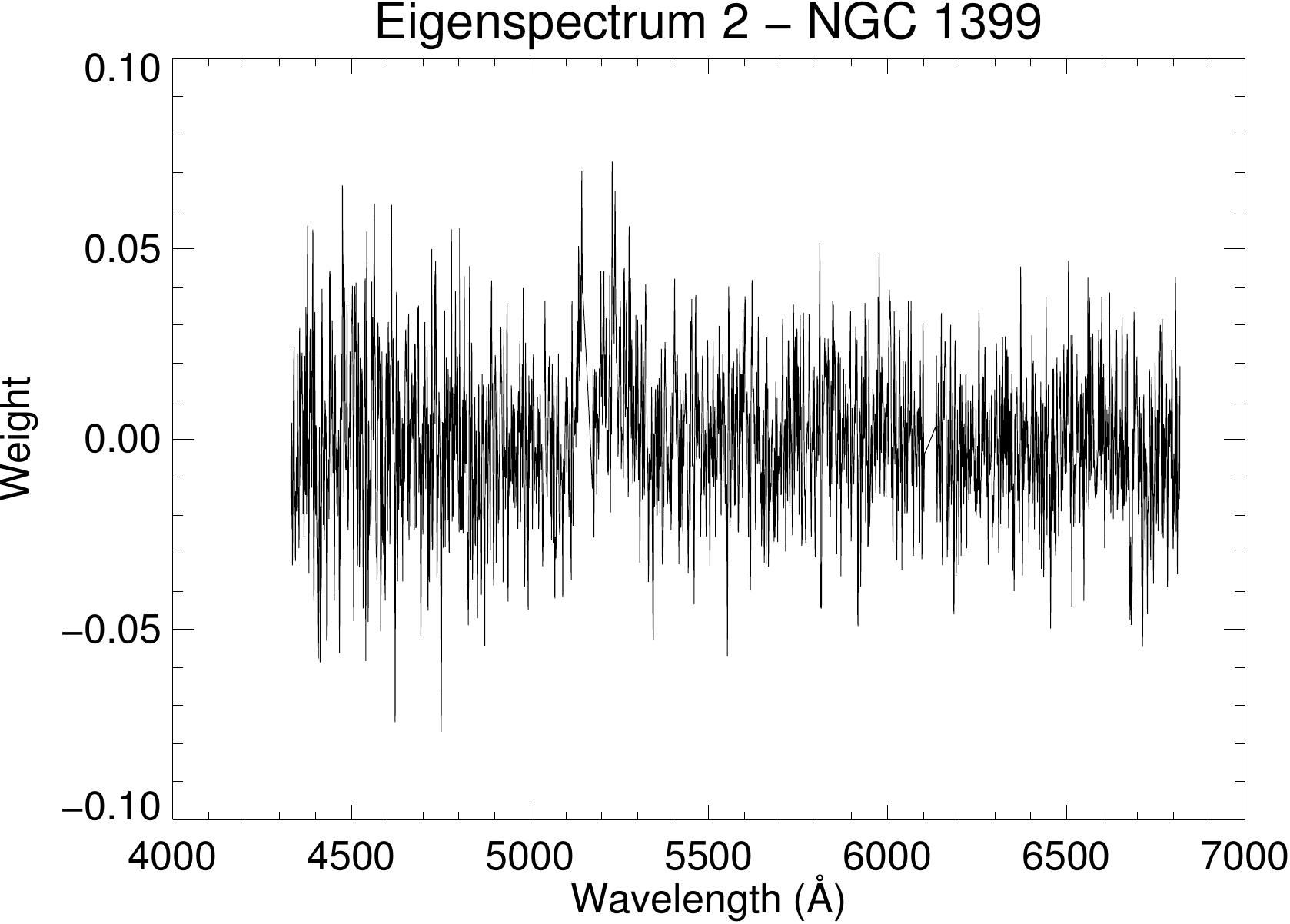}
\includegraphics[scale=0.4]{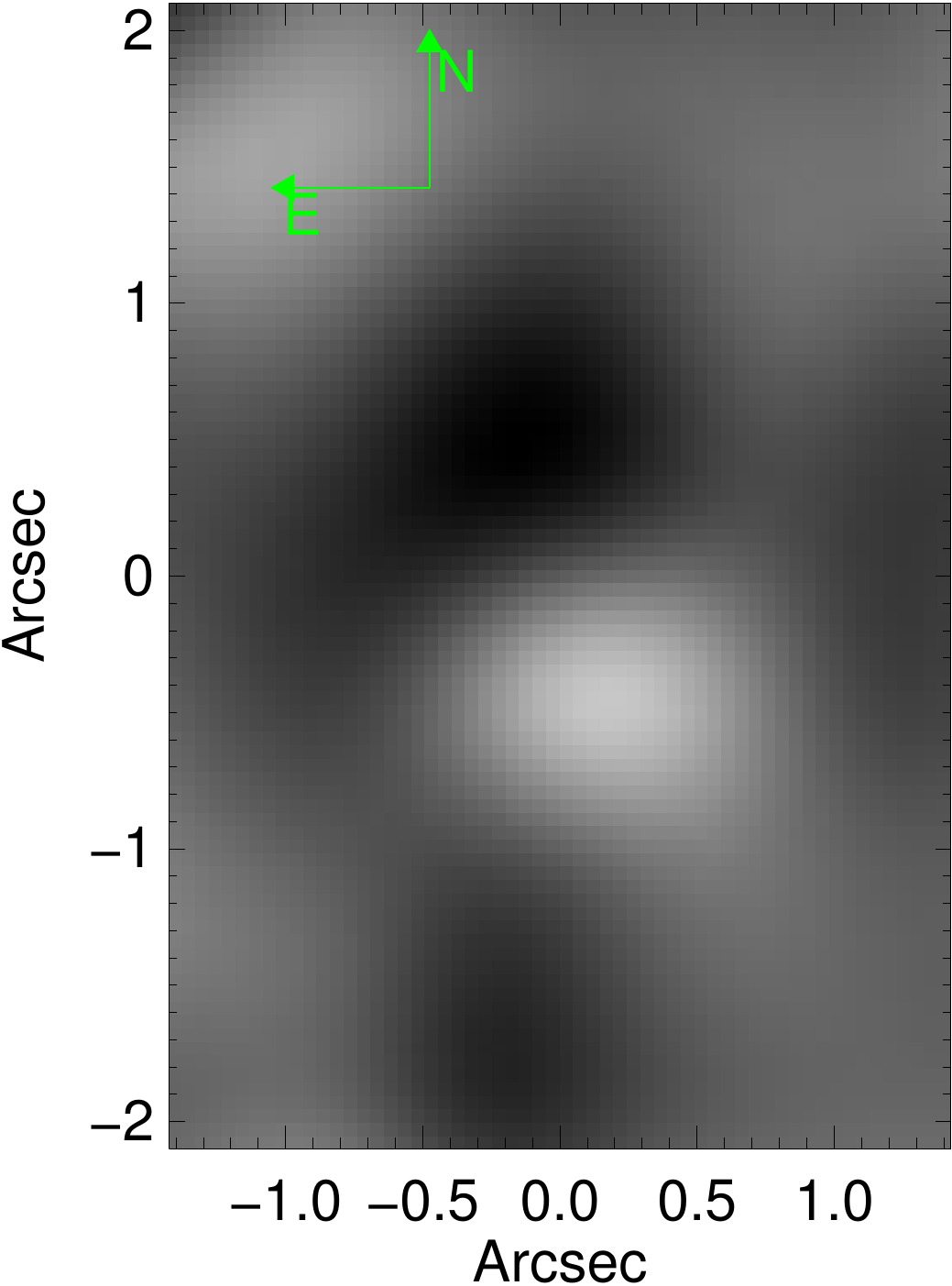}
\vspace{1cm}
\hspace{3cm}
\includegraphics[scale=0.4]{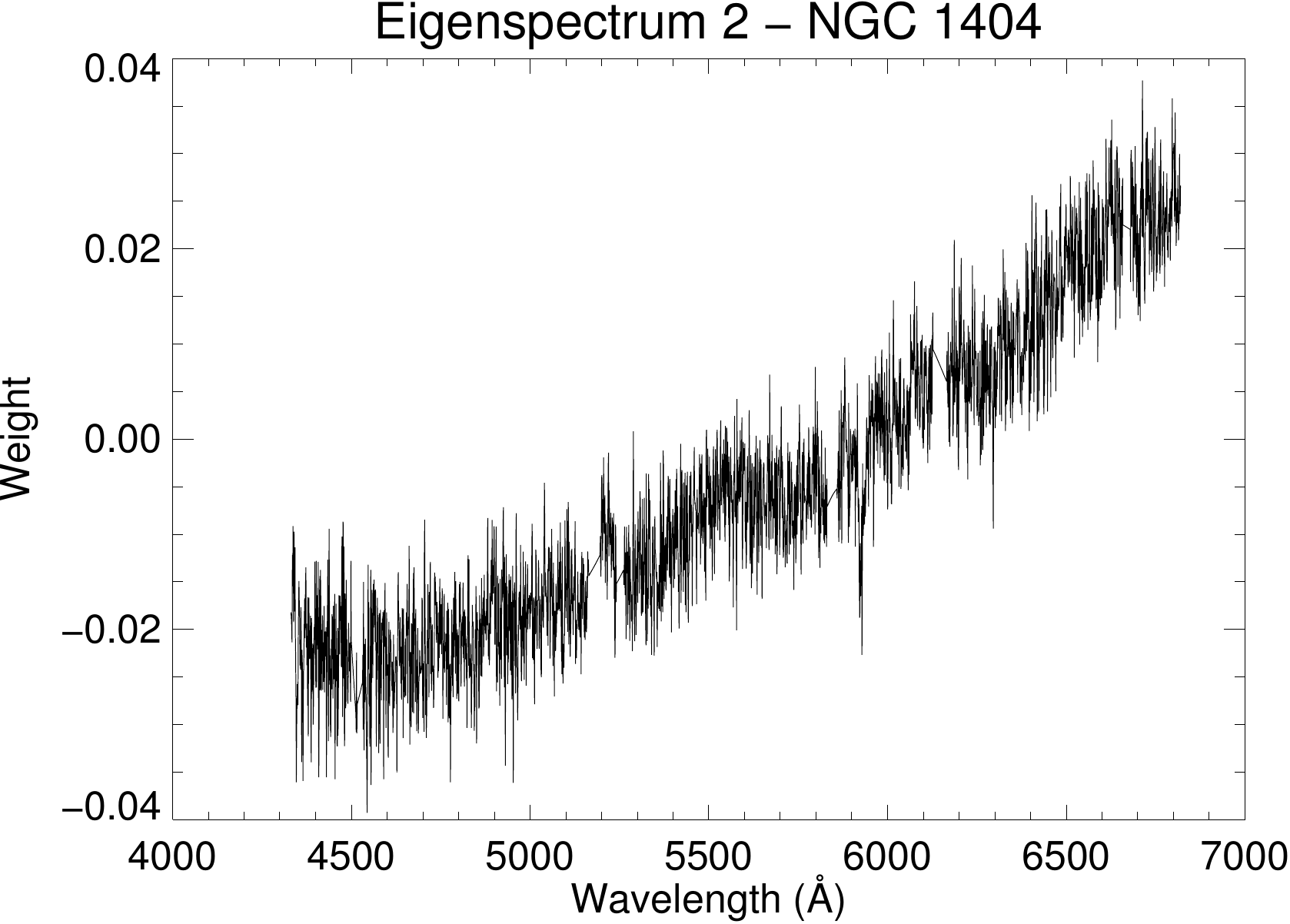}
\caption{Tomogram and eigenspectrum 2 of NGC 1399 and NGC 1404.} \label{eigen2_n1399_n1404}
\end{figure*}

\begin{figure*}
	\begin{center}
		\hspace{-0.5cm}
		\includegraphics[scale=0.3]{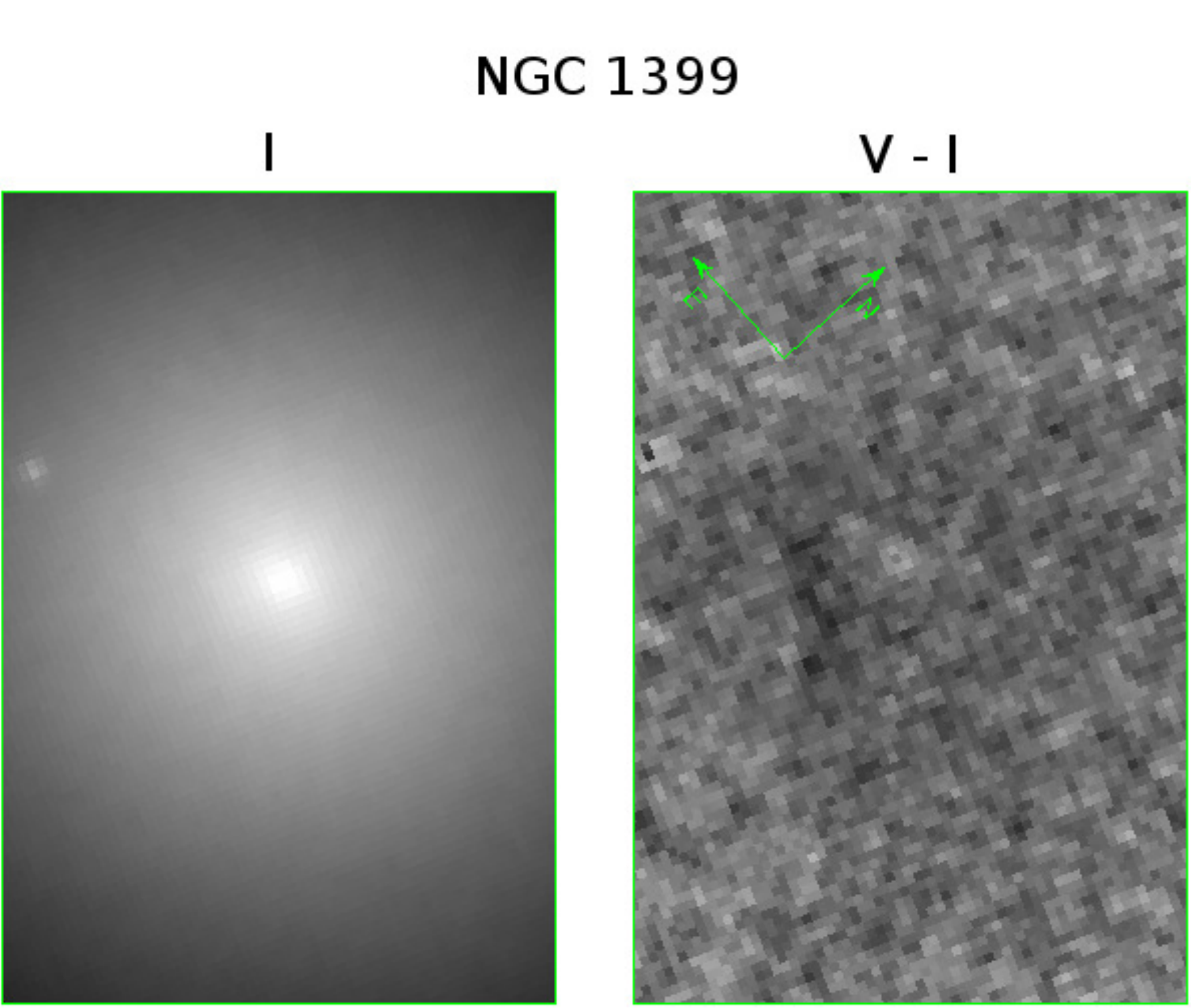}
		\hspace{1.0cm}
		\includegraphics[scale=0.3]{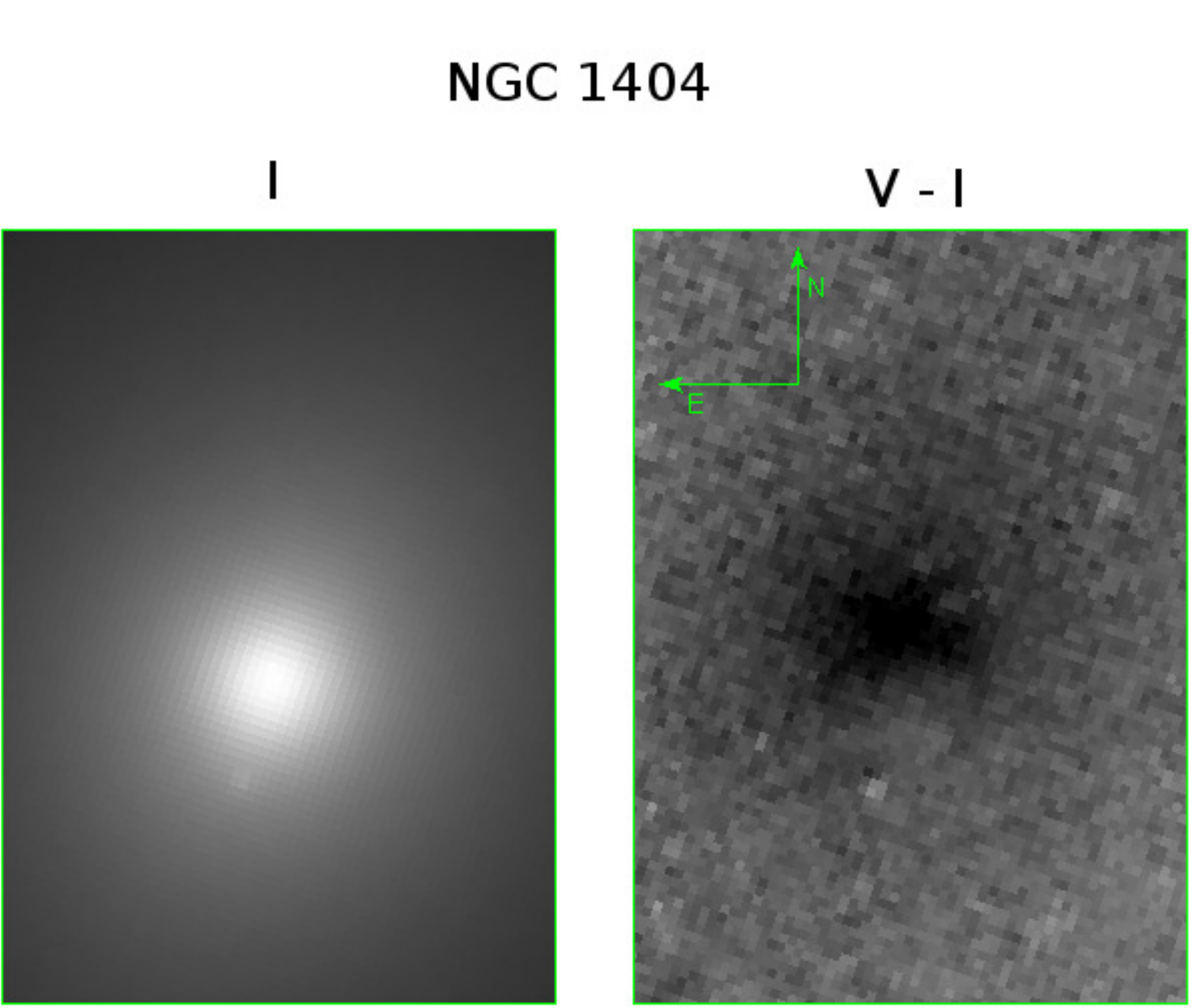}
	\caption{\textit{HST} images for NGC 1399 and NGC 1404. Note that, in NGC 1404, a reddening is detected in the V - I image. In this galaxy, eigenspectrum 2 shows an anti correlation between the red and the blue regions of the continuum, in addition to the Na D absorption lines. \label{HST_n1399_n1404}}
	
	\end{center}
\end{figure*}

Despite the absence of emission lines, we show the second eigenspectra and tomograms for NGC 1399 and NGC 1404 in Fig. \ref{eigen2_n1399_n1404}. Note that eigenspectrum 2 in NGC 1399 is dominated by noise, while in NGC 1404 an anti correlation between the red region and the blue region of the continuum is detected. Moreover, the Na D lines are also present in eigenspectrum 2 of NGC 1404. This means that NGC 1404 also has a reddening along the FOV caused by the ISM. In Fig \ref{HST_n1399_n1404}, we show \textit{HST} images of both objects. In NGC 1404, the \textit{V - I} image reveals such extinction detected by PCA Tomography. In NGC 1399, no sign of extinction is detected in the \textit{V - I} image.

\subsection{Stellar and gas kinematic tomograms and eigenspectra} \label{tom_disc}

In galaxy IC 1459, the gas and stellar kinematics are correlated with each other in eigenvector 3, which resulted from PCA Tomography applied to the 4825-6800 \AA\ spectral range (Fig. \ref{results_IC1459}). It would be appropriate if both components could be unveiled in distinct PCA analysis. Thus, we applied PCA Tomography to the spectral range bounded by 6250 and 6800 \AA\ (hereafter the red region) to detect the kinematics associated only with the gas component, since the variance in this region is dominated by the [O I]$\lambda \lambda$6300, 6363, H$\alpha$+[N II]$\lambda \lambda$6548, 6583 and [S II]$\lambda \lambda$6716, 6731 emission lines. On the other hand, for the stellar kinematics, PCA Tomography was applied to the 5100-5800 \AA\ spectral range (hereafter the yellow region), whose variance is dominated by the stellar absorption lines. Note that we excluded the Na D line from the yellow region, since this line may be seriously affected by the ISM. Table \ref{kinematical_eigenvalues} shows the variance related to both stellar and gaseous kinematics for each galaxy of the sample.

PCA Tomography applied to the red region of the data cubes revealed gas kinematics in seven galaxies of the sample, in addition to IC 1459. Anti correlations between the red and the blue wings of the [O I]$\lambda \lambda$6300, 6363, H$\alpha$+[N II]$\lambda \lambda$6548, 6583 and [S II]$\lambda \lambda$6716, 6731 emission lines were detected in the eigenspectra from the red region. Figs. \ref{fig_gas_disc_1}, \ref{fig_gas_disc_2} and \ref{fig_gas_disc_3} show these eigenspectra for seven galaxies, together with the eigenspectra from the red region that represent the AGN in these objects. Their respective tomograms are characterized by a bipolar structure, whereby each pole is anti correlated with the other. In Figs. \ref{fig_gas_disc_1}, \ref{fig_gas_disc_2} and \ref{fig_gas_disc_3}, the red poles in the tomograms are related to the projections of the red wings of the emission lines, while the blue poles are projections of the blue wings of the same emission lines. This means that, when compared to the galactic nuclei, the red poles in the tomograms are redshifted components, while the blue poles are blueshifted components. 

\begin{figure*}
\begin{center}
\includegraphics[scale=0.4]{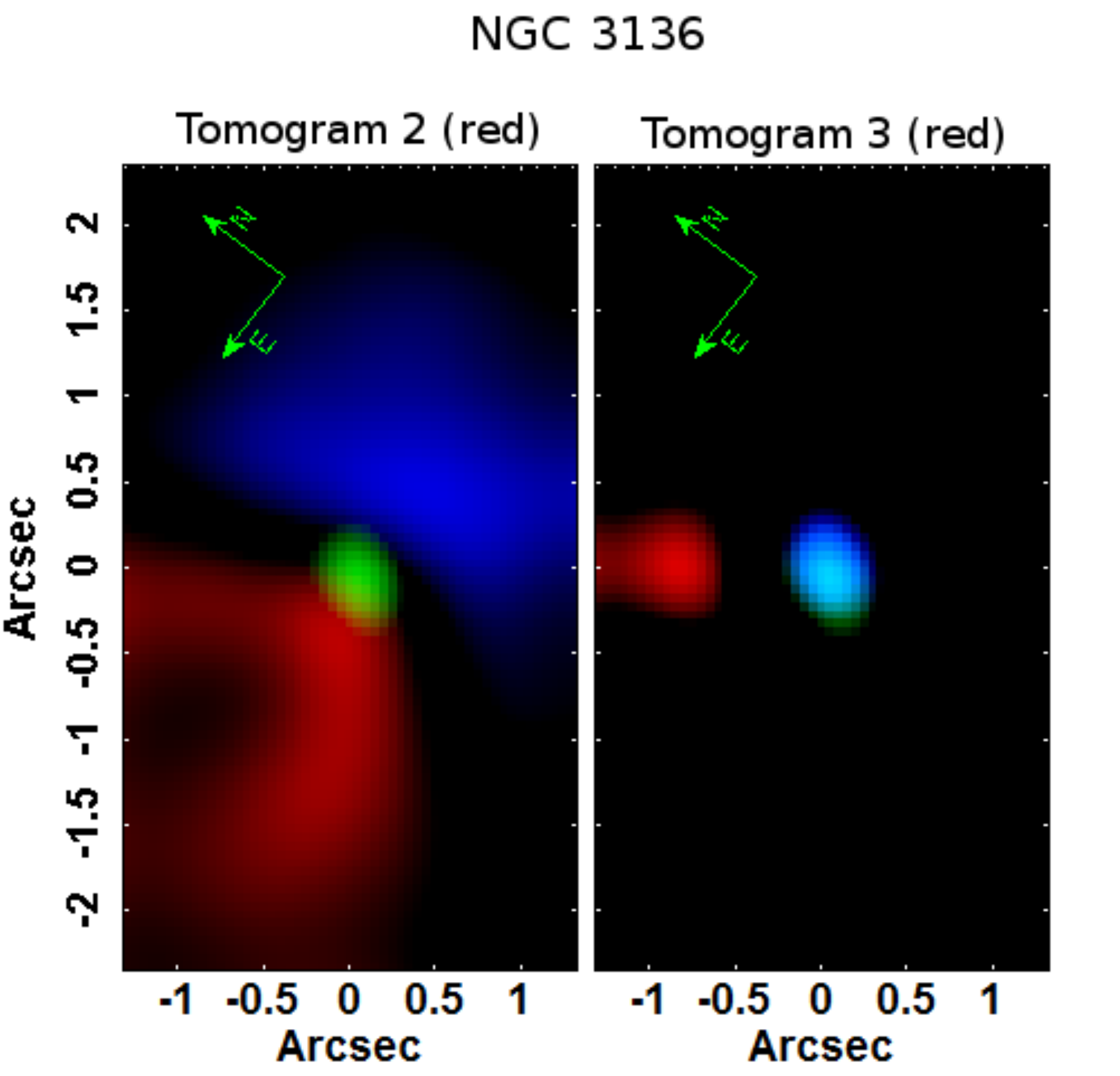}
\includegraphics[width=8cm,height=6cm]{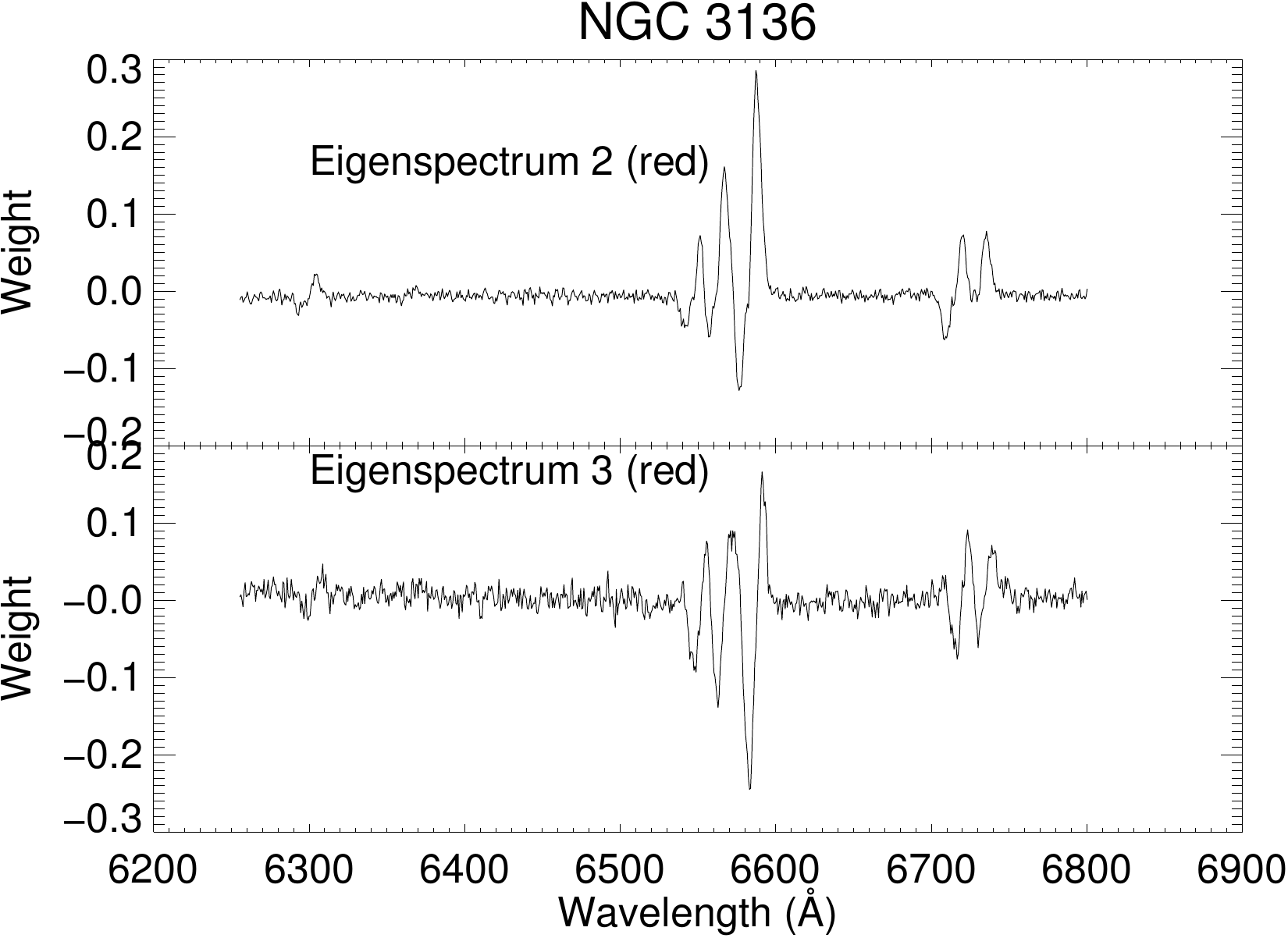}
\vspace{1cm}
\caption{Tomograms and eigenspectra 2 and 3 resulted from PCA Tomography applied to the red region of the data cube of NGC 3136. Both eigenvectors correspond to a gas kinematics, but in distinct objects. In tomogram 2, two extended features are unveiled: one is blueshifted and the other is redshifted when compared to the centre of the galaxy. On the other hand, tomogram 3 shows two point-like objects. In this case, the source related to the projection of the blue wing of the emission lines from eigenspectrum 3 is at the same location of the point-like object which was interpreted as a LINER.\label{tomogram_N3136_2}}
\end{center}
\end{figure*}

Eigenspectra 2 and 3 from PCA Tomography applied to the red region of the data cube of NGC 3136, shown in Fig. \ref{tomogram_N3136_2}, show anti correlations between the red and blue wings of the emission lines. However, their respective tomograms revealed distinct structures. The second tomogram showed two extended structures anti correlated with each other. When compared to the centre of NGC 3136, one of these structures is blueshifted and the other is redshifted. The hypothesis of an ionized disc is not necessarily discarded, but both structures are more extended when compared to the bipolar structures observed in the others galaxies of the sample. In tomogram 3, two point-like structures anti correlated with each other were detected, given that the source related to the projection of the blue wing of eigenspectrum 3 is located at the same position of the point-like object revealed by eigenspectrum 2 of the PCA Tomography applied to the whole data cube and that is shown in green in both images of the Fig. \ref{tomogram_N3136_2}.

Note that the gas kinematics is shown in eigenspectra 2 from the red region for four cases, while the AGN is detected in eigenspectra 3 from the red region of these objects. It happens because in the red region of the data cubes of these galaxies, the variance associated with the kinematics is higher than the variances related to the AGNs. In IC 1459, NGC 2663 and NGC 4546, the AGN is shown in eigenspectra 2 from PCA Tomography applied to the red region and the gas kinematics in eigenspectra 3. It is worth mentioning that both the AGNs of NGC 4546 and IC 1459 have the higher luminosities among the sample (Paper II). In the case of NGC 2663, this may be associated with the fact that the velocity of the gas rotation is quite low (Paper III).

The P.A. of the bipolar structures, shown in table \ref{tab_PA_pca}, were determined by first measuring the positions of the highest and the lowest weights of the kinematic tomograms, since these regions are deeply affected by the anti correlation between the red and the blue wings of the lines. Then, we fitted a linear function that passes through both points. The angle of this line with respect to the north is the P.A. of the bipolar structure. The errors were estimated with a Monte Carlo simulation, by measuring a hundred times the positions of the highest and the lowest weights of the tomograms after adding random noise, which is given by a Gaussian with $\sigma$ equal to the uncertainty related to each weight of the tomogram. According to \citet{1997ApJ...475..173H}, the uncertainty associated with these weights is equal to the error $\sigma_\lambda$ of the spectral intensity. We estimated $\sigma_\lambda$ $\sim$ 2.1$\times$10$^{-18}$ erg s$^{-1}$ cm$^{-2}$ \AA$^{-1}$ for the data cubes from the GS-2008A-Q-51 programme and $\sigma_\lambda$ $\sim$ 5$\times$10$^{-19}$ erg s$^{-1}$ cm$^{-2}$ \AA$^{-1}$ for the data cubes from the GS-2008B-Q-21 programme\footnote{The details of these estimates will be shown in Paper II}. We propose that ionized gas discs are present in the internal region of six galaxies of the sample, in an analogy with IC 1459. For NGC 3136, an ionization cone is more likely, since tomogram 2 of the PCA Tomography applied to the red region of the data cube of this galaxy reveals a more extended structure when compared to the gaseous kinematic tomograms of the other galaxies of the sample. Also regarding this procedure, NGC 1399 and NGC 1404 do not show any sign of emission lines.

When PCA Tomography was applied to the yellow region of the data cubes, kinematic signatures related to a stellar disc were unveiled in six galaxies of the sample, in addition to IC 1459. In this case, the red and the blue wings of the stellar lines are anti correlated with each other, as shown in Figs. \ref{fig_stellar_disc_1}, \ref{fig_stellar_disc_2} and \ref{fig_stellar_disc_3}. Since the anti correlation between both red and blue wings of the stellar lines is not so obvious as it is for the gaseous lines, we cross-correlated the eigenspectra from the yellow region possessing the stellar kinematic signature with the eigenspectra from the yellow region that represent the bulges of these seven galaxies. A sinusoidal curve is expected for this procedure, since one of the wings has to be correlated with the stellar lines and the other wing must be anti correlated with the stellar lines. In fact, in Fig. \ref{correlacao_cruzada}, we see this behaviour for the seven objects with the stellar kinematics detected. Although the \textit{x}-axis of Fig. \ref{correlacao_cruzada} is given by velocity, no reliable information on this issue may be taken from the cross-correlation. Again, the anti correlation between the blue and the red wings of the stellar lines are affected by the displacement of the spectra to the blue or to the red and by the intensity of the stellar continuum. Moreover, since we are cross-correlating the kinematics eigenspectrum with their respective eigenspectra 1 (that represent the stellar component of the galaxies), the stellar velocity dispersion also affects the results of the cross-correlations. Just like the gas disc, the tomograms are characterized by a bipolar structure and are shown in Figs. \ref{fig_stellar_disc_1}, \ref{fig_stellar_disc_2} and \ref{fig_stellar_disc_3}. The P.A. of the stellar structures are presented in table \ref{tab_PA_pca}. Here, we also assumed that the kinematic centre of the stellar components are given by the position of the AGNs, except for NGC 1404, where we assumed that the centre was given by the central region of the galactic bulge detected in the tomogram 1 resulting from PCA Tomography applied to the yellow region of the data cube. These results indicate that there is a stellar component in rotation around the respective nuclei of these six galaxies. In NGC 1399, NGC 2663 and NGC 3136, no stellar rotation was detected.

Starting out from the hypothesis that these six galaxies have gas and stellar discs, we found that both components are corotating in NGC 1380 and ESO 208 G-21 and counter-rotating in IC 1459 and NGC 7097. In NGC 4546 and IC 5181, these components seem to rotate in distinct planes.

\begin{table}
 \scriptsize
 \begin{center}
 \begin{tabular}{@{}lcc}
  \hline
  Galaxy name & $\Lambda_{cin\ gas}$ (\%) & $\Lambda_{cin\ stellar}$ (\%)  \\
  \hline
  ESO 208 G-21 & 0.19 & 0.094  \\
  IC 1459 & 0.38 & 0.038 \\
  IC 5181 & 0.19 & 0.027 \\
  NGC 1380 & 0.18 & 0.0073 \\
  NGC 1399 &  & $<$ 0.0088\\
  NGC 1404 &  & 0.0030\\
  NGC 2663 & 0.05 & $<$ 0.040 \\
  NGC 3136 &  0.15$^*$ &$<$ 0.011\\
  NGC 4546 & 0.13 & 0.026 \\
  NGC 7097 &  0.20 & 0.0064\\
  \hline
 \end{tabular}
 \caption{Eigenvalues related to the gas and stellar kinematics in the galaxies of the sample. For NGC 3136, the eigenvalues associated with the kinematics of the extended gas component are shown, although it is not interpreted as a gas disc.  \label{kinematical_eigenvalues}
}
 \end{center}
\end{table}

\renewcommand{\thefigure}{\arabic{figure}\alph{subfigure}}
\setcounter{subfigure}{1}

\begin{figure*}
\begin{center}
\includegraphics[scale=0.4]{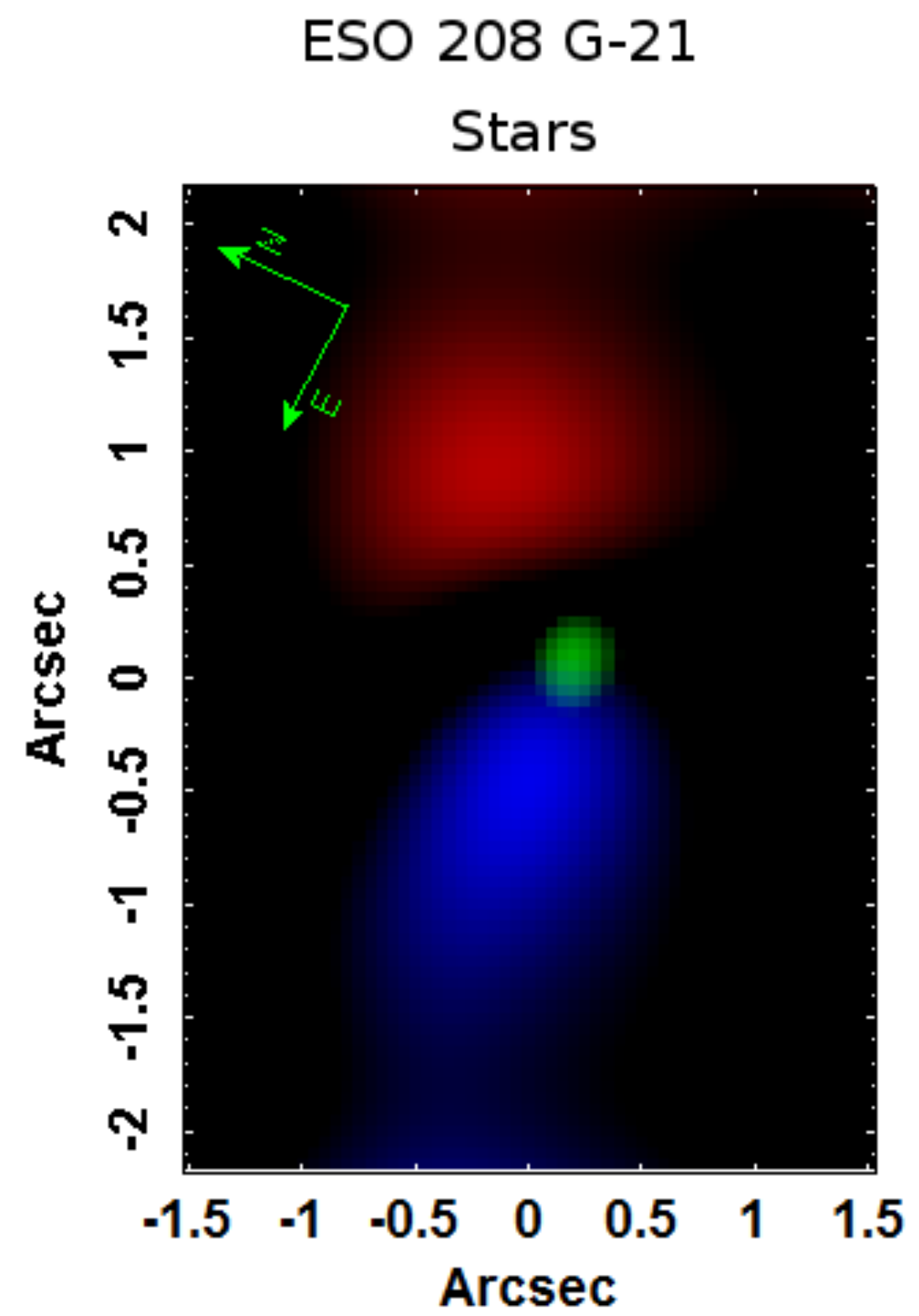}
\includegraphics[width=70mm,height=60mm]{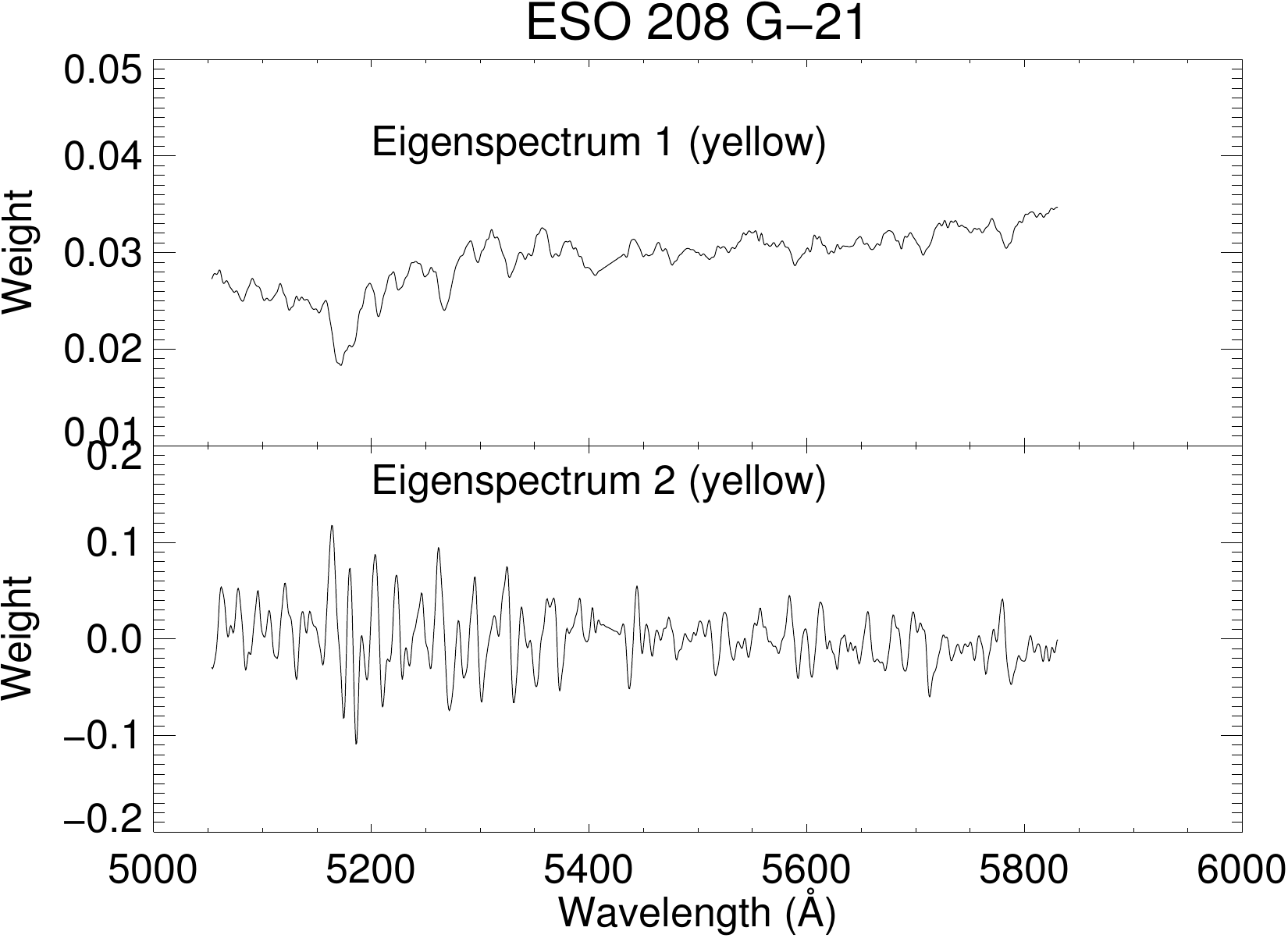}
\vspace{0.5cm}

\includegraphics[scale=0.4]{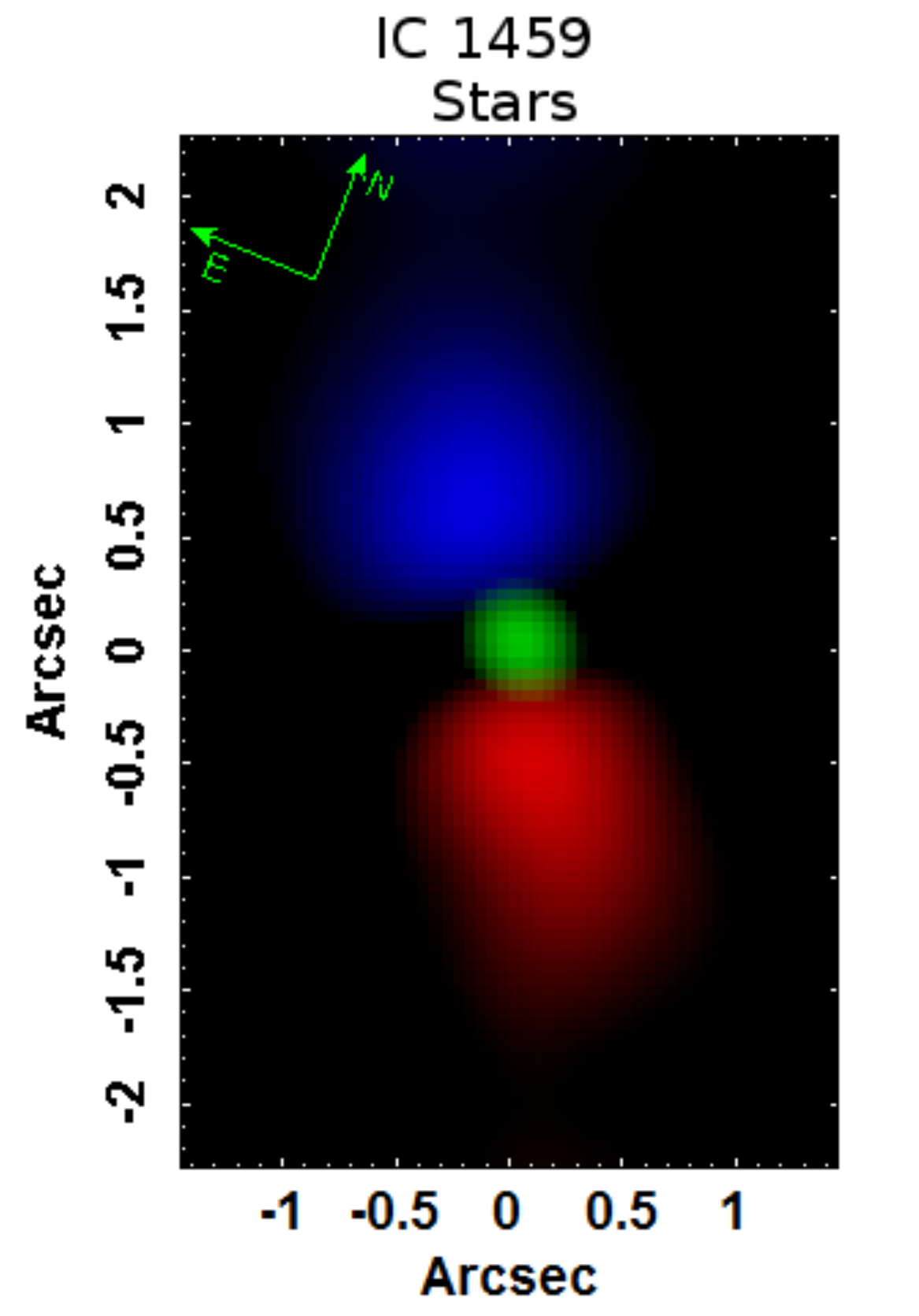}
\includegraphics[width=70mm,height=60mm]{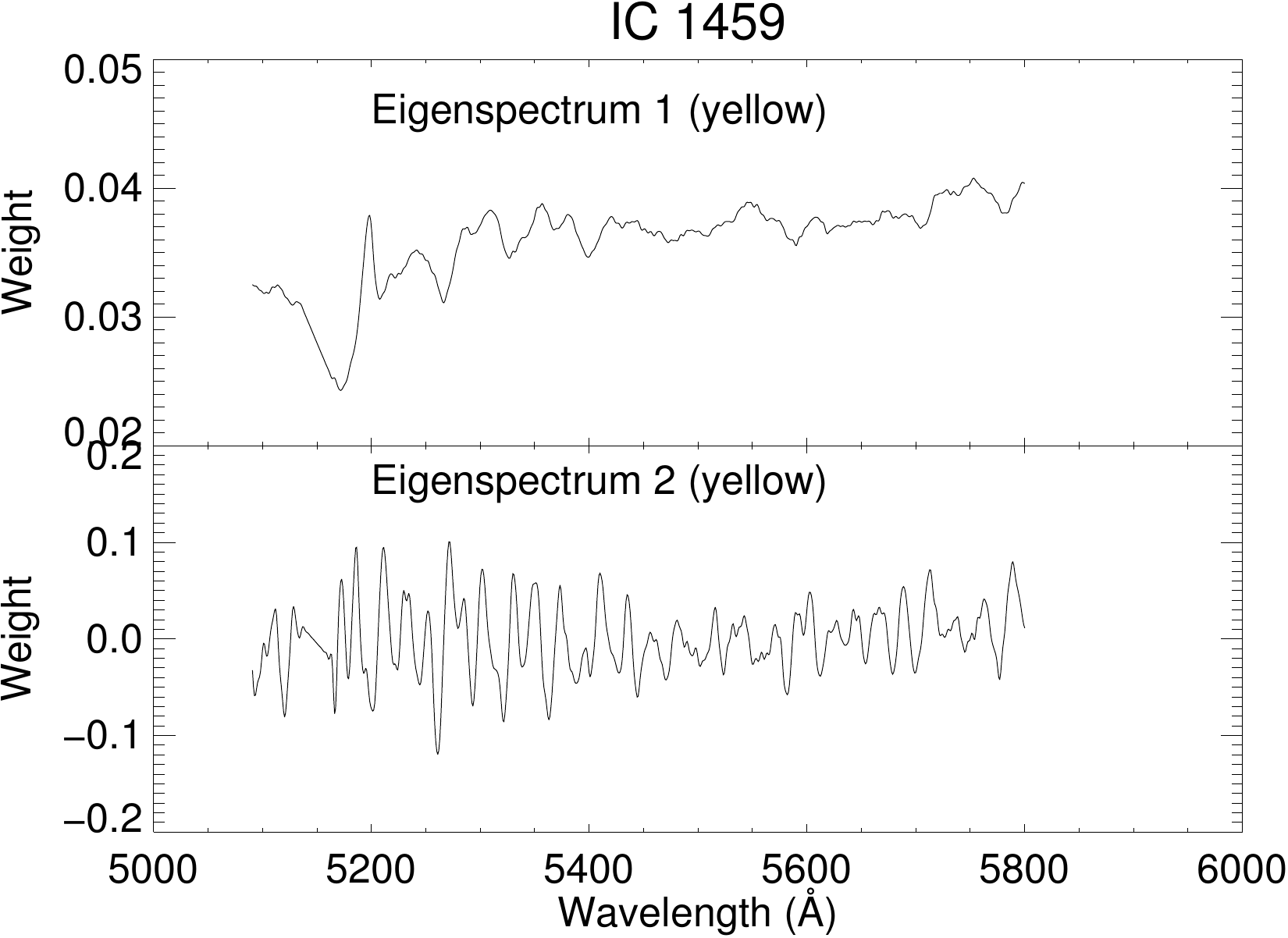}
\vspace{0.5cm}

\includegraphics[scale=0.4]{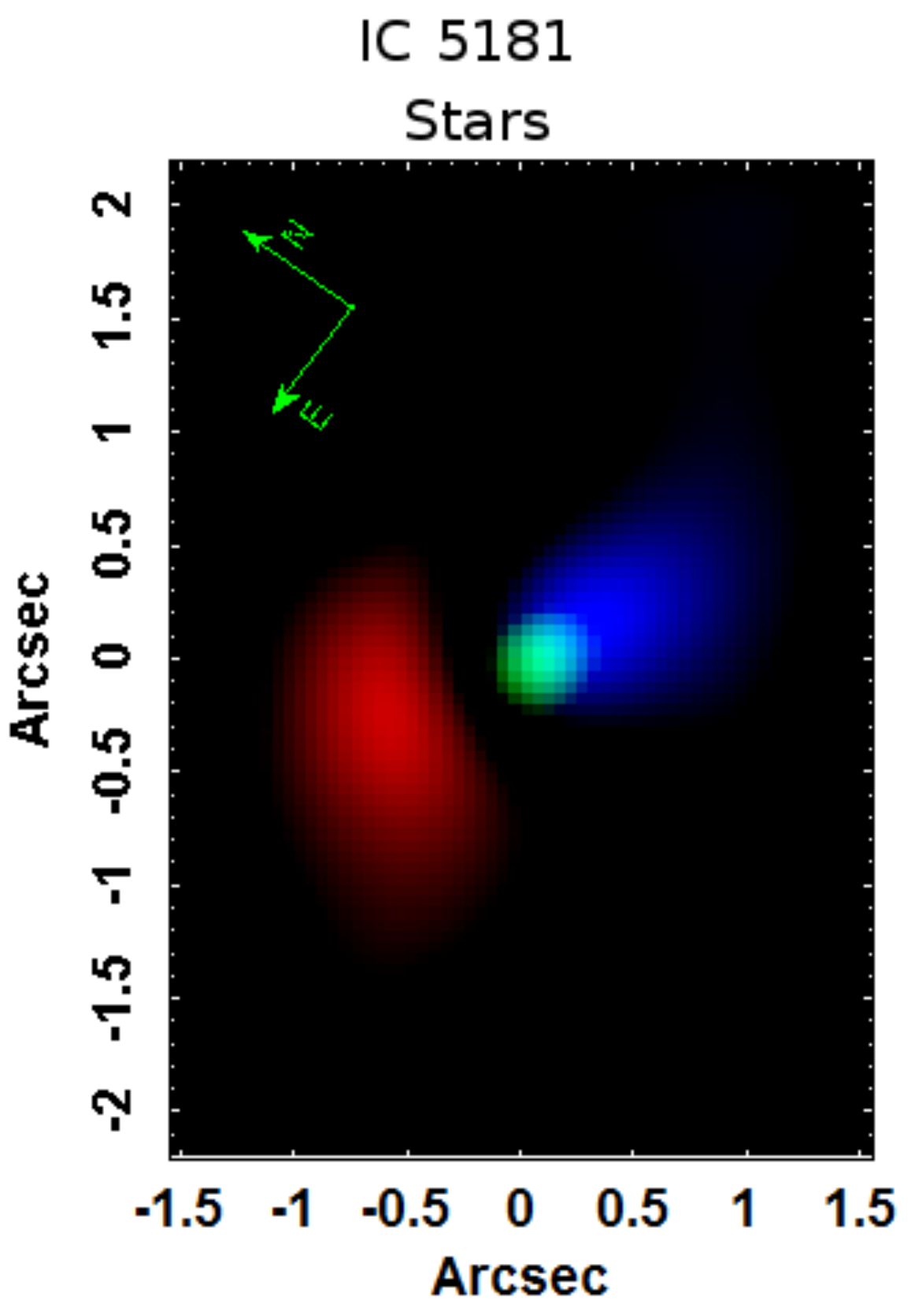}
\includegraphics[width=70mm,height=60mm]{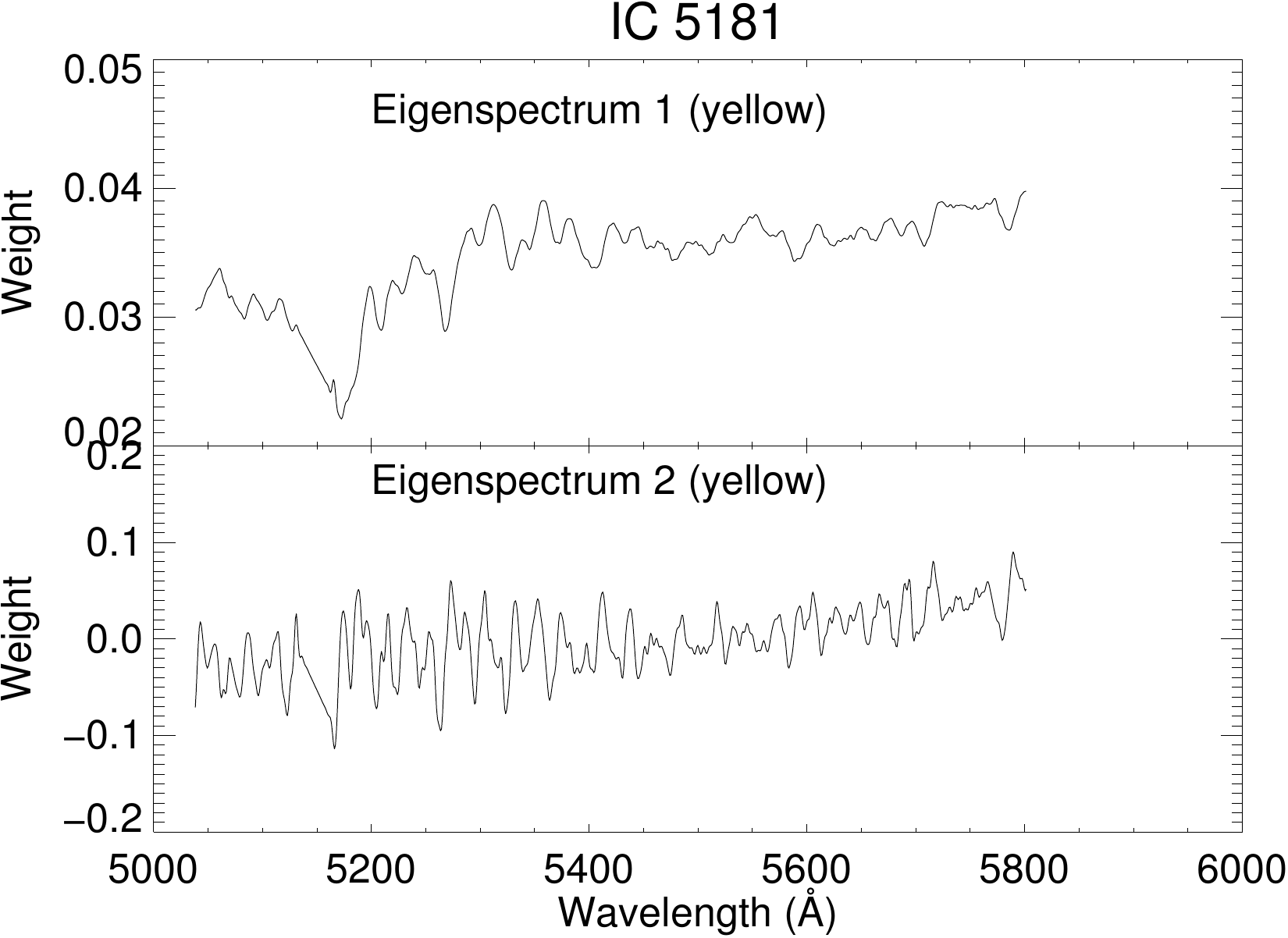}
\vspace{0.5cm}

\caption{PCA Tomography applied to the yellow region of the data cubes for the galaxies ESO 208 G-21, IC 1459 and IC 5181. In the tomograms shown above, regions blueshifted (redshifted) when compared to the galactic centres are shown in blue (red). Both regions were detected in the same tomogram, where each colour is a pole belonging to the bipolar structure. The tomograms that correspond to the AGNs (Fig. \ref{autoespectros_AGN}) are shown in green. In the eigenspectra, the kinematic signature is characterized by the anti correlation between the red and blue wings of the absorption lines, although this anti correlation is not so clear for the stellar kinematics as it is for the gaseous kinematics. 
\label{fig_stellar_disc_1}
}
\end{center}
\end{figure*}

\addtocounter{figure}{-1}
\addtocounter{subfigure}{1}

\begin{figure*}
\begin{center}
\includegraphics[scale=0.4]{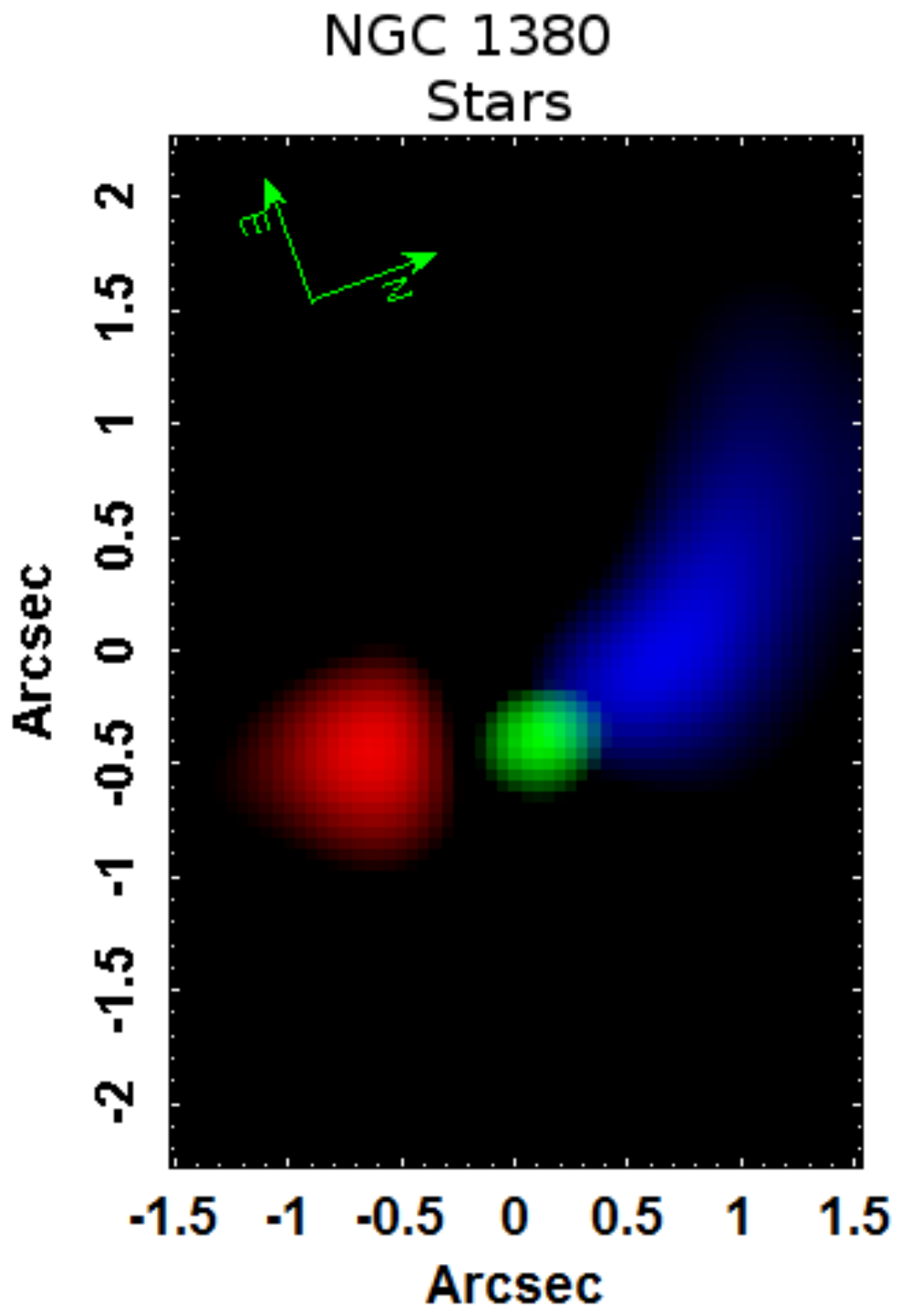}
\includegraphics[width=70mm,height=60mm]{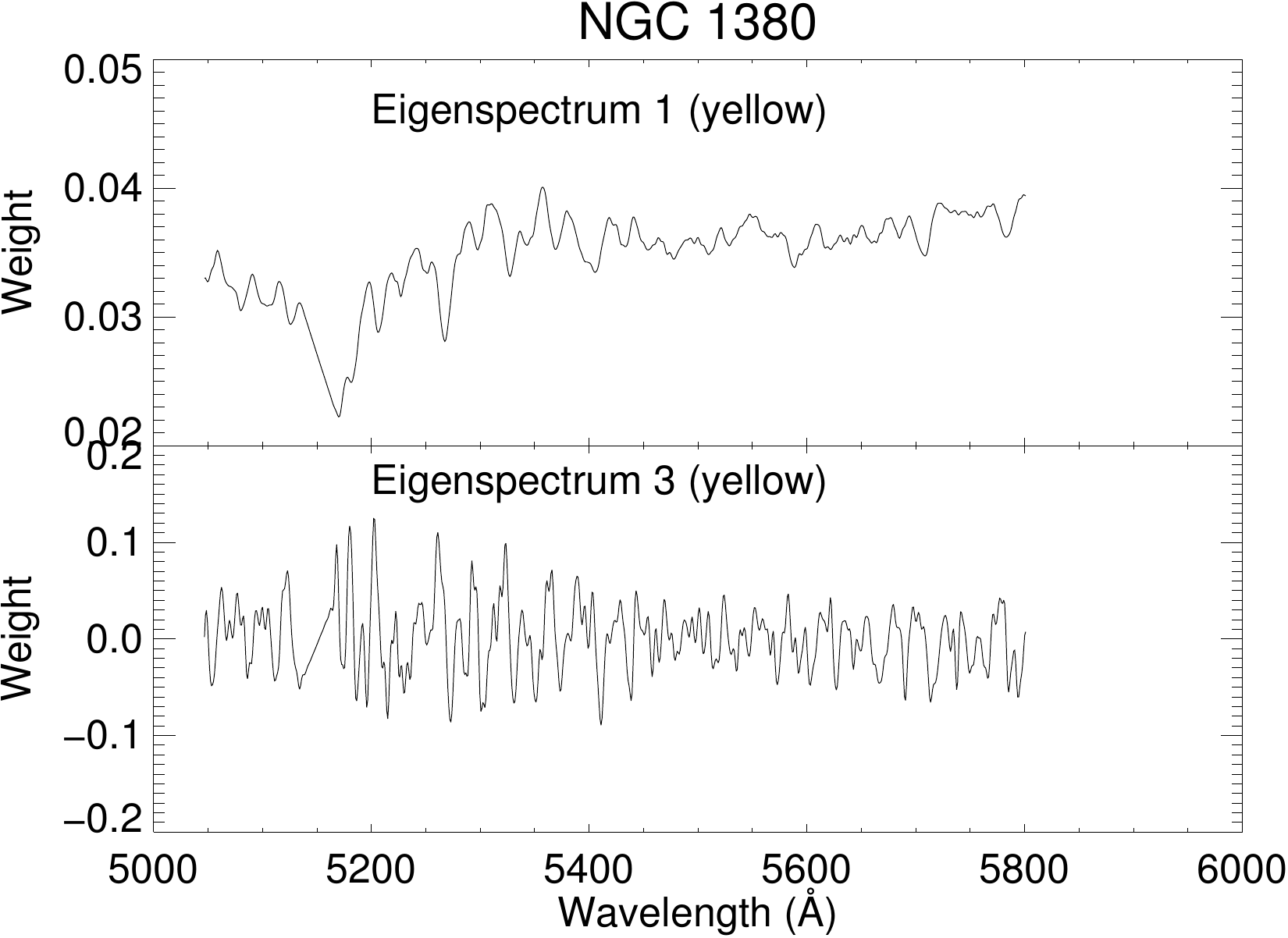}
\vspace{0.5cm}

\includegraphics[scale=0.4]{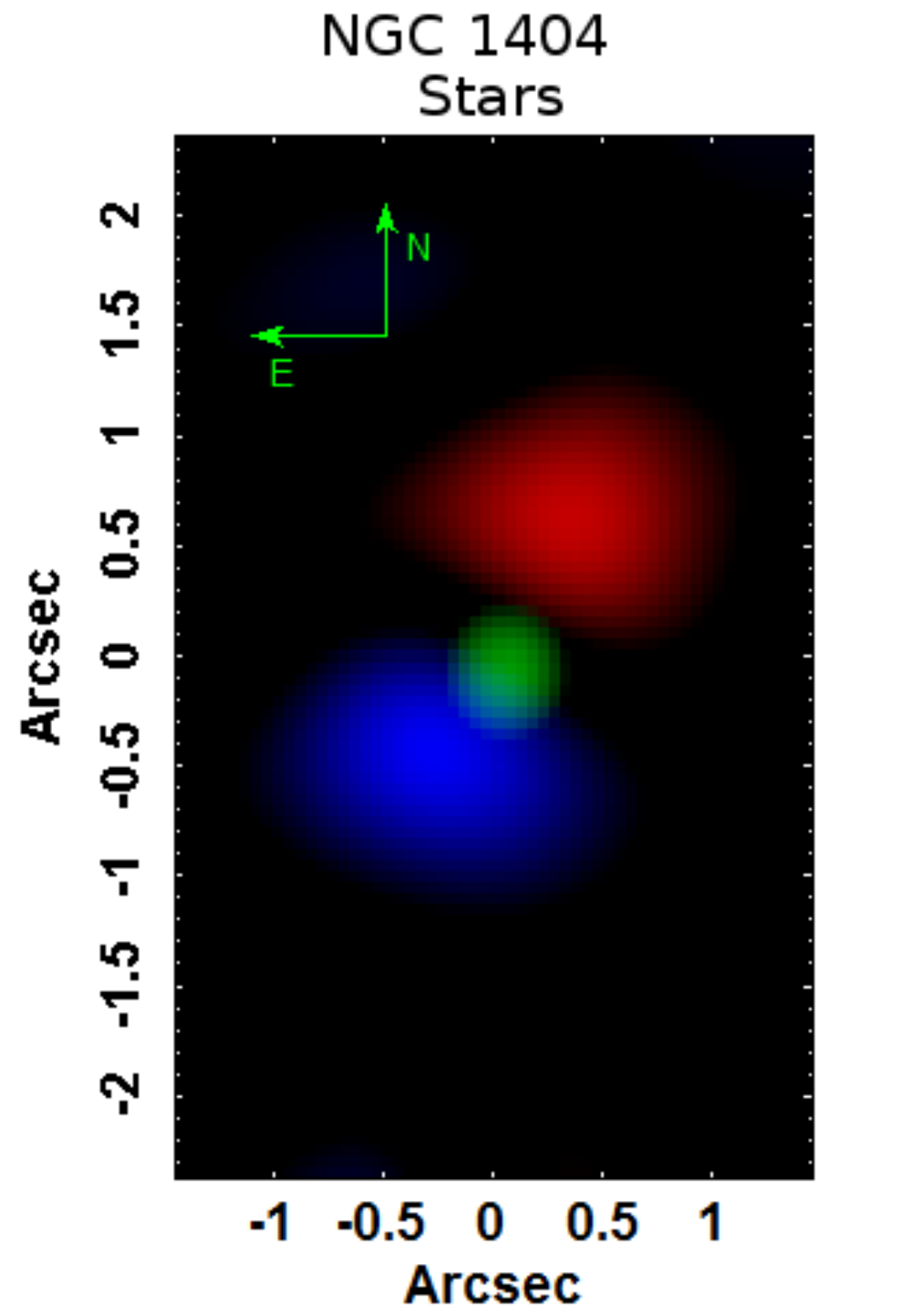}
\includegraphics[width=70mm,height=60mm]{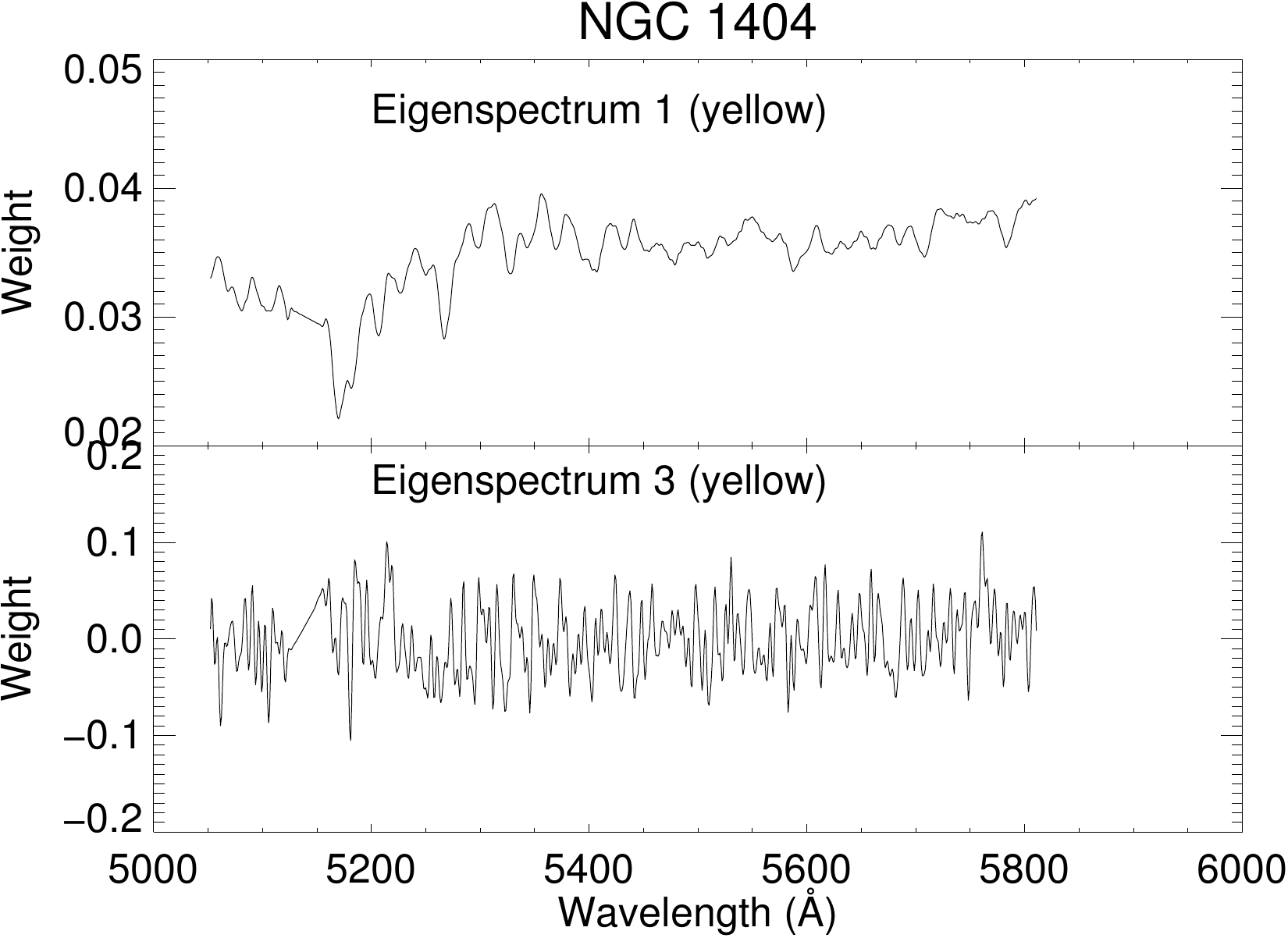}
\vspace{0.5cm}

\includegraphics[scale=0.4]{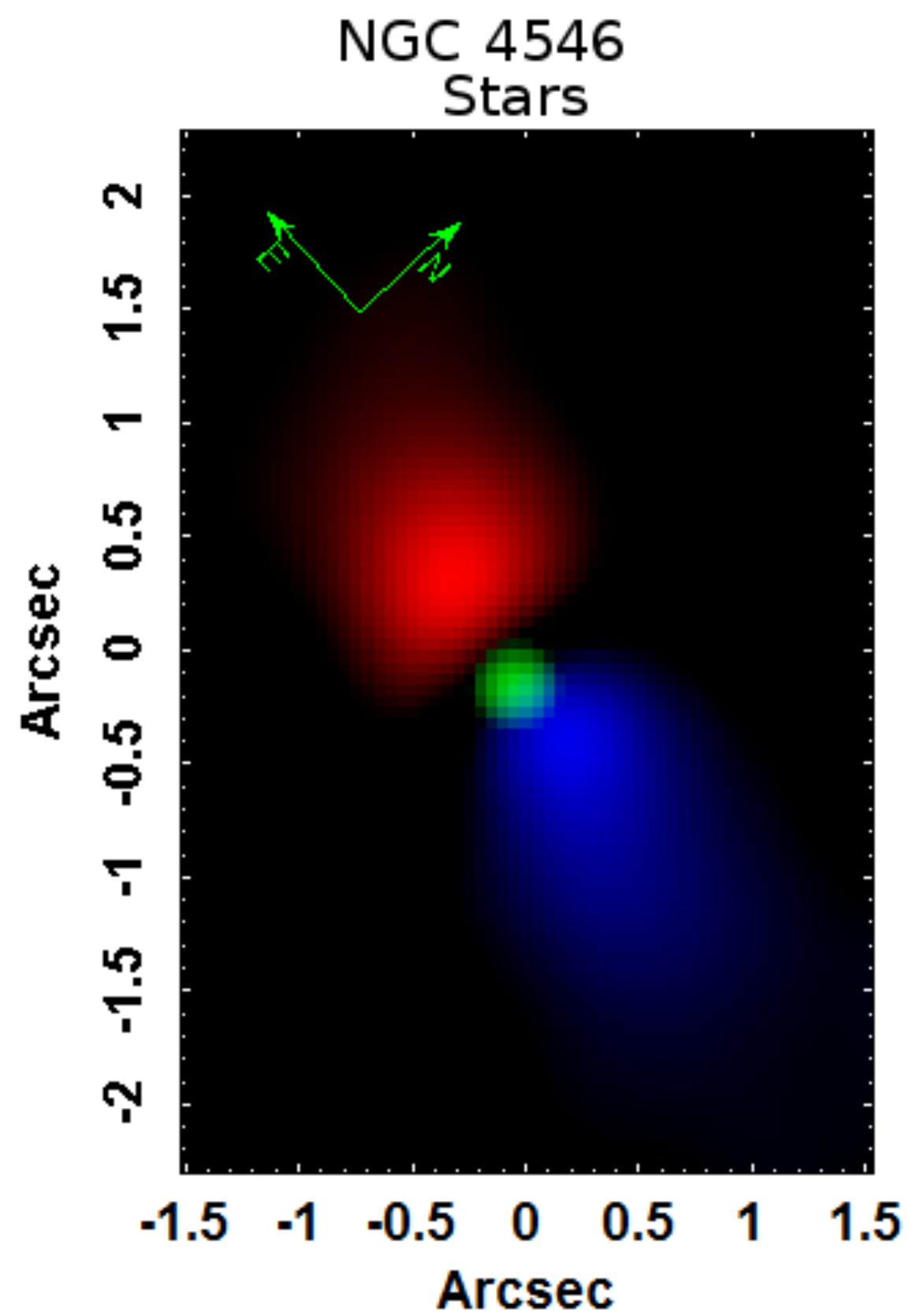}
\includegraphics[width=70mm,height=60mm]{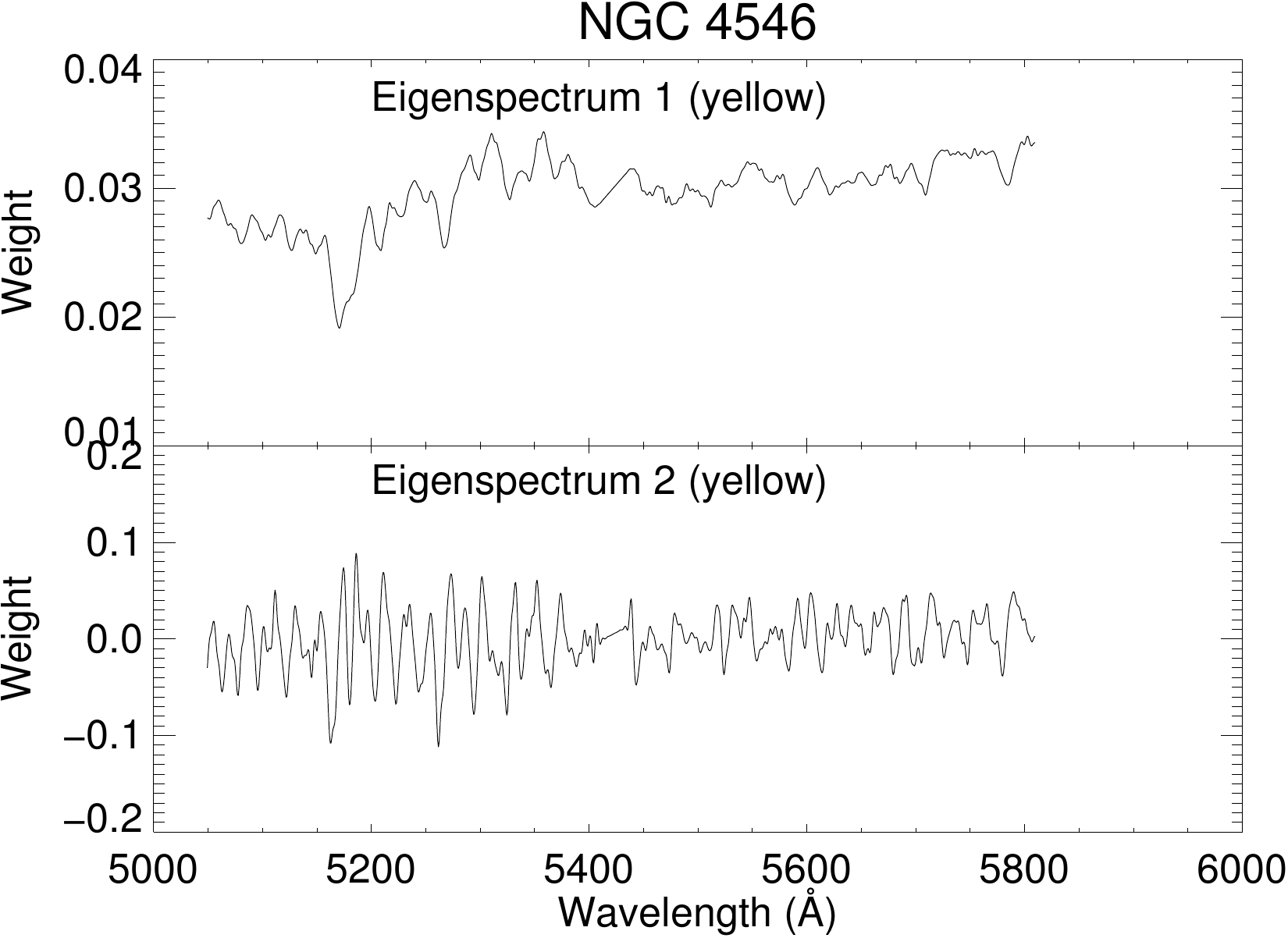}
\vspace{0.5cm}

\caption{The same as in Fig. \ref{fig_stellar_disc_1} for the galaxies NGC 1380, NGC 1404 and NGC 4546.  
\label{fig_stellar_disc_2}
}
\end{center}
\end{figure*}

\addtocounter{figure}{-1}
\addtocounter{subfigure}{1}

\begin{figure*}
\begin{center}
\includegraphics[scale=0.4]{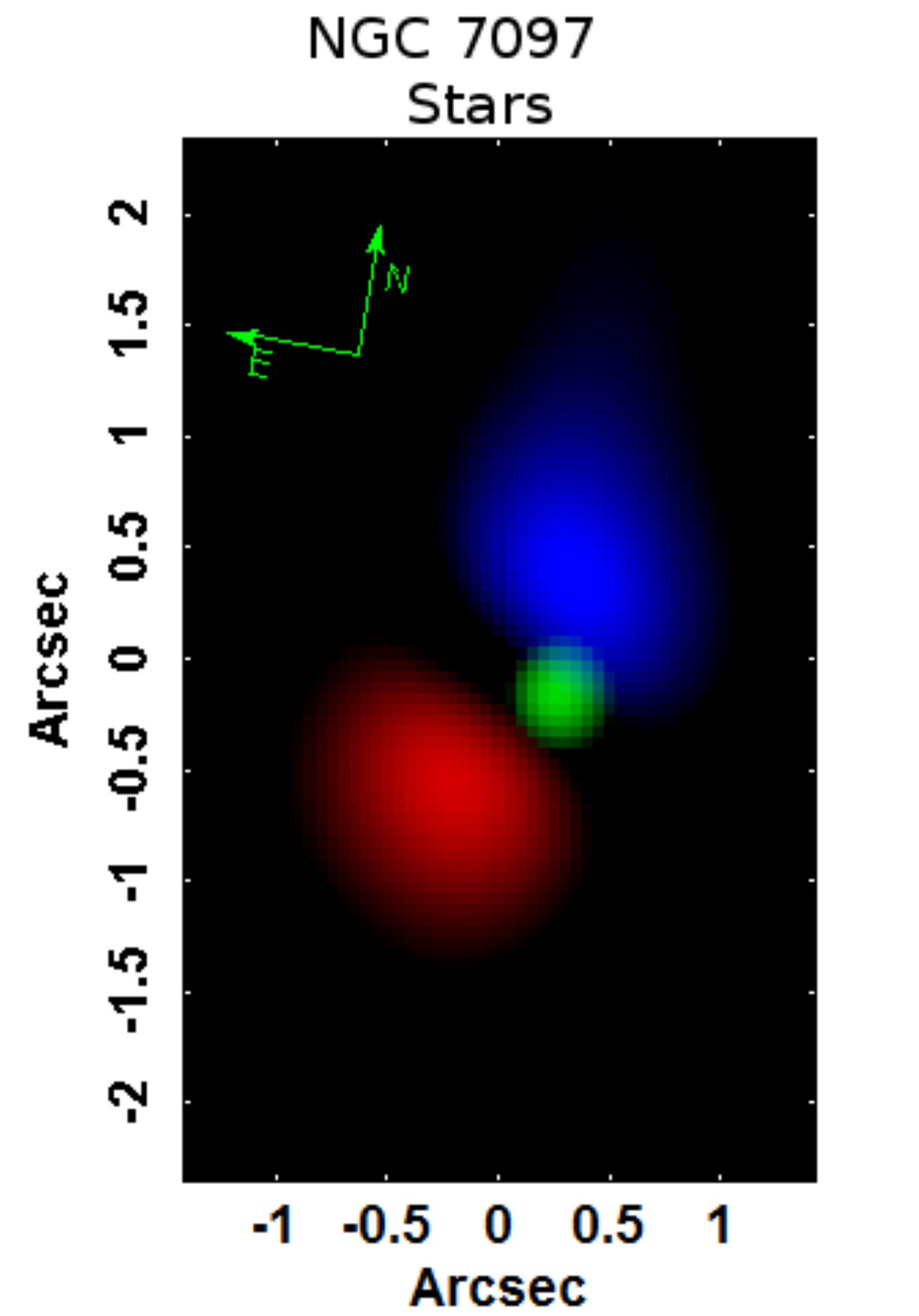}
\includegraphics[width=70mm,height=60mm]{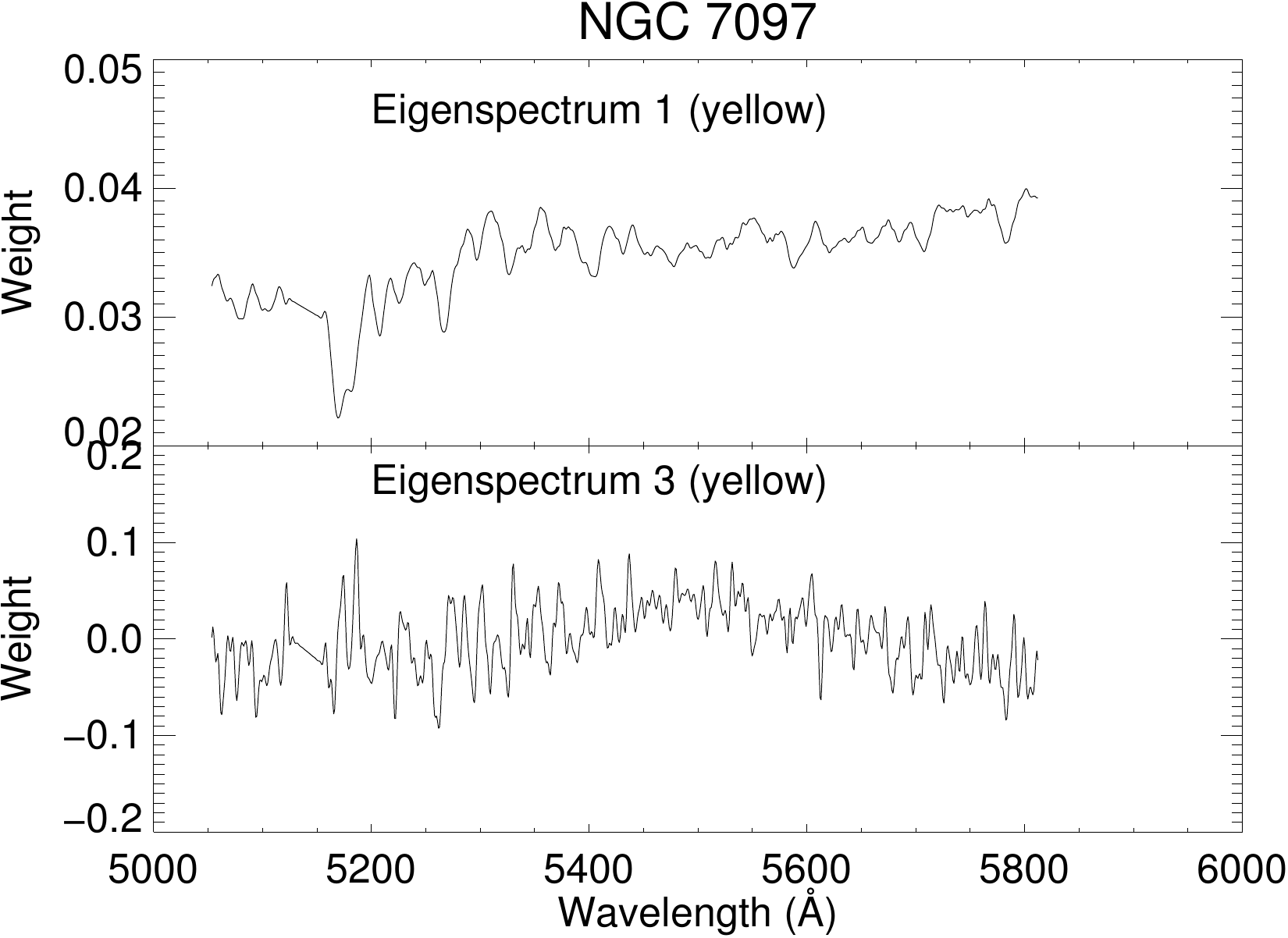}
\vspace{0.5cm}

\caption{The same as in Fig. \ref{fig_stellar_disc_1} for the galaxy NGC 7097. 
\label{fig_stellar_disc_3}
}
\end{center}
\end{figure*}

\renewcommand{\thefigure}{\arabic{figure}}


\begin{figure*}
\begin{center}
\includegraphics[scale=0.35]{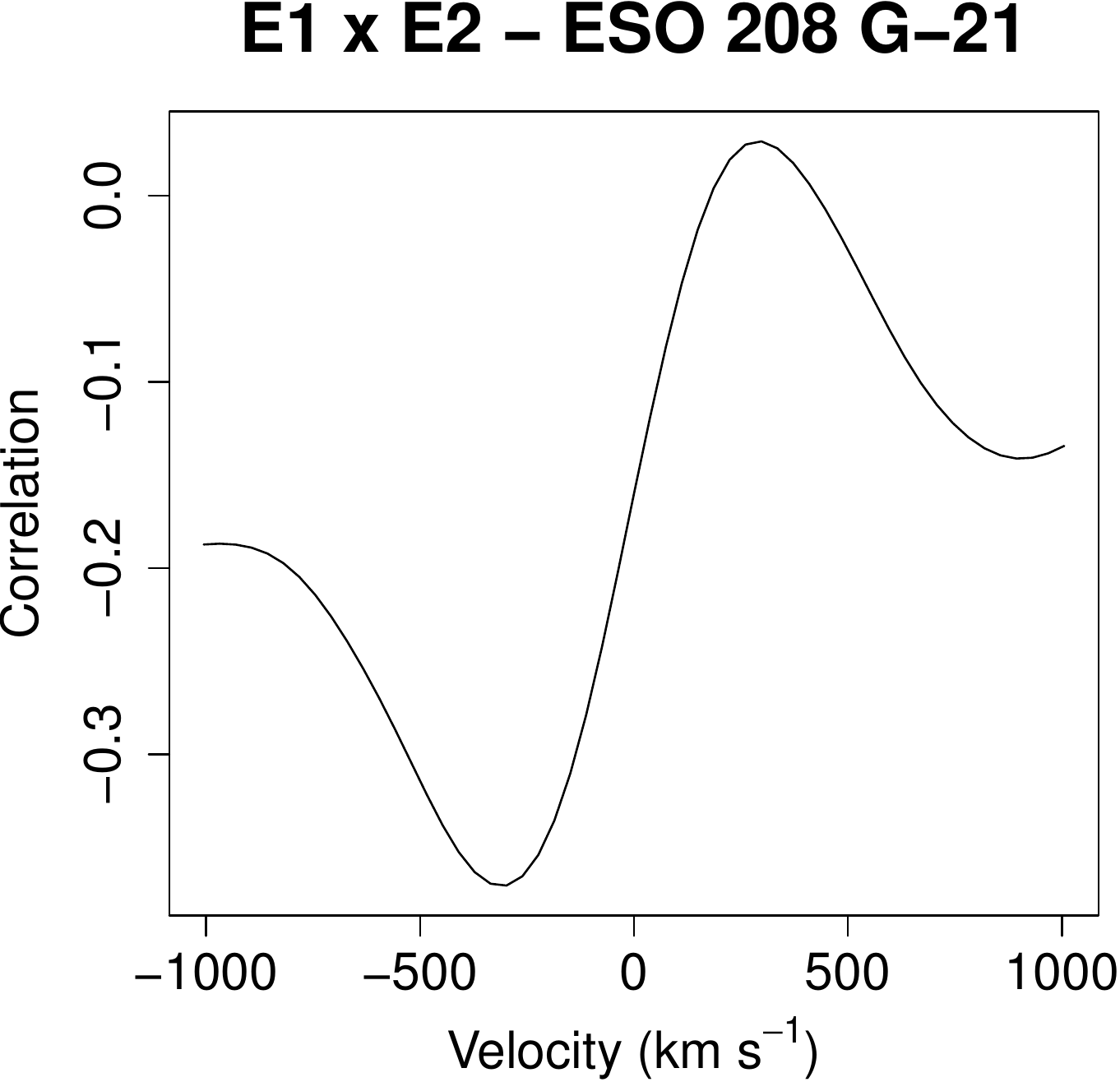}
\includegraphics[scale=0.35]{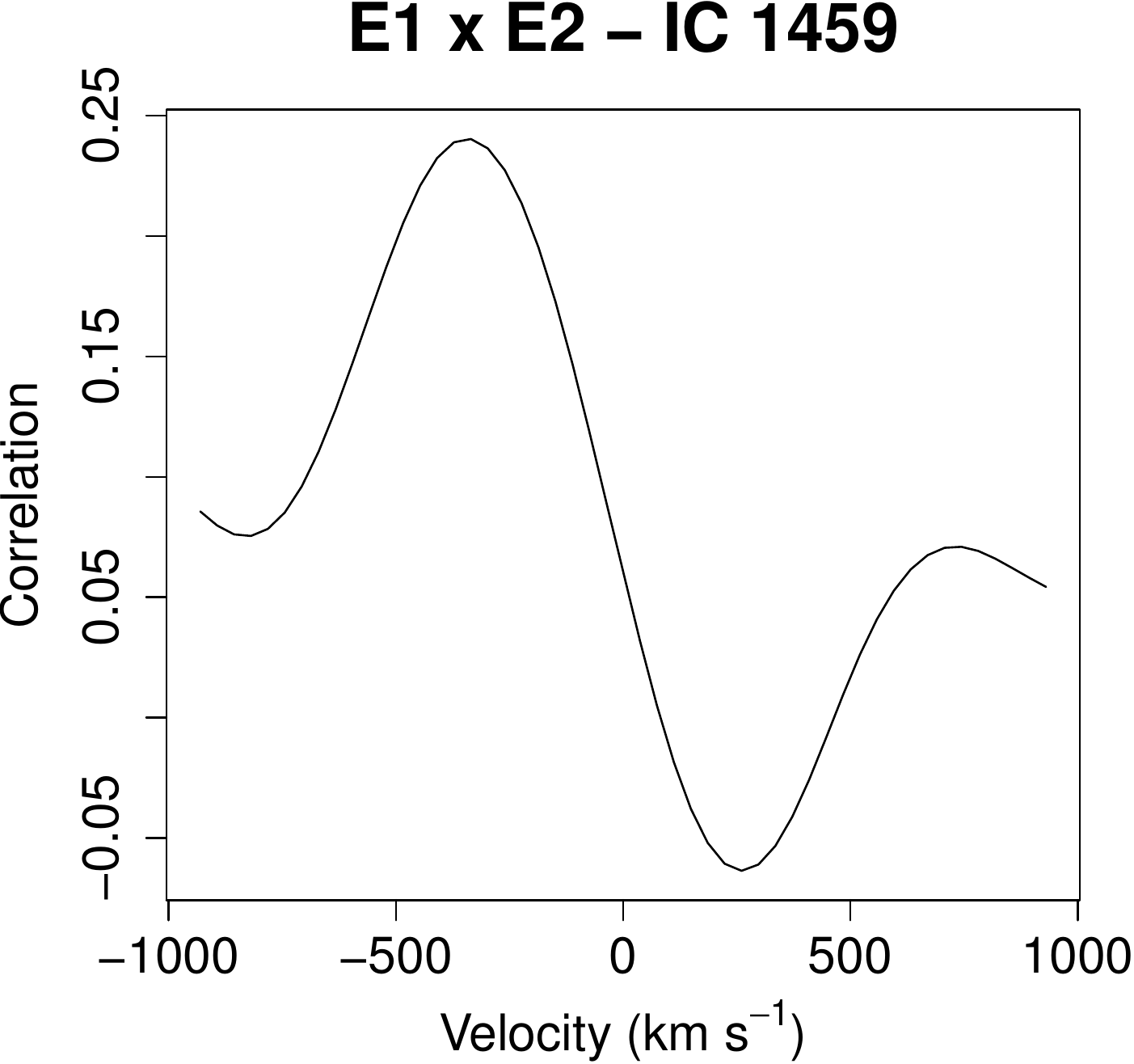}
\includegraphics[scale=0.35]{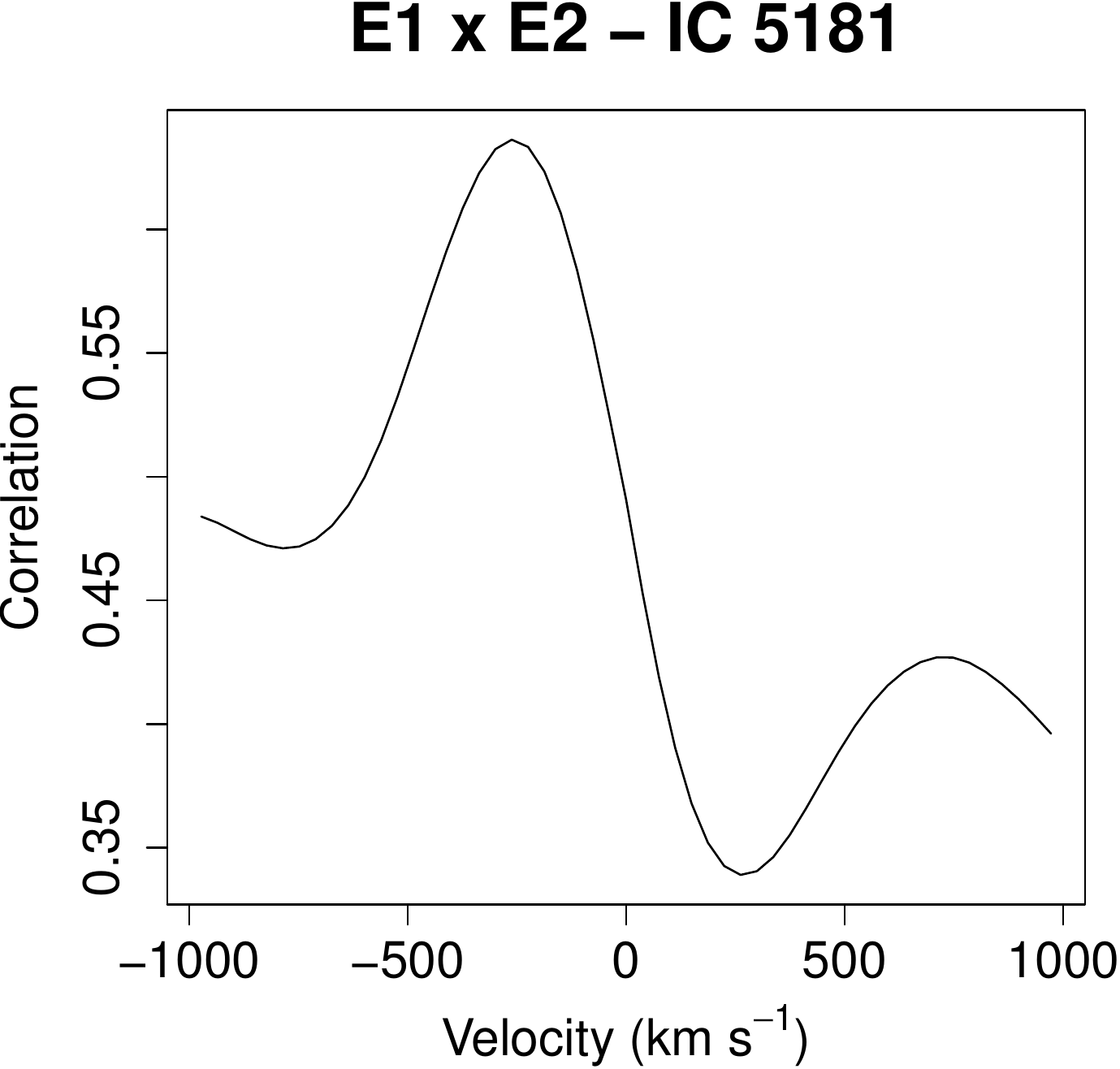}
\includegraphics[scale=0.35]{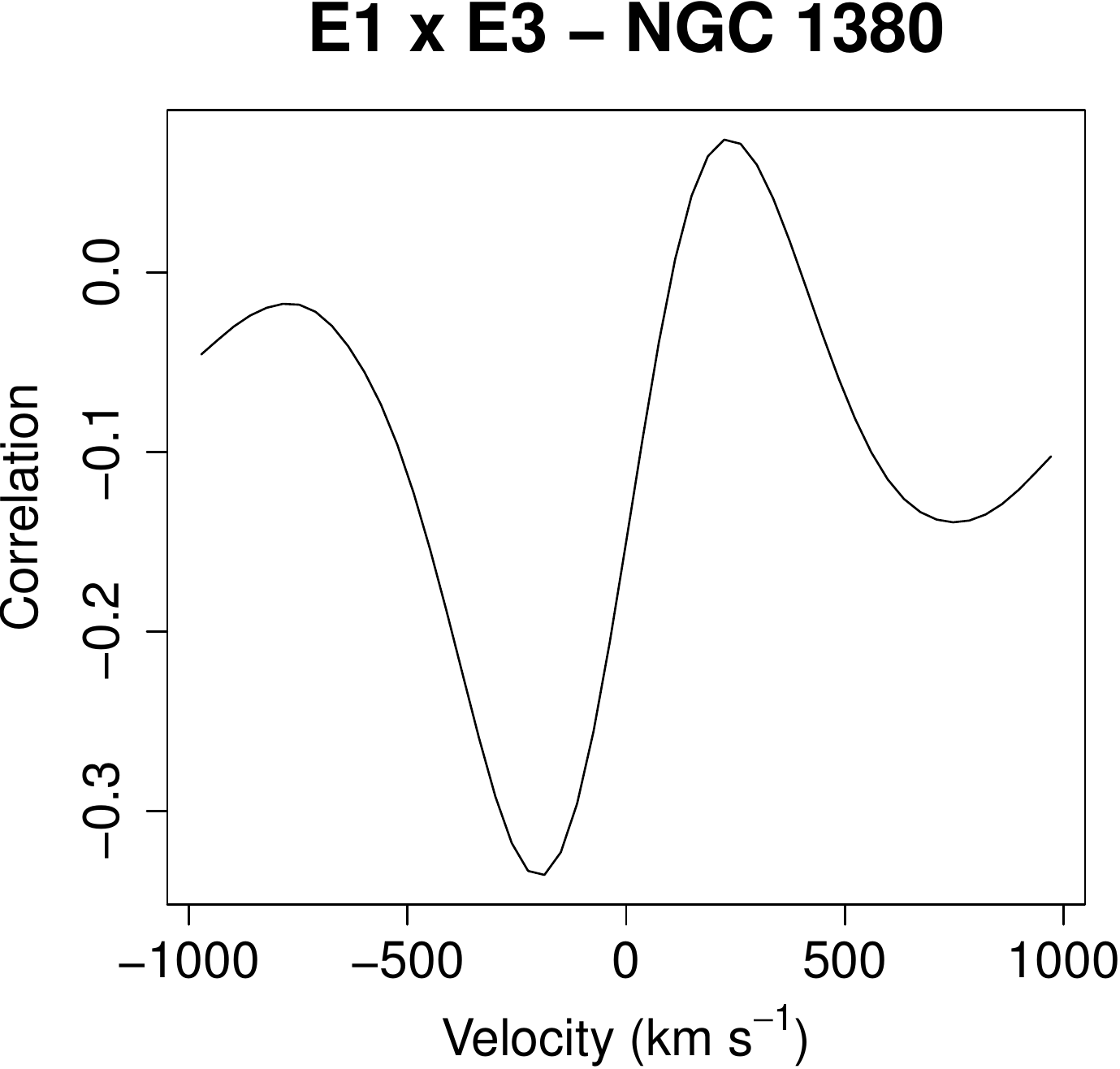}
\includegraphics[scale=0.35]{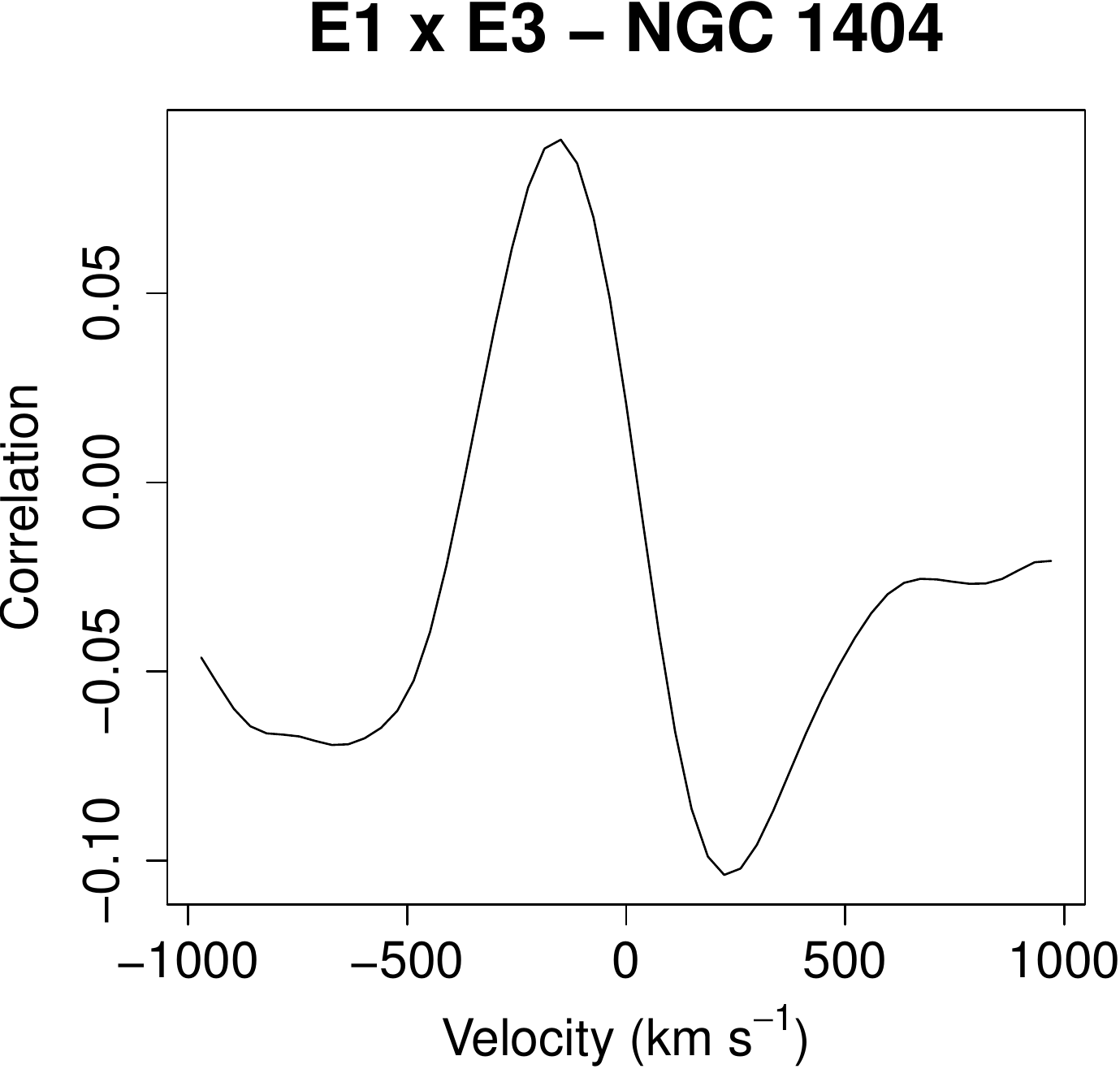}
\includegraphics[scale=0.35]{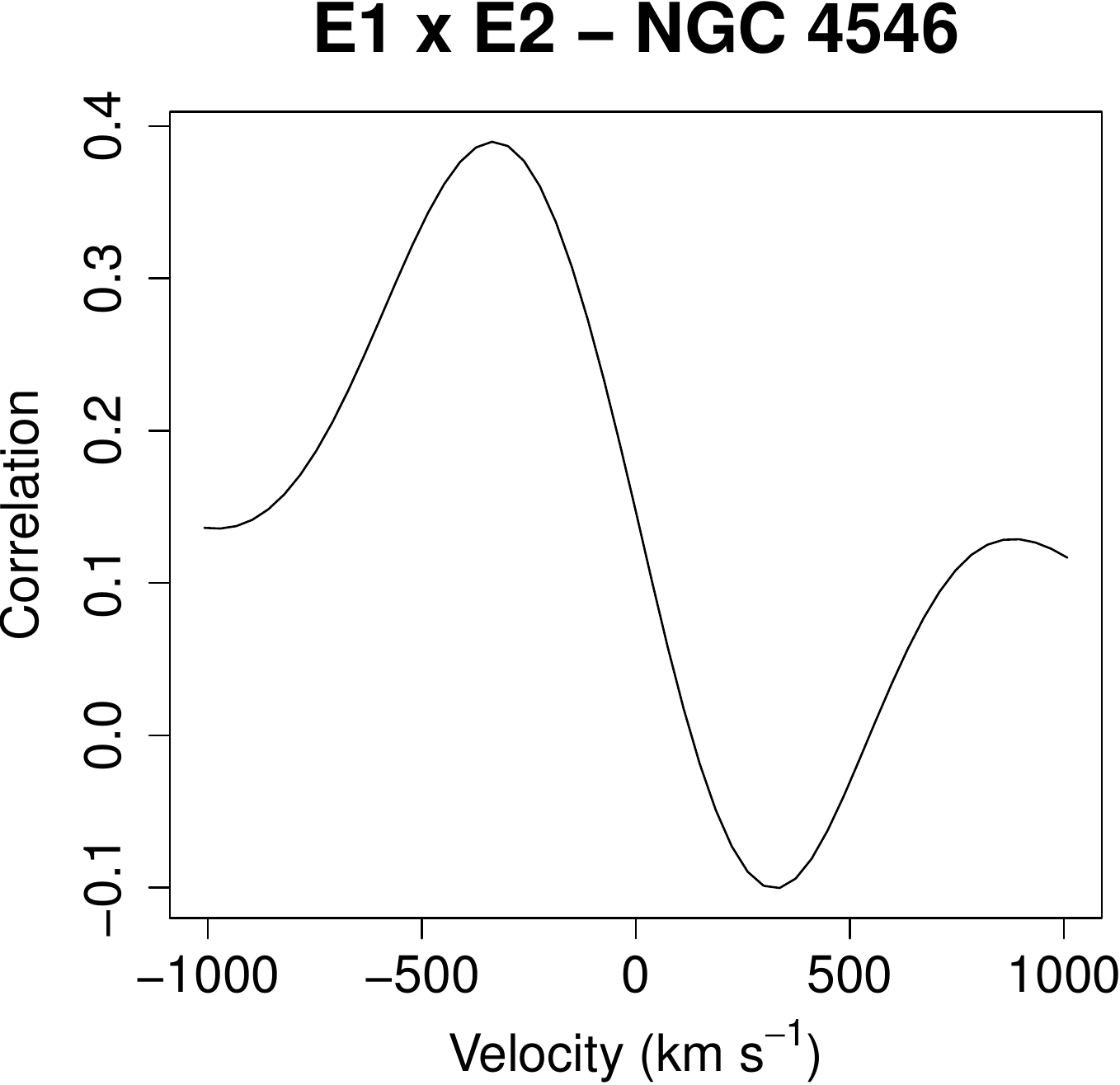}
\includegraphics[scale=0.35]{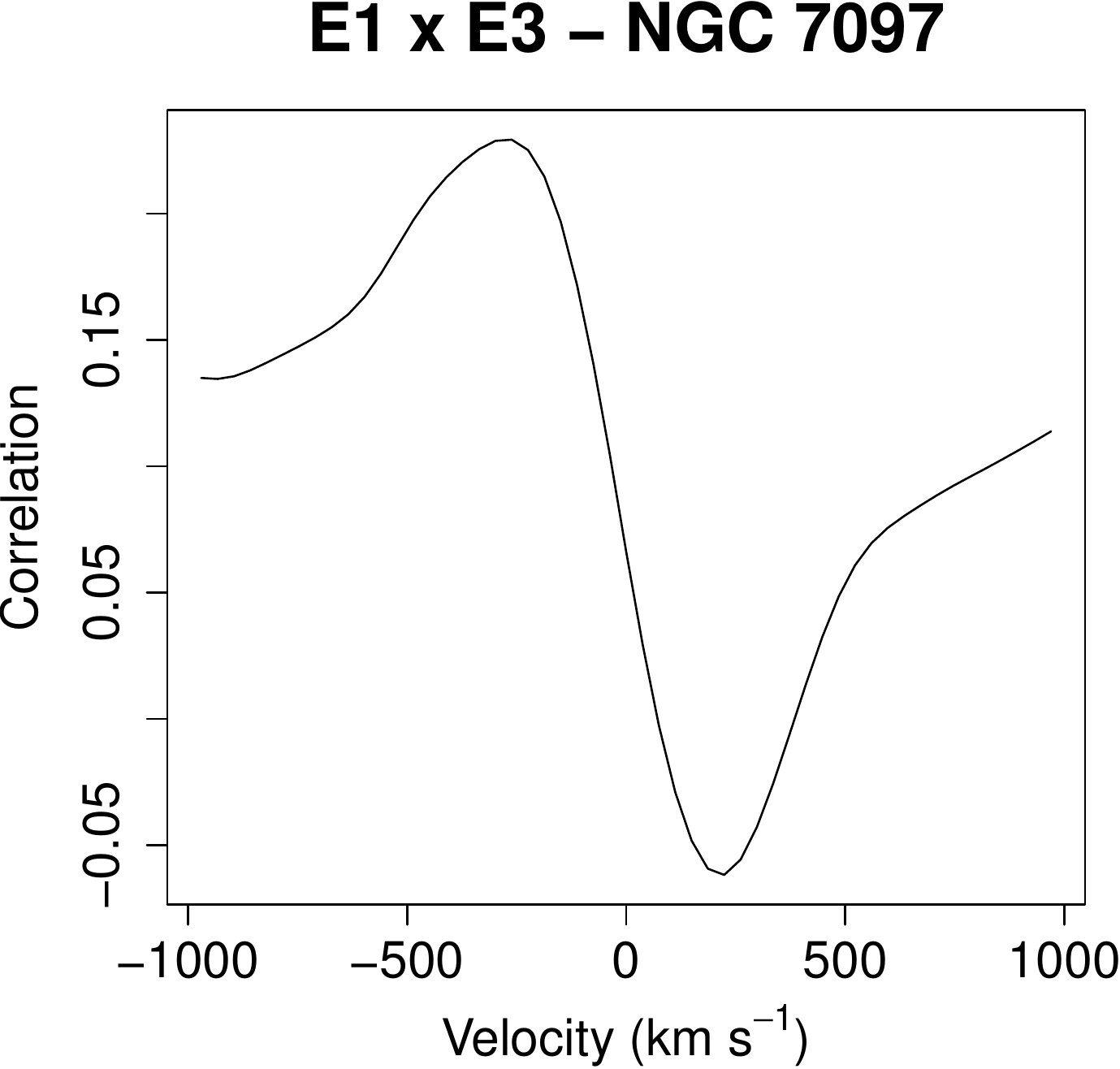}

\caption{Cross-correlation between the eigenspectra that represent stellar kinematics and their respective eigenspectra that represent the bulges of the galaxies (eigenspectrum 1 in all cases). Since the stellar kinematics is interpreted as an anti correlation between the red and the blue wings of the stellar lines, one of the wings should be correlated with the stellar lines in eigenspectrum 1 and the other wing should be anti correlated with the stellar lines in eigenspectrum 1, resulting in a sinusoidal curve for the cross-correlation. Although the \textit{x}-axis is given by the velocity, no reliable information about this feature may be extracted from these figures, since they may be affected by the radial velocity, the intensity of the stellar component and the stellar velocity dispersion. \label{correlacao_cruzada}
}
\end{center}
\end{figure*}

\begin{table}
 \scriptsize
 \begin{center}
 \begin{tabular}{@{}lccc}
  \hline
  Galaxy name & P.A.$_{stellar}$ & P.A.$_{gas}$ &P.A.$_{stellar}$ - P.A.$_{gas}$ \\
  \hline
  ESO 208 G-21 & -57$^o \pm$4$^o$ & -73$^o \pm$3$^o$ & 16$^o$ \\
  NGC 1380 & 177$^o \pm$4$^o$ & -175$^o \pm$4$^o$ & 8$^o$ \\
  IC 1459 & -143$^o \pm$3$^o$ & 39$^o \pm$3$^o$ & -182$^o$ \\
  NGC 7097 & 159$^o \pm$6$^o$ & -9$^o \pm$3$^o$ & 168$^o$ \\
  IC 5181 & 61$^o \pm$4$^o$ & -21$^o \pm$2$^o$ & 82$^o$ \\
  NGC 4546 & 82$^o \pm$3$^o$ & -149$^o \pm$3$^o$ & -129$^o$ \\
  NGC 2663 & - & -120$^o \pm$5$^o$ & - \\
  NGC 1404 & -32$^o \pm$6$^o$ & - & - \\
  NGC 3136 (E) & - & 98$^o \pm$7$^o$ & - \\
  NGC 3136 (P) & - & 37$^o \pm$2$^o$ &  -\\
 
  \hline
 \end{tabular}
 \caption{P.A. of the gas and stellar discs of the galaxies of the sample. In NGC 3136, (E) is related to the extended structures and (P) is related to the point-like structures.  \label{tab_PA_pca}
}
 \end{center}
\end{table}

\section{Discussions and conclusions} \label{sec:conc}

PCA Tomography applied to data cubes in the optical range is a recently developed technique for information extraction \citep{1997ApJ...475..173H,2009MNRAS.395...64S} and noise treatment (\citealt{2009MNRAS.395...64S}; Steiner et al. in preparation). The interpretation of the results is, frequently, subtle. Observations in other wavelengths are, often, important to reliably interpret the data.

In this work, we showed that the eigenvector 2, related to data cubes of galaxies with emission lines, may be interpreted as dominated by correlations caused by an AGN. Thus, a new and powerful methodology for this kind of research is unveiled. Splitting the data cubes in two spectral regions, one dominated by the variance of absorption lines (5150-5800 \AA), of stellar origin, and the other dominated by the variance of the emission lines (6250-6850 \AA), allows us to study the stellar or gaseous kinematics. With this methodology, we are able to identify stellar and gaseous discs and to conclude that, notwithstanding a small sample, the angular moment of both structures is uncoupled. This finding is important in discussing the origin of the circumnuclear gas in ETGs. One should be caution, however, with this conclusion, as some of the disc-like features could actually be contaminated by outflows. 

\citet{1986AJ.....91.1062P} observed 9 of 10 galaxies from our sample and only in NGC 1399 and NGC 1404 they did not detect any sign of the [N II] and H$\alpha$ emission lines. In both objects, PCA Tomography did not reveal any sign of emission lines either. NGC 4546 was the only galaxy of the sample which was not observed by \citet{1986AJ.....91.1062P}; however, \citet{1987ApJ...318..531G} related emission lines in the central region of this object.

Eigenspectra 2 presented in this work have shown line ratios typical of LINERs for seven galaxies of the sample. For ESO 208 G-21, it was not possible to detect the H$\beta$ and [O III] emission lines. However, low-ionization lines like [O I], [N II], and [S II], when compared to H$\alpha$, have typical intensities of LINERs in this object. The detection of extended kinematic features in eight galaxies of the sample suggests that the circumnuclear gas emission is quite common. Since the eigenspectra related to the gas kinematics are characterized by anti correlations between the red and blue wings of the emission lines, it is not possible to estimate their line ratios. However, the observed eigenspectra show strong correlations in the low ionization lines, typical of LINERs.

Point-like sources unveiled by the tomograms related to the eigenspectra 2 support the hypothesis that the LINERs seen in nuclear regions of the eight galaxies are, indeed, photoionized by an AGN. Moreover, in IC 1459, the eigenspectrum 2 showed evidences of the featureless continuum emitted by the AGN. One could argue that if these galaxies have an AGN, broad components from the BLR should be appearing in some emission lines, particularly in H$\alpha$. However, the detection of this component should be carried out by carefully decomposing the nuclear emission lines. Since even in eigenspectra 2 there might be kinematic effects superposed on the correlation between the emission lines (see, for instance, the case of NGC 3136 and, less obviously, in IC 5181, especially in the [O III]$\lambda$5007 line of this object), it is not recommended to decompose the emission lines when manipulating eigenspectra. Radio observations have supported the presence of an AGN in NGC 2663 and NGC 7097 \citep{1994MNRAS.269..928S}. \citet{2010MNRAS.402.2187S} argued that an AGN must be responsible for the H$\beta$ emission in the nuclear region of NGC 4546. The evidences of an AGN in IC 1459 were discussed in Section \ref{case_IC1459}. With regard to the others galaxies of the sample, a careful analysis of their nuclear spectra will be performed in Paper II. 

In two galaxies (IC 1459 and NGC 7097), the gas and the stellar discs are counter-rotating in their central regions. This has already been discussed in the literature (IC 1459 - \citealt{1988ApJ...327L..55F,2002ApJ...578..787C}; NGC 7097 - \citealt{1986ApJ...305..136C}). In NGC 1380 and ESO 208 G-21, both structures are corotating. In two other objects (IC 5181 and NGC 4546), the rotating planes are not coincident. Curiously, in both galaxies the stellar discs seem to be asymmetric, as revealed by PCA Tomography. In NGC 4546, \citet{2006MNRAS.366.1151S} measured the difference between the P.A. of the gaseous and stellar discs and found P.A.$_{stellar}$ - P.A.$_{gas}$ = -144$^o$, which matches our calculation of -129$^o$ (see table \ref{tab_PA_pca}).

Although it has been proposed that the detected gas kinematics would indicate gas discs in seven galaxies, one should not discard outright that these structures may be related to ionization cones. The co-rotation between the gas and stellar kinematics seen in NGC 1380 and ESO 208 G-21 is compatible with the idea of a gas disc, with an internal origin (e.g. stellar mass loss). In IC 1459, NGC 7097 and NGC 4546, the gas kinematics have the same P.A. as the gas discs found in regions more distant from the nuclei of these galaxies \citep{2002ApJ...578..787C,1986ApJ...305..136C,2006MNRAS.366.1151S}. In fact, for NGC 7097, \citet{2011ApJ...734L..10R} proposed that an ionization cone is observed in the perpendicular direction of the gas disc. Since tomograms represent weights that involve intensities and kinematics, a more extended emission indicates that, in NGC 3136, tomogram 2 of the red spectral region (Fig. \ref{tomogram_N3136_2}) reveals an emission with an opening angle that is wider than in other galaxies. This suggests that the emission might not be associated with a disc and possibly represents ionization cones or, perhaps, a combination of both, disc and cones.

In the case of NGC 2663, no evidences of gaseous disc have been published before. In IC 5181, \citet{2013A&A...560A..14P} also found that the stellar and the gaseous components are orthogonally rotating. Ionization cones may be present in both cases. In IC 5181, the difference between the P.A. of the stellar and gaseous kinematics is $\sim$ 112.5$^o$, which means that both components do not rotate in the same direction. In this case, it is also not possible to claim that the circumnuclear region of gas is really a disc. Maybe an ionization cone in regions closer to the nucleus is important and may have an effect on the tomogram, since the most internal regions are more intense and, therefore, have greater weight in the tomograms. Thus, the gas kinematic analysis, along with other methods (e.g. radial velocity extraction or the application of PCA Tomography to synthetic discs or ionization cones), is important to confirm the nature of this component. The extended emission will be revisited in Paper III.

It is worth mentioning that, with PCA Tomography, one is able to obtain information about the kinematics of an object in a few minutes, while extracting the stellar velocity map with standard fitting techniques (e.g. ppxf - \citealt{2004PASP..116..138C}) takes several hours for a typical GMOS data cube. Also the presence of AGN activity is revealed in a faster way than subtracting the stellar components of all spectra of a data cube. Moreover, tomograms and eigenspectra are quite useful to have a first qualitative look in data cubes. Besides AGN activity and kinematics, PCA Tomography may be also very helpful in detecting other and previously unsuspected physical phenomena. For example, in NGC 7097, only with PCA Tomography, \citet{2011ApJ...734L..10R} proposed that the light of the AGN is reflected in our line of sight by an ionization cone. If this scenario may be recovered for others objects, one may select, more effectively, a sample galaxies to be observed by using spectropolarimetry. Although we highlighted the importance of PCA Tomography to extract preliminary information, we actually propose to analyse data cubes using both PCA Tomography and standard techniques, since it may be complementary. In the case of this work, Papers II and III will provide the complement of this Paper I. PCA Tomography is also an efficient way of summarizing the information in a data cube. Usually, $\sim$ 5 eigenvectors/tomograms explain more than 99\% of the variance. Finally, PCA is a nonparametric method and, therefore, does not depend on previous assumptions. The interpretation, however, may frequently be quite subtle.

Below, we summarize our findings.

\begin{itemize}
	\item The PCA Tomography method is efficient to detect LINER emissions, either nuclear or circumnuclear, in massive galaxies with previously known emission lines. This methodology is also effective to detect stellar discs. 
	\item All objects from our sample, previously selected as having emission lines, may be classified as nuclear LINERs, photoionized by an AGN and, also, as presenting circumnuclear emission.
	\item We found that in IC 1459, the Fe5270 stellar absorption feature is correlated with the AGN emission lines, suggesting the presence of a featureless continuum.
	\item We found that the nuclear region, expressed in eigenspectra 2, is usually reddened and, when so, associated with interstellar Na D absorption, indicating the presence of dust and neutral gas.
	\item In two galaxies (NGC 1399 and NGC 1404) whose emission lines were not previously known, PCA Tomography also did not detect any sign of ionized gas.
	\item Six of eight galaxies with an AGN detected by PCA Tomography also present a circumnuclear emission of gas with disc-like features.
	\item The elliptical galaxy NGC 3136 shows intense emission lines and may be classified as a LINER, but it has a complex structure. With PCA Tomography, we detected two point-like objects and features that seem to be co-existing disc and ionization cones. 
	\item In seven galaxies of the sample (four S0 and three E), the stellar component has a disc-like kinematical signature.
	\item In two galaxies, the stellar and gaseous discs are in co-rotation, whereas in other two they are counter-rotating; the discs appear not to be aligned in two objects. This decoupling of the angular momentum of the gas and of the stars suggests that the origin of the gas is not internal and argues against the idea that the gas and stellar ionization sources are spatially close. 
	\item NGC 2663 shows only a gas kinematics (disc or ionization cone), but does not have a stellar disc. On the other hand, NGC 1404 has only a stellar disc. 

\end{itemize}

\section*{Acknowledgements}

Some of the data presented in this paper were obtained from the Mikulski Archive for Space Telescopes (MAST). STScI is operated by the Association of Universities for Research in Astronomy, Inc., under NASA contract NAS5-26555. Support for MAST for non-\textit{HST} data is provided by the NASA Office of Space Science via grant NNX09AF08G and by other grants and contracts. This paper is based on observations obtained at the Gemini Observatory, which is operated by the Association of Universities for Research in Astronomy, Inc., under a cooperative agreement with the NSF on behalf of the Gemini partnership: the National Science Foundation (United States), the National Research Council (Canada), CONICYT (Chile), the Australian Research Council (Australia), Minist\'{e}rio da Ci\^{e}ncia, Tecnologia e Inova\c{c}\~{a}o (Brazil) and Ministerio de Ciencia, Tecnolog\'{i}a e Innovaci\'{o}n Productiva (Argentina). IRAF is distributed by the National Optical Astronomy Observatory, which is operated by the Association of Universities for Research in Astronomy (AURA) under cooperative agreement with the National Science Foundation. We also acknowledge the usage of the HyperLeda data base (http://leda.univ-lyon1.fr). This research has made use of the NASA/IPAC Extragalactic Database (NED), which is operated by the Jet Propulsion Laboratory, California Institute of Technology, under contract with the National Aeronautics and Space Administration.

T.V.R and R.B.M. also acknowledge FAPESP for the financial support under grants 2008/06988-0 (T.V.R.), 2012/21350-7 (T.V.R.) and 2012/02262-8 (R.B.M.). We also thank the anonymous referee for valuable suggestions that improved the quality of this paper.

\bibliographystyle{mn2e}
\bibliography{bibliografia}

\begin{thebibliography}{}

\bibitem[\protect\citeauthoryear{{Allington-Smith}, {Murray}, {Content},
  {Dodsworth}, {Davies}, {Miller}, {Jorgensen}, {Hook}, {Crampton} \&
  {Murowinski}}{{Allington-Smith} et~al.}{2002}]{2002PASP..114..892A}
{Allington-Smith} J.,  {Murray} G.,  {Content} R.,  {Dodsworth} G.,  {Davies}
  R.,  {Miller} B.~W.,  {Jorgensen} I.,  {Hook} I.,  {Crampton} D.,
  {Murowinski} R.,  2002, \pasp, 114, 892

\bibitem[\protect\citeauthoryear{{Annibali}, {Bressan}, {Rampazzo},
  {Zeilinger}, {Vega} \& {Panuzzo}}{{Annibali}
  et~al.}{2010}]{2010A&A...519A..40A}
{Annibali} F.,  {Bressan} A.,  {Rampazzo} R.,  {Zeilinger} W.~W.,  {Vega} O.,
   {Panuzzo} P.,  2010, \aap, 519, A40

\bibitem[\protect\citeauthoryear{{Antonucci}}{{Antonucci}}{1993}]{1993ARA&A..3%
1..473A}
{Antonucci} R.,  1993, \araa, 31, 473

\bibitem[\protect\citeauthoryear{{Antonucci} \& {Miller}}{{Antonucci} \&
  {Miller}}{1985}]{1985ApJ...297..621A}
{Antonucci} R.~R.~J.,  {Miller} J.~S.,  1985, \apj, 297, 621

\bibitem[\protect\citeauthoryear{{Baldwin}, {Phillips} \&
  {Terlevich}}{{Baldwin} et~al.}{1981}]{1981PASP...93....5B}
{Baldwin} J.~A.,  {Phillips} M.~M.,    {Terlevich} R.,  1981, \pasp, 93, 5

\bibitem[\protect\citeauthoryear{{Barth}, {Filippenko} \& {Moran}}{{Barth}
  et~al.}{1999}]{1999ApJ...525..673B}
{Barth} A.~J.,  {Filippenko} A.~V.,    {Moran} E.~C.,  1999, \apj, 525, 673

\bibitem[\protect\citeauthoryear{{Barth} \& {Shields}}{{Barth} \&
  {Shields}}{2000}]{2000PASP..112..753B}
{Barth} A.~J.,  {Shields} J.~C.,  2000, \pasp, 112, 753

\bibitem[\protect\citeauthoryear{{Binette}, {Magris}, {Stasi{\'n}ska} \&
  {Bruzual}}{{Binette} et~al.}{1994}]{1994A&A...292...13B}
{Binette} L.,  {Magris} C.~G.,  {Stasi{\'n}ska} G.,    {Bruzual} A.~G.,  1994,
  \aap, 292, 13

\bibitem[\protect\citeauthoryear{{B{\"o}nsch} \& {Potulski}}{{B{\"o}nsch} \&
  {Potulski}}{1998}]{1998Metro..35..133B}
{B{\"o}nsch} G.,  {Potulski} E.,  1998, Metrologia, 35, 133

\bibitem[\protect\citeauthoryear{{Caldwell}, {Kirshner} \&
  {Richstone}}{{Caldwell} et~al.}{1986}]{1986ApJ...305..136C}
{Caldwell} N.,  {Kirshner} R.~P.,    {Richstone} D.~O.,  1986, \apj, 305, 136

\bibitem[\protect\citeauthoryear{{Cappellari} \& {Emsellem}}{{Cappellari} \&
  {Emsellem}}{2004}]{2004PASP..116..138C}
{Cappellari} M.,  {Emsellem} E.,  2004, \pasp, 116, 138

\bibitem[\protect\citeauthoryear{{Cappellari}, {Verolme}, {van der Marel},
  {Kleijn}, {Illingworth}, {Franx}, {Carollo} \& {de Zeeuw}}{{Cappellari}
  et~al.}{2002}]{2002ApJ...578..787C}
{Cappellari} M.,  {Verolme} E.~K.,  {van der Marel} R.~P.,  {Kleijn} G.~A.~V.,
  {Illingworth} G.~D.,  {Franx} M.,  {Carollo} C.~M.,    {de Zeeuw} P.~T.,
  2002, \apj, 578, 787

\bibitem[\protect\citeauthoryear{{Carollo}, {Franx}, {Illingworth} \&
  {Forbes}}{{Carollo} et~al.}{1997}]{1997ApJ...481..710C}
{Carollo} C.~M.,  {Franx} M.,  {Illingworth} G.~D.,    {Forbes} D.~A.,  1997,
  \apj, 481, 710

\bibitem[\protect\citeauthoryear{{Cid Fernandes}, {Stasi{\'n}ska}, {Mateus} \&
  {Vale Asari}}{{Cid Fernandes} et~al.}{2011}]{2011MNRAS.413.1687C}
{Cid Fernandes} R.,  {Stasi{\'n}ska} G.,  {Mateus} A.,    {Vale Asari} N.,
  2011, \mnras, 413, 1687

\bibitem[\protect\citeauthoryear{{Czerny} \& {Hryniewicz}}{{Czerny} \&
  {Hryniewicz}}{2011}]{2011A&A...525L...8C}
{Czerny} B.,  {Hryniewicz} K.,  2011, \aap, 525, L8

\bibitem[\protect\citeauthoryear{{de Vaucouleurs}, {de Vaucouleurs}, {Corwin}
  Jr., {Buta}, {Paturel} \& {Fouque}}{{de Vaucouleurs}
  et~al.}{1991}]{1991trcb.book.....D}
{de Vaucouleurs} G.,  {de Vaucouleurs} A.,  {Corwin} Jr. H.~G.,  {Buta} R.~J.,
  {Paturel} G.,    {Fouque} P.,  1991, {Third Reference Catalogue of Bright
  Galaxies}

\bibitem[\protect\citeauthoryear{{de Zeeuw}, {Bureau}, {Emsellem}, {Bacon},
  {Carollo}, {Copin}, {Davies}, {Kuntschner}, {Miller}, {Monnet}, {Peletier} \&
  {Verolme}}{{de Zeeuw} et~al.}{2002}]{2002MNRAS.329..513D}
{de Zeeuw} P.~T.,  {Bureau} M.,  {Emsellem} E.,  {Bacon} R.,  {Carollo} C.~M.,
  {Copin} Y.,  {Davies} R.~L.,  {Kuntschner} H.,  {Miller} B.~W.,  {Monnet} G.,
   {Peletier} R.~F.,    {Verolme} E.~K.,  2002, \mnras, 329, 513

\bibitem[\protect\citeauthoryear{{Elitzur} \& {Shlosman}}{{Elitzur} \&
  {Shlosman}}{2006}]{2006ApJ...648L.101E}
{Elitzur} M.,  {Shlosman} I.,  2006, \apjl, 648, L101

\bibitem[\protect\citeauthoryear{{Eracleous}, {Hwang} \& {Flohic}}{{Eracleous}
  et~al.}{2010}]{2010ApJ...711..796E}
{Eracleous} M.,  {Hwang} J.~A.,    {Flohic} H.~M.~L.~G.,  2010, \apj, 711, 796

\bibitem[\protect\citeauthoryear{{Fabbiano}, {Elvis}, {Markoff},
  {Siemiginowska}, {Pellegrini}, {Zezas}, {Nicastro}, {Trinchieri} \&
  {McDowell}}{{Fabbiano} et~al.}{2003}]{2003ApJ...588..175F}
{Fabbiano} G.,  {Elvis} M.,  {Markoff} S.,  {Siemiginowska} A.,  {Pellegrini}
  S.,  {Zezas} A.,  {Nicastro} F.,  {Trinchieri} G.,    {McDowell} J.,  2003,
  \apj, 588, 175

\bibitem[\protect\citeauthoryear{{Ferland} \& {Netzer}}{{Ferland} \&
  {Netzer}}{1983}]{1983ApJ...264..105F}
{Ferland} G.~J.,  {Netzer} H.,  1983, \apj, 264, 105

\bibitem[\protect\citeauthoryear{{Filippenko}}{{Filippenko}}{1982}]{1982PASP..%
.94..715F}
{Filippenko} A.~V.,  1982, \pasp, 94, 715

\bibitem[\protect\citeauthoryear{{Filippenko} \& {Halpern}}{{Filippenko} \&
  {Halpern}}{1984}]{1984ApJ...285..458F}
{Filippenko} A.~V.,  {Halpern} J.~P.,  1984, \apj, 285, 458

\bibitem[\protect\citeauthoryear{{Franx} \& {Illingworth}}{{Franx} \&
  {Illingworth}}{1988}]{1988ApJ...327L..55F}
{Franx} M.,  {Illingworth} G.~D.,  1988, \apjl, 327, L55

\bibitem[\protect\citeauthoryear{Fukunaga}{Fukunaga}{1990}]{fukunaga}
Fukunaga K.,  1990, {Statistical Pattern Recognition}.
Second Edition, ACADEMIC PRESS INC.

\bibitem[\protect\citeauthoryear{{Galletta}}{{Galletta}}{1987}]{1987ApJ...318.%
.531G}
{Galletta} G.,  1987, \apj, 318, 531

\bibitem[\protect\citeauthoryear{{Gonzales} \& {Woods}}{{Gonzales} \&
  {Woods}}{2008}]{2008gonzaleswoods}
{Gonzales} R.,  {Woods} R.,  2008, {Digital Image Processing}

\bibitem[\protect\citeauthoryear{{Gonz{\'a}lez-Mart{\'{\i}}n}, {Masegosa},
  {M{\'a}rquez}, {Guainazzi} \&
  {Jim{\'e}nez-Bail{\'o}n}}{{Gonz{\'a}lez-Mart{\'{\i}}n}
  et~al.}{2009}]{2009A&A...506.1107G}
{Gonz{\'a}lez-Mart{\'{\i}}n} O.,  {Masegosa} J.,  {M{\'a}rquez} I.,
  {Guainazzi} M.,    {Jim{\'e}nez-Bail{\'o}n} E.,  2009, \aap, 506, 1107

\bibitem[\protect\citeauthoryear{{Halpern} \& {Steiner}}{{Halpern} \&
  {Steiner}}{1983}]{1983ApJ...269L..37H}
{Halpern} J.~P.,  {Steiner} J.~E.,  1983, \apjl, 269, L37

\bibitem[\protect\citeauthoryear{{Heckman}}{{Heckman}}{1980}]{1980A&A....87..1%
52H}
{Heckman} T.~M.,  1980, \aap, 87, 152

\bibitem[\protect\citeauthoryear{{Heyer} \& {Schloerb}}{{Heyer} \&
  {Schloerb}}{1997}]{1997ApJ...475..173H}
{Heyer} M.~H.,  {Schloerb} F.~P.,  1997, \apj, 475, 173

\bibitem[\protect\citeauthoryear{{Ho}}{{Ho}}{2008}]{2008ARA&A..46..475H}
{Ho} L.~C.,  2008, \araa, 46, 475

\bibitem[\protect\citeauthoryear{{Ho}, {Filippenko} \& {Sargent}}{{Ho}
  et~al.}{1993}]{1993ApJ...417...63H}
{Ho} L.~C.,  {Filippenko} A.~V.,    {Sargent} W.~L.~W.,  1993, \apj, 417, 63

\bibitem[\protect\citeauthoryear{{Ho}, {Filippenko} \& {Sargent}}{{Ho}
  et~al.}{2003}]{2003ApJ...583..159H}
{Ho} L.~C.,  {Filippenko} A.~V.,    {Sargent} W.~L.~W.,  2003, \apj, 583, 159

\bibitem[\protect\citeauthoryear{{Ho}, {Filippenko}, {Sargent} \& {Peng}}{{Ho}
  et~al.}{1997}]{1997ApJS..112..391H}
{Ho} L.~C.,  {Filippenko} A.~V.,  {Sargent} W.~L.~W.,    {Peng} C.~Y.,  1997,
  \apjs, 112, 391

\bibitem[\protect\citeauthoryear{{Hook}, {J{\o}rgensen}, {Allington-Smith},
  {Davies}, {Metcalfe}, {Murowinski} \& {Crampton}}{{Hook}
  et~al.}{2004}]{2004PASP..116..425H}
{Hook} I.~M.,  {J{\o}rgensen} I.,  {Allington-Smith} J.~R.,  {Davies} R.~L.,
  {Metcalfe} N.,  {Murowinski} R.~G.,    {Crampton} D.,  2004, \pasp, 116, 425

\bibitem[\protect\citeauthoryear{{Kim}}{{Kim}}{1989}]{1989ApJ...346..653K}
{Kim} D.-W.,  1989, \apj, 346, 653

\bibitem[\protect\citeauthoryear{{Kormendy}, {Bender}, {Ajhar}, {Dressler},
  {Faber}, {Gebhardt}, {Grillmair}, {Lauer}, {Richstone} \&
  {Tremaine}}{{Kormendy} et~al.}{1996}]{1996ApJ...473L..91K}
{Kormendy} J.,  {Bender} R.,  {Ajhar} E.~A.,  {Dressler} A.,  {Faber} S.~M.,
  {Gebhardt} K.,  {Grillmair} C.,  {Lauer} T.~R.,  {Richstone} D.,
  {Tremaine} S.,  1996, \apjl, 473, L91

\bibitem[\protect\citeauthoryear{{Lucy}}{{Lucy}}{1974}]{1974AJ.....79..745L}
{Lucy} L.~B.,  1974, \aj, 79, 745

\bibitem[\protect\citeauthoryear{{Magorrian}, {Tremaine}, {Richstone},
  {Bender}, {Bower}, {Dressler}, {Faber}, {Gebhardt}, {Green}, {Grillmair},
  {Kormendy} \& {Lauer}}{{Magorrian} et~al.}{1998}]{1998AJ....115.2285M}
{Magorrian} J.,  {Tremaine} S.,  {Richstone} D.,  {Bender} R.,  {Bower} G.,
  {Dressler} A.,  {Faber} S.~M.,  {Gebhardt} K.,  {Green} R.,  {Grillmair} C.,
  {Kormendy} J.,    {Lauer} T.,  1998, \aj, 115, 2285

\bibitem[\protect\citeauthoryear{Menezes, Steiner \& Ricci}{Menezes
  et~al.}{2013}]{2041-8205-765-2-L40}
Menezes R.~B.,  Steiner J.~E.,    Ricci T.~V.,  2013, \apjl, 765, L40

\bibitem[\protect\citeauthoryear{{Menezes}, {Steiner} \& {Ricci}}{{Menezes}
  et~al.}{2014}]{2014MNRAS.438.2597M}
{Menezes} R.~B.,  {Steiner} J.~E.,    {Ricci} T.~V.,  2014, \mnras, 438, 2597

\bibitem[\protect\citeauthoryear{{Menezes}, {Steiner}, {Ricci} \&
  {Oliveira}}{{Menezes} et~al.}{2010}]{2010IAUS..267..123M}
{Menezes} R.~B.,  {Steiner} J.~E.,  {Ricci} T.~V.,    {Oliveira} A.~S.,  2010,
  in IAU Symposium Vol.~267 of IAU Symposium, {The Active Nucleus in NGC 4579}.
pp 123--123

\bibitem[\protect\citeauthoryear{{Nicastro}}{{Nicastro}}{2000}]{2000ApJ...530L%
..65N}
{Nicastro} F.,  2000, \apjl, 530, L65

\bibitem[\protect\citeauthoryear{{Osterbrock} \& {Ferland}}{{Osterbrock} \&
  {Ferland}}{2006}]{2006agna.book.....O}
{Osterbrock} D.~E.,  {Ferland} G.~J.,  2006, {Astrophysics of gaseous nebulae
  and active galactic nuclei}.
2nd.~ed.~by D.E.~Osterbrock and G.J.~Ferland.~Sausalito, CA: University Science
  Books, 2006

\bibitem[\protect\citeauthoryear{{Paturel}, {Petit}, {Prugniel}, {Theureau},
  {Rousseau}, {Brouty}, {Dubois} \& {Cambr{\'e}sy}}{{Paturel}
  et~al.}{2003}]{2003A&A...412...45P}
{Paturel} G.,  {Petit} C.,  {Prugniel} P.,  {Theureau} G.,  {Rousseau} J.,
  {Brouty} M.,  {Dubois} P.,    {Cambr{\'e}sy} L.,  2003, \aap, 412, 45

\bibitem[\protect\citeauthoryear{{Phillips}}{{Phillips}}{1979}]{1979ApJ...227L%
.121P}
{Phillips} M.~M.,  1979, \apjl, 227, L121

\bibitem[\protect\citeauthoryear{{Phillips}, {Jenkins}, {Dopita}, {Sadler} \&
  {Binette}}{{Phillips} et~al.}{1986}]{1986AJ.....91.1062P}
{Phillips} M.~M.,  {Jenkins} C.~R.,  {Dopita} M.~A.,  {Sadler} E.~M.,
  {Binette} L.,  1986, \aj, 91, 1062

\bibitem[\protect\citeauthoryear{{Pizzella}, {Morelli}, {Corsini}, {Dalla
  Bont{\`a}} \& {Cesetti}}{{Pizzella} et~al.}{2013}]{2013A&A...560A..14P}
{Pizzella} A.,  {Morelli} L.,  {Corsini} E.~M.,  {Dalla Bont{\`a}} E.,
  {Cesetti} M.,  2013, \aap, 560, A14

\bibitem[\protect\citeauthoryear{{Prugniel} \& {Heraudeau}}{{Prugniel} \&
  {Heraudeau}}{1998}]{1998A&AS..128..299P}
{Prugniel} P.,  {Heraudeau} P.,  1998, \aaps, 128, 299

\bibitem[\protect\citeauthoryear{{Ricci}, {Steiner} \& {Menezes}}{{Ricci}
  et~al.}{2011}]{2011ApJ...734L..10R}
{Ricci} T.~V.,  {Steiner} J.~E.,    {Menezes} R.~B.,  2011, \apjl, 734, L10

\bibitem[\protect\citeauthoryear{{Richardson}}{{Richardson}}{1972}]{1972JOSA..%
.62...55R}
{Richardson} W.~H.,  1972, Journal of the Optical Society of America
  (1917-1983), 62, 55

\bibitem[\protect\citeauthoryear{{Riffel}, {Riffel}, {Ferrari} \&
  {Storchi-Bergmann}}{{Riffel} et~al.}{2011}]{2011MNRAS.416..493R}
{Riffel} R.,  {Riffel} R.~A.,  {Ferrari} F.,    {Storchi-Bergmann} T.,  2011,
  \mnras, 416, 493

\bibitem[\protect\citeauthoryear{{Sarzi}, {Falc{\'o}n-Barroso}, {Davies},
  {Bacon}, {Bureau}, {Cappellari}, {de Zeeuw}, {Emsellem}, {Fathi},
  {Krajnovi{\'c}}, {Kuntschner}, {McDermid} \& {Peletier}}{{Sarzi}
  et~al.}{2006}]{2006MNRAS.366.1151S}
{Sarzi} M.,  {Falc{\'o}n-Barroso} J.,  {Davies} R.~L.,  {Bacon} R.,  {Bureau}
  M.,  {Cappellari} M.,  {de Zeeuw} P.~T.,  {Emsellem} E.,  {Fathi} K.,
  {Krajnovi{\'c}} D.,  {Kuntschner} H.,  {McDermid} R.~M.,    {Peletier} R.~F.,
   2006, \mnras, 366, 1151

\bibitem[\protect\citeauthoryear{{Sarzi}, {Shields}, {Schawinski} \& {et
  al.}}{{Sarzi} et~al.}{2010}]{2010MNRAS.402.2187S}
{Sarzi} M.,  {Shields} J.~C.,  {Schawinski} K.,    {et al.} 2010, \mnras, 402,
  2187

\bibitem[\protect\citeauthoryear{{Schnorr M{\"u}ller}, {Storchi-Bergmann},
  {Riffel}, {Ferrari}, {Steiner}, {Axon} \& {Robinson}}{{Schnorr M{\"u}ller}
  et~al.}{2011}]{2011MNRAS.413..149S}
{Schnorr M{\"u}ller} A.,  {Storchi-Bergmann} T.,  {Riffel} R.~A.,  {Ferrari}
  F.,  {Steiner} J.~E.,  {Axon} D.~J.,    {Robinson} A.,  2011, \mnras, 413,
  149

\bibitem[\protect\citeauthoryear{{Skrutskie}, {Cutri}, {Stiening} \& {et
  al.}}{{Skrutskie} et~al.}{2006}]{2006AJ....131.1163S}
{Skrutskie} M.~F.,  {Cutri} R.~M.,  {Stiening}   {et al.} 2006, \aj, 131, 1163

\bibitem[\protect\citeauthoryear{{Slee}, {Sadler}, {Reynolds} \&
  {Ekers}}{{Slee} et~al.}{1994}]{1994MNRAS.269..928S}
{Slee} O.~B.,  {Sadler} E.~M.,  {Reynolds} J.~E.,    {Ekers} R.~D.,  1994,
  \mnras, 269, 928

\bibitem[\protect\citeauthoryear{{Sparks}, {Wall}, {Thorne}, {Jorden}, {van
  Breda}, {Rudd} \& {Jorgensen}}{{Sparks} et~al.}{1985}]{1985MNRAS.217...87S}
{Sparks} W.~B.,  {Wall} J.~V.,  {Thorne} D.~J.,  {Jorden} P.~R.,  {van Breda}
  I.~G.,  {Rudd} P.~J.,    {Jorgensen} H.~E.,  1985, \mnras, 217, 87

\bibitem[\protect\citeauthoryear{{Stasi{\'n}ska}, {Vale Asari}, {Cid
  Fernandes}, {Gomes}, {Schlickmann}, {Mateus}, {Schoenell}, {Sodr{\'e}} Jr. \&
  {Seagal Collaboration}}{{Stasi{\'n}ska} et~al.}{2008}]{2008MNRAS.391L..29S}
{Stasi{\'n}ska} G.,  {Vale Asari} N.,  {Cid Fernandes} R.,  {Gomes} J.~M.,
  {Schlickmann} M.,  {Mateus} A.,  {Schoenell} W.,  {Sodr{\'e}} Jr. L.,
  {Seagal Collaboration} 2008, \mnras, 391, L29

\bibitem[\protect\citeauthoryear{{Steiner}, {Menezes}, {Ricci} \&
  {Oliveira}}{{Steiner} et~al.}{2009}]{2009MNRAS.395...64S}
{Steiner} J.~E.,  {Menezes} R.~B.,  {Ricci} T.~V.,    {Oliveira} A.~S.,  2009,
  \mnras, 395, 64

\bibitem[\protect\citeauthoryear{{van Dokkum}}{{van
  Dokkum}}{2001}]{2001PASP..113.1420V}
{van Dokkum} P.~G.,  2001, \pasp, 113, 1420

\bibitem[\protect\citeauthoryear{{Veilleux} \& {Osterbrock}}{{Veilleux} \&
  {Osterbrock}}{1987}]{1987ApJS...63..295V}
{Veilleux} S.,  {Osterbrock} D.~E.,  1987, \apjs, 63, 295

\bibitem[\protect\citeauthoryear{{Verdoes Kleijn}, {van der Marel}, {Carollo}
  \& {de Zeeuw}}{{Verdoes Kleijn} et~al.}{2000}]{2000AJ....120.1221V}
{Verdoes Kleijn} G.~A.,  {van der Marel} R.~P.,  {Carollo} C.~M.,    {de Zeeuw}
  P.~T.,  2000, \aj, 120, 1221

\bibitem[\protect\citeauthoryear{{Yan} \& {Blanton}}{{Yan} \&
  {Blanton}}{2012}]{2012ApJ...747...61Y}
{Yan} R.,  {Blanton} M.~R.,  2012, \apj, 747, 61

\end{thebibliography}

\appendix

\section{Complementary data treatment} \label{tratamento_dados_complementares}

After the basic reduction process shown in Section \ref{sec:observation_data_reduction}, the data cubes still possess noises that must be suppressed in order to proceed with a more accurate analysis. The methods used in the removal of high- and low- frequency noise are briefly discussed below. In addition, the correction of the differential atmospheric refraction effect, which also affects our data cubes, is also described.  

\subsection{High spectral and spatial frequency noise removal}

\begin{figure*}
\begin{center}
\includegraphics[scale=0.30]{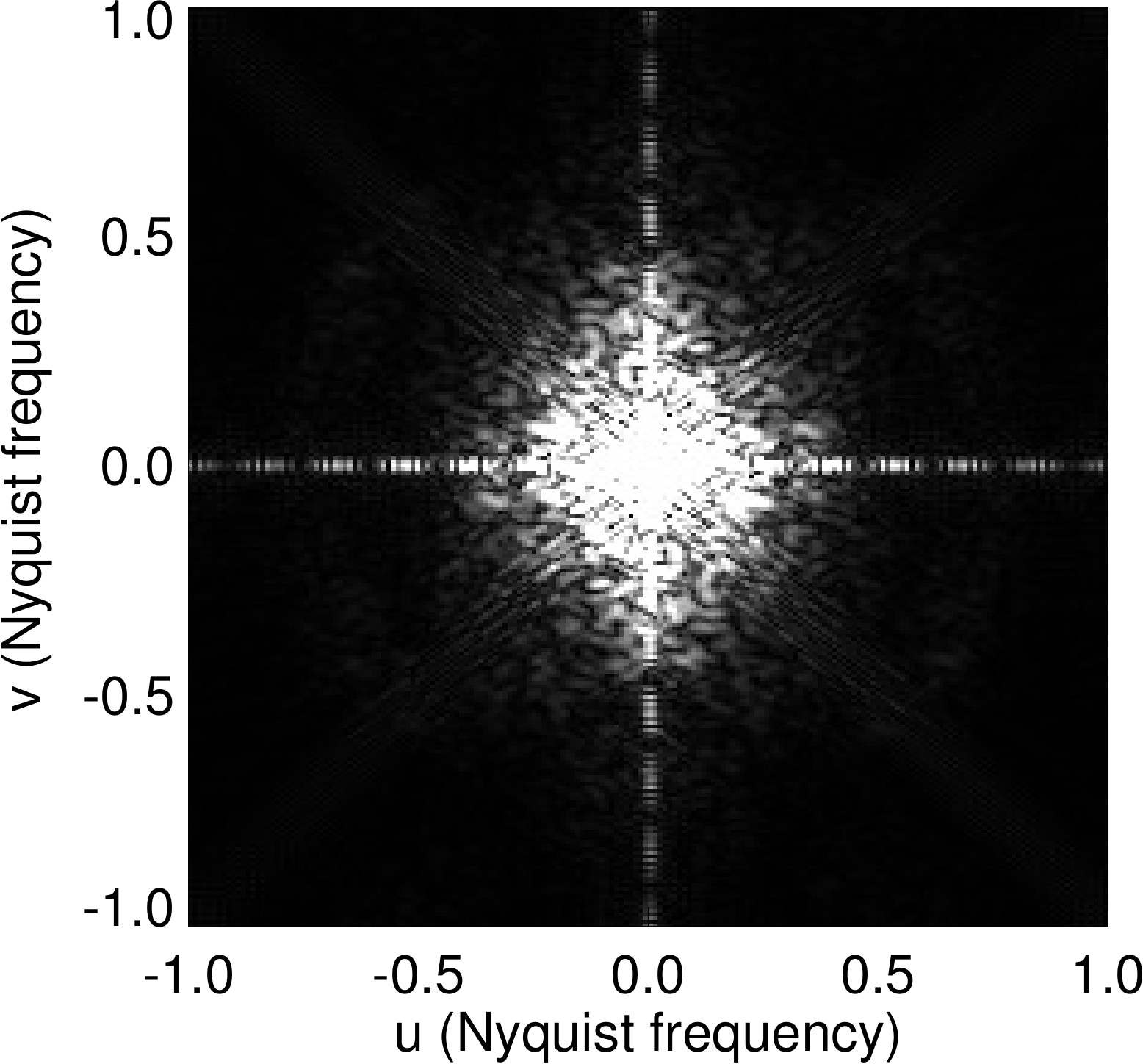}
\hspace{0cm}
\vspace{0.3cm}
\includegraphics[scale=0.30]{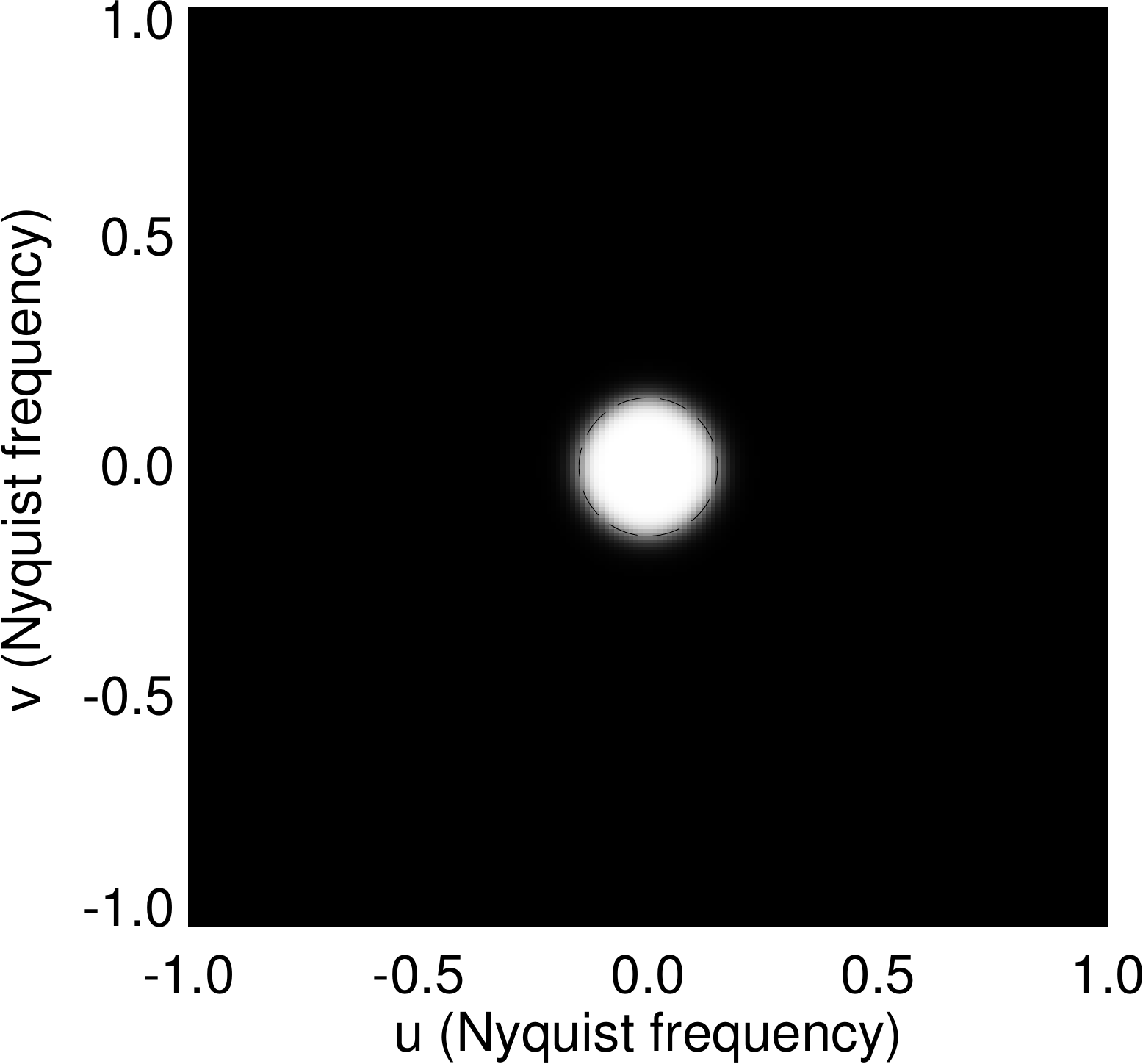}
\includegraphics[scale=0.30]{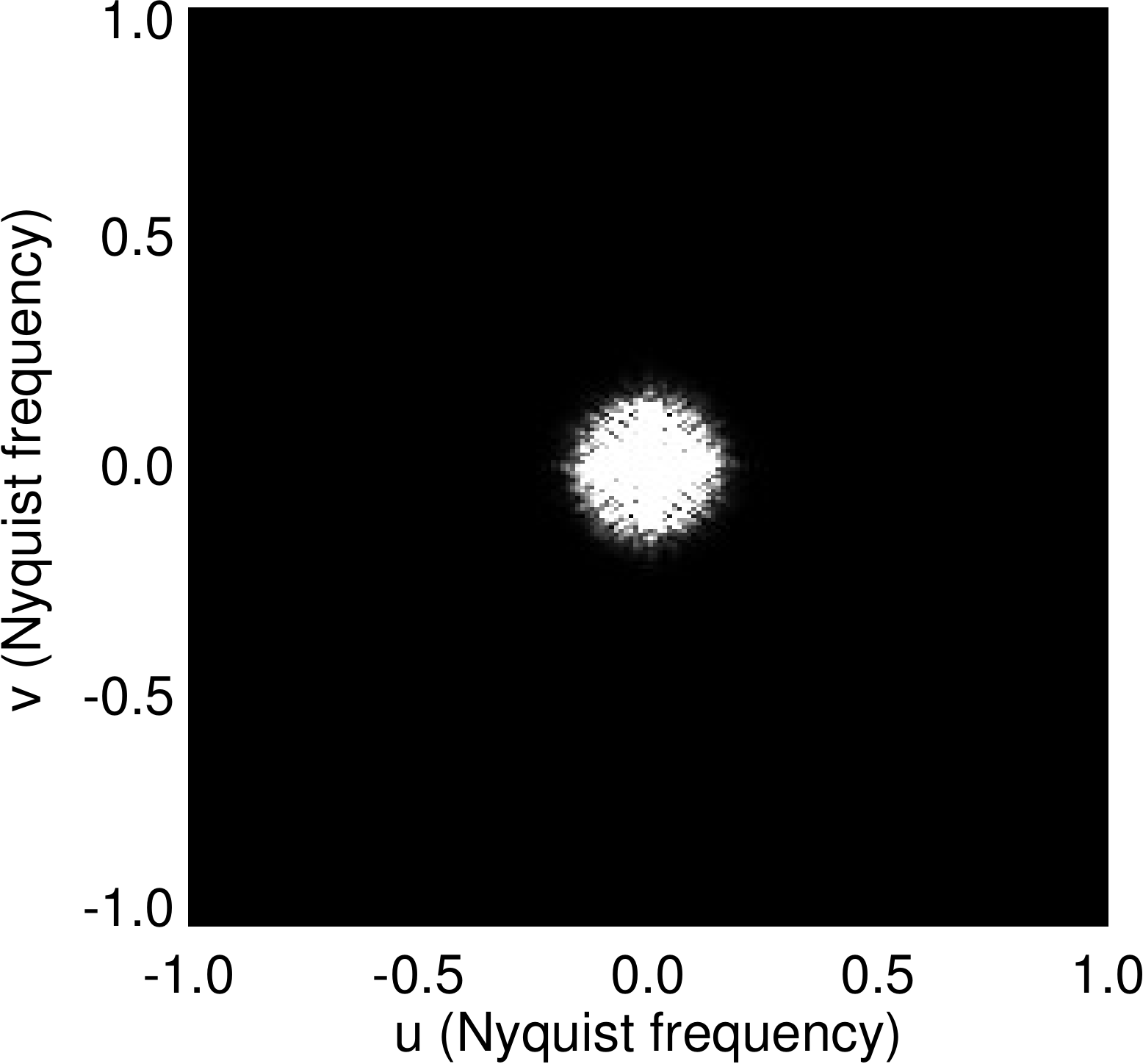}
\end{center}
\caption{Left: Fourier transform of the central image from the data cube of IC 1459. Centre: Butterworth filter with $n$ = 6 and $a = b =$ 0.15$F_{NY}$. Right: Fourier transform of the central image from the data cube of IC 1459 multiplied by the Butterworth filter. In all images, $u_c$ = $v_c$ = 0$F_{NY}$. The traced circle in black has a radius of 0.15$F_{NY}$ and represents the points where $H(u,v)$ = 0.5. In the central region of the filter, $H(u,v) \sim$ 1 and in its extremities, $H(u,v) \sim$ 0. The inverse Fourier transform of the figure on the right results in the central image of the data cube of IC 1459, free from high frequency noises.  \label{butterworth_images}
}
\end{figure*}

The spatial high-frequency noise contained in the data cubes was removed with a Butterworth filter $H(u,v)$ \citep{2008gonzaleswoods}, which was multiplied to the Fourier transform of each image of the data cubes. The filter equation is given by

\begin{equation}
H(u,v) = \frac{1}{1+\left[\sqrt{ \left(\frac{u-u_c}{a}\right)^2+\left(\frac{v-v_c}{b}\right)^2} \right]^{2n}},
	\label{butterworth_filter}
\end{equation}
where $u$ and $v$ are the spatial frequencies related to the $x$ and $y$ dimensions of the data cubes, respectively, relative to the $u_c$ and $v_c$ frequencies, $a$ and $b$ are the cutoff frequencies and $n$ is the filter index. The filter parameters used for all the galaxies of the sample were $n$ = 6 and $a$ = $b$ = 0.15 $F_{NY}$, where $F_{NY}$ is the Nyquist frequency of the spatial dimension. The filter is shown in Fig. \ref{butterworth_images}. These values of $a$ and $b$ were chosen in a manner that only artefacts with frequencies higher that the PSF frequency are removed. If one supposes that the spatial PSF is given by a Gaussian function, then the full width at half-maximum (FWHM) of the PSFs in Fourier space, in spaxels$^{-1}$, is given by

\begin{equation}
	FWHM_{FT}(PSF) = \frac{4ln(2)\phi}{\pi FWHM_{SP}(PSF)},
	\label{FT_FWHM_equation}
\end{equation}
where $\phi$ is the size of each spaxel, in arcsec, and $FWHM_{SP}(PSF)$ corresponds to the FWHM of the PSF in the spatial dimension of the data cubes, also in arcsec. If $F_{NY}$ = 0.5 spaxel$^{-1}$ and $\phi$ = 0.05 (see section \ref{sec:observation_data_reduction}), then for $FWHM_{SP}(PSF)$ $\sim$ 0.7arcsec (observation with the best seeing, see table \ref{sampledescription}), $FWHM_{FT}(PSF)$ $\sim$ 0.13$F_{NY}$. Values of $a$ and $b$ $>$ 0.13 $F_{NY}$ assure that the Butterworth filtering in all data cubes of the sample removes only artefacts with a frequency higher than that of the PSF. In fact, we performed a subtraction between the original and the filtered data cubes and an average image of the result showed only features with frequencies higher than the PSF (see Fig. \ref{fig_original_filtradas}).

For the removal of high-frequency noises in the spectra of the data cubes, a spectral Butterworth filter $H(k)$ was used. It is given by

\begin{equation}
	H(k) = \frac{1}{1+\left[\sqrt{\left(\frac{k-k_c}{l}\right)^2} \right]^{2n}}. \label{spectral_butterworth}
\end{equation}
Note that equation \ref{spectral_butterworth} is quite similar to equation \ref{butterworth_filter}, but, in this case, the $k$ parameter corresponds to the spectral frequency\footnote{One should be careful that $k$ is the sampling frequency of the spectra. It has nothing to do with the wave frequency.} of the data cubes, where $l$ is the cutoff frequency. We used $n$ = 6 and $l$ = 0.40$F_{NY}$ in all galaxies. The $l$ values were chosen with the same logic of the spatial dimensions, i.e. removing only frequencies higher than the frequency of a resolution element of the spectra.

\subsection{Instrumental fingerprint: identification and removal}
\label{fingerprint_removal}

\begin{figure*}
	\includegraphics[height=5cm, width=16cm, angle=0]{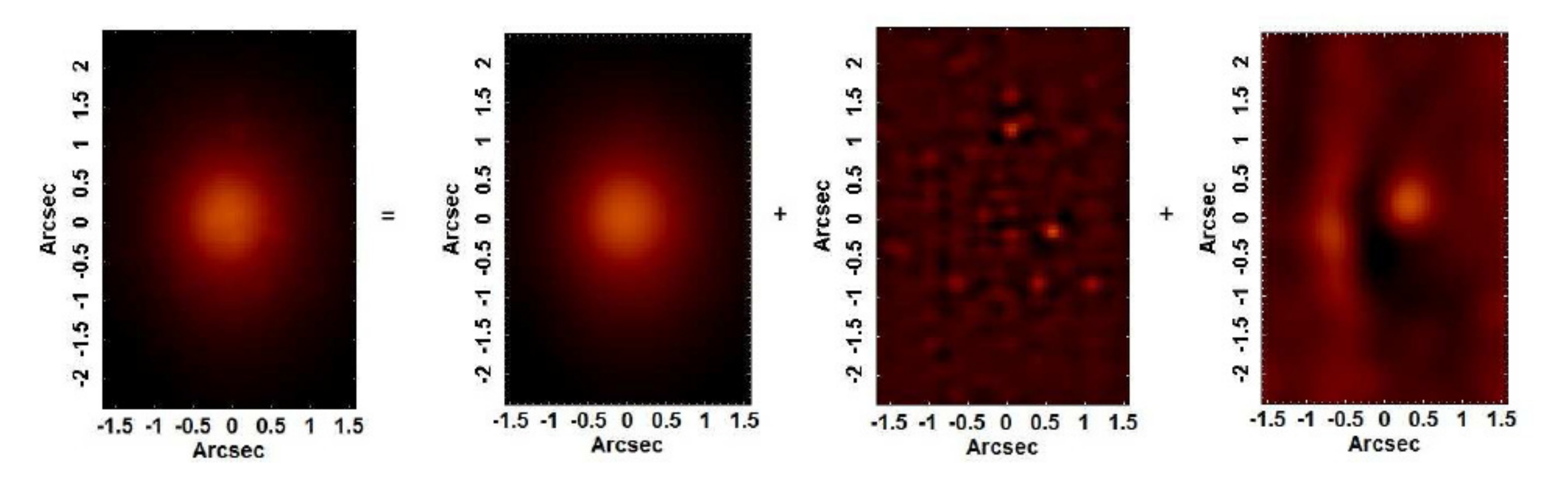}
	\caption{Average image from an original data cube is equal to the average image of the cleaned data cube plus the high frequency noise, removed with Butterworth filtering, plus the instrumental fingerprint. The analysis performed in a clean data cube is far more accurate.	\label{fig_original_filtradas}
}
\end{figure*}

After the removal of the high-frequency noises, the data cubes still have an instrumental signature of low spatial and spectral frequency. In order to remove this feature, we applied PCA Tomography to each data cube of the sample. The tomograms of this instrumental defect are given by vertical strips, while the eigenspectra show a low-frequency correlation between the wavelengths. A feature suppression of these eigenvectors is not recommendable, because they may contain correlations of high-frequency features between the wavelengths, which may be related to real characteristics, such as stellar kinematics. So, we fitted a low-order cubic spline to each eigenspectrum related to the noise in order to account only for the low-frequency correlations between the wavelengths. Using the fit results as eigenspectra, together with the original noise tomograms, we performed a feature enhancement of the fingerprint. In fact, we built a data cube that contained only the fingerprint. The final step was to subtract the fingerprint data cube from the galaxy data cube. Fig. \ref{fig_original_filtradas} shows an example of how an average image of a given data cube looks like after the removal of the high- and low-frequency artefacts. This procedure must be performed in each data cube, individually, since the variances associated with the fingerprint may vary for different data cubes.  

A complete approach about to procedure will be presented by Steiner et al. (in preparation). 

\subsection{Correction of the differential atmospheric refraction effect} \label{sec:refracao}

The Gemini telescope does not provide an atmospheric dispersion corrector. Thus, data cubes obtained with the GMOS-IFU instrument contain features caused by the differential atmospheric refraction effect, which is characterized mainly by spatial shifts along the FOV as a function of the wavelength. We calculated the shifts of each image of the data cubes relative to the redder image of the data cubes using the equations proposed by \citet{1998Metro..35..133B} to describe the light dispersion for each wavelength caused by the Earth's atmosphere, and the equations proposed by \citet{1982PASP...94..715F} to project these dispersions on the spatial dimension of the data cubes. Thus, it was possible to shift each image of the data cubes to positions that would minimize the differential refraction effect. 

\end{document}